# A sulfonitride transparent conductive thin film with ultra-high refractive index

Eugène Bertin[a*], Shima Kadkhodazadeh[a], José María Castillo-Robles[b], Finja Tadge[a], Alba Pérez Millan[a], Anat Itzhak[a], Javier Sanz Rodrigo[a], Manuel Dillenz[b], Juan Maria García Lastra[b], Søren Raza[c], Ivano E. Castelli[b], Andrea Crovetto[a*]

[a] *National Centre for Nano Fabrication and Characterization (DTU Nanolab), Technical University of Denmark*

[b] *Department of Energy Conversion and Storage (DTU Energy), Technical University of Denmark*

[c] *Department of Physics (DTU Physics), Technical University of Denmark*

**corresponding authors*

## Abstract

With the rise of AI-assisted materials screening, extraordinary properties are now frequently predicted in experimentally uncharted material systems, highlighting the need to develop new synthesis methods for unconventional materials beyond the classic bulk powder form. Here, we establish the first thin-film growth route for any metal sulfonitride compound by realizing $Zr_2SN_2$ films with a rare and compelling combination of optical and electrical properties. $Zr_2SN_2$ is transparent across most of the visible range while exhibiting a very high average refractive index of 2.95 in the visible, exceeding expectations based on conventional refractive index-bandgap scaling. Importantly, the same $Zr_2SN_2$ film shows degenerate n-type conductivity with carrier density above $10^{20}$ $cm^{-3}$ and intragrain mobility above 8 $cm^2V^{-1}s^{-1}$, approaching those of established transparent conductive oxides. $Zr_2SN_2$ thus demonstrates that strong light-matter interaction, optical transparency and electrical conductivity can be reconciled within a single material platform, revealing a new class of high-refractive-index transparent conductors.

## 1 Introduction

Inorganic compounds with two chemically distinct and experimentally challenging anions, such as sulfonitrides, phosphoiodides, and arsenoselenides, occupy a largely uncharted region of materials space, yet their mixed-anion frameworks offer an unusually broad range of tunable bonding characters [1–3]. The lack of established growth routes for these compounds limits their experimental realization and leaves their predicted properties difficult to validate. The problem is aggravated when the application of the predicted properties is in fields such as optics or electronics, which usually require materials in the form of thin films rather than the bulk crystals preferred by solid-state synthetic chemists.

An intriguing case is the sulfonitride $Zr_2SN_2$ (Figure 1(a,b)), an earth-abundant and non-toxic material expected to combine high ambipolar electrical conductivity and optical transparency in the visible [4] with ultra-high refractive index [5,6] according to computational methods. Despite its relatively narrow bandgap of 1.45 eV, indirect and dipole-forbidden transitions are predicted to shift optical absorption to above 3.0 eV [4], giving the material an unexpectedly wide transparency window. At the same time, $Zr_2SN_2$ retains the low carrier effective masses typical of narrow gap materials [7] and shows negligible defect compensation for both shallow donors and shallow acceptors, a hallmark of high ambipolar dopability [8,4]. Remarkably, a very large refractive index is also predicted in $Zr_2SN_2$ despite its wide transparency range [5,6]. While this unusual mix of optical and electrical properties has been computationally predicted, they have never been experimentally measured.

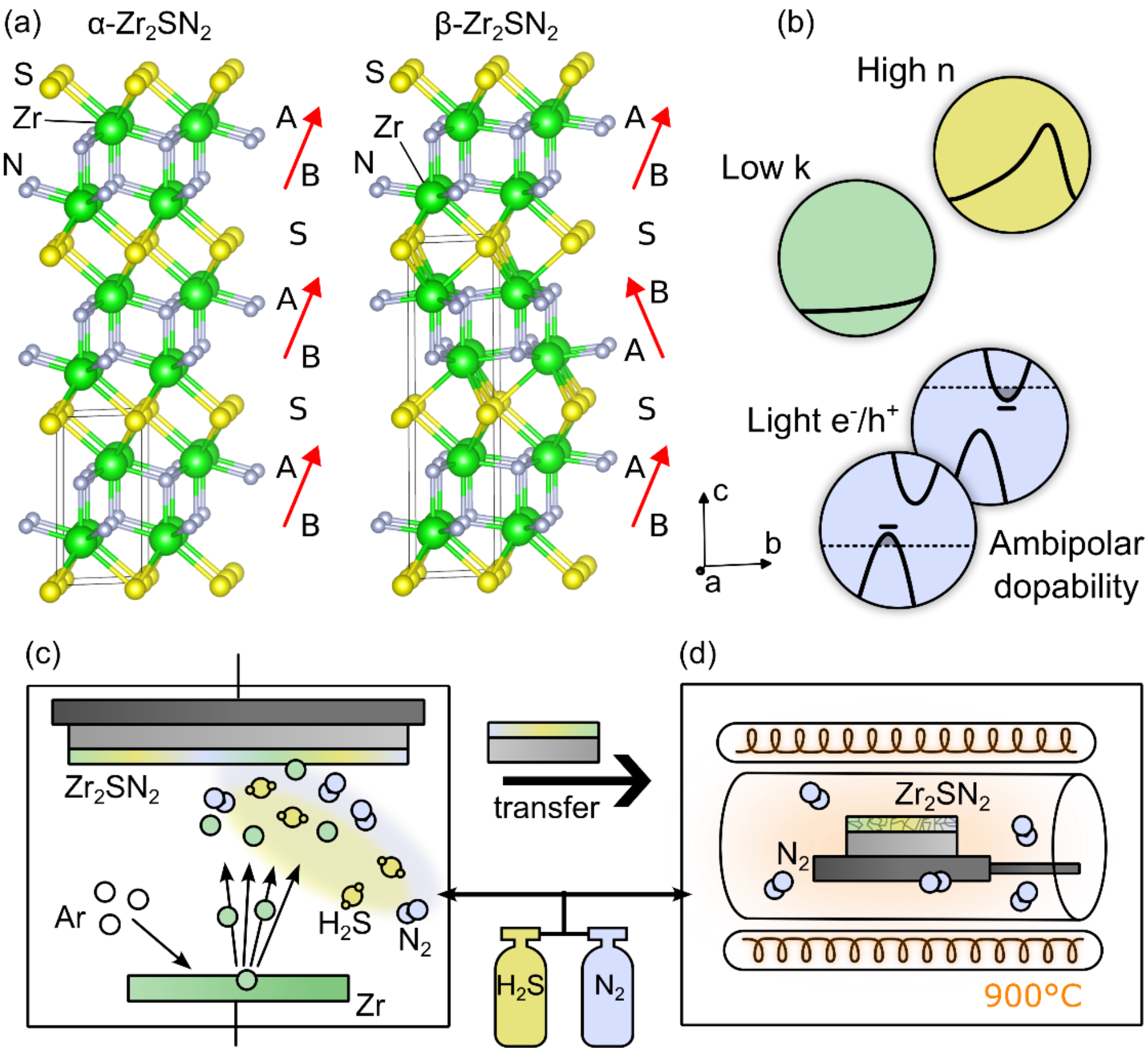

*Figure 1: (a) Crystal structure of α- and β-$Zr_2SN_2$ along the a-axis. (b) The unique combination of optical and electrical properties predicted for β-$Zr_2SN_2$. (c,d) A two-step thin-film synthesis workflow consisting of (c) sputtering of metallic Zr in a $H_2S/N_2$ reactive gas mixture to grow an amorphous film followed by (d) annealing in $N_2$ at 900 °C to crystallize the film.*

$Zr_2SN_2$ has only been synthesized as a powder using an anion metathesis reaction between ZrNCl and $A_2S$ (A = Na, K, Rb) [9]. Two structurally similar $Zr_2SN_2$ hexagonal phases, α- and β-$Zr_2SN_2$, were reported at different synthesis temperatures (Table S1). Both structures consist of a stack of $Zr_2N_2$ layers separated by S interlayers (Figure 1(a)) but differ in their stacking sequence along the c-axis. Synthesis of metal sulfonitrides in general has been limited to a few bulk powders or single crystals [10–12]. Thin-film synthesis requires S and N sources with precisely balanced chemical potentials under O-free conditions, which are difficult to achieve in conventional thin-film processes. In this work, we develop a two-step thin-film synthesis route (Figure 1(c,d)) to grow air-stable, polycrystalline $Zr_2SN_2$ and experimentally demonstrate the coexistence of the predicted optical and electrical properties. This is both the first demonstration of a thin film sulfonitride, and the first sulfonitride growth route only employing elements of the targeted compound (plus hydrogen) in the synthesis precursors.

# 2 Results

## 2.1 A synthesis route for $Zr_2SN_2$ films

Amorphous Zr-S-N films are deposited by double-gas reactive sputtering at substrate temperature of 465 °C and working pressure of 2 mTorr (Figure S13-Figure S14). Elevated temperatures and low pressures minimize the O impurity content in the film, down to below 2% (Methods 5.1, Table S5, Table S6). We measure a slightly S-rich composition with respect to the ideal 2:1:2 stoichiometry (Figure 2 ("As grown"), Table S7, Table S9).

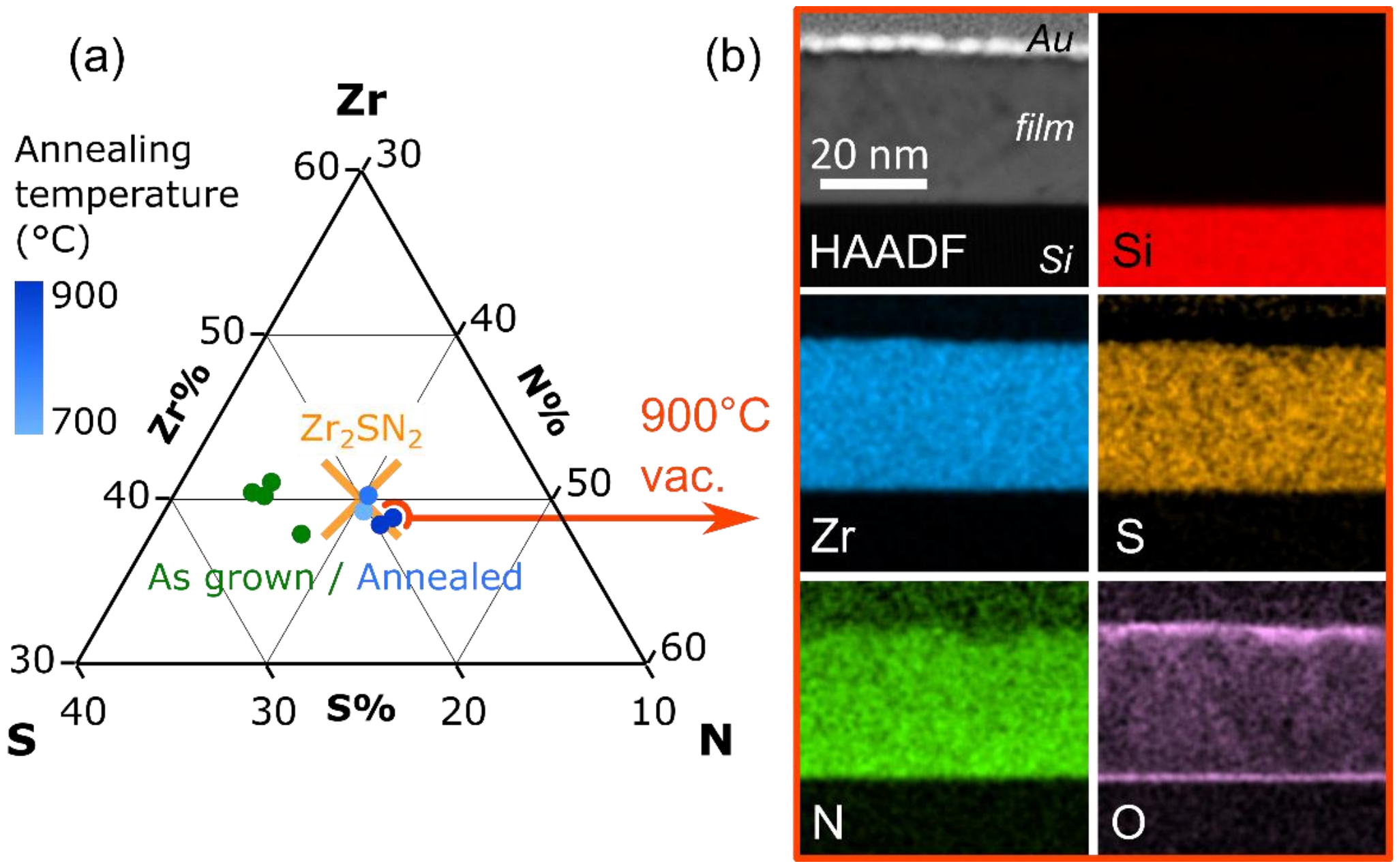

*Figure 2: (a) Chemical composition of as grown Zr-S-N films (green markers) and films annealed between 700-900 °C (blue markers) alongside the stoichiometric $Zr_2SN_2$ composition (orange cross, ternary plot center). (b) High angle annular dark field (HAADF) STEM-EDX cross sectional chemical composition maps of the film deposited on Si and annealed at 900 °C in vacuum. A gold layer is deposited on the film during lamella preparation to avoid unwanted charging of the sample.*

To induce crystallization, the amorphous films are annealed at 700 to 900 °C in $N_2$ atmosphere at 500 Torr (see Methods 5.1). Energy-dispersive X-ray spectroscopy (EDX) comparison before and after annealing (Figure 2(a), Table S9), shows a distinct shift towards the 2:1:2 stoichiometry, with a small S loss and N enrichment for all temperatures. Volatile species in excess or deficit relative to the $Zr_2SN_2$ stoichiometry are efficiently exchanged with the gas phase during annealing.

After a 900 °C thermal treatment at lower pressure (0.2 Torr, labelled "vac", vacuum), the deviation from stoichiometry remains low with Zr/N and S/N ratios within 10% of the nominal composition, and the lowest O content (< 9%) among all the annealed films (Table S9). Therefore, the $Zr_2SN_2$ composition is stable at temperatures up to 900 °C at low pressures, even in the absence of a S source in the gas phase. Spatially resolved composition mapping using scanning transmission electron microscopy (STEM) EDX (see Methods 5.1) reveals that Zr, S and N are homogeneously distributed (Figure 2(b)), with no signs of phase segregation of binary sulfide or nitrides, and with O mainly confined to the top and bottom film surfaces.

X-ray diffraction (XRD) patterns recorded in the θ/2θ geometry for the different annealing temperatures and pressures (Figure 3(a), Figure S17), show that annealing in $N_2$ enables crystallization of the amorphous films. The crystallization is evident from 800 °C and up, where a broad XRD feature from the amorphous film becomes a set of defined diffraction peaks, the strongest peaks matching the (00l) planes of α- or β-$Zr_2SN_2$. In STEM cross-sectional images (Figure 4) a clear change in contrast can be seen from the as-grown amorphous film to the annealed film, with crystal grains evident in the latter.

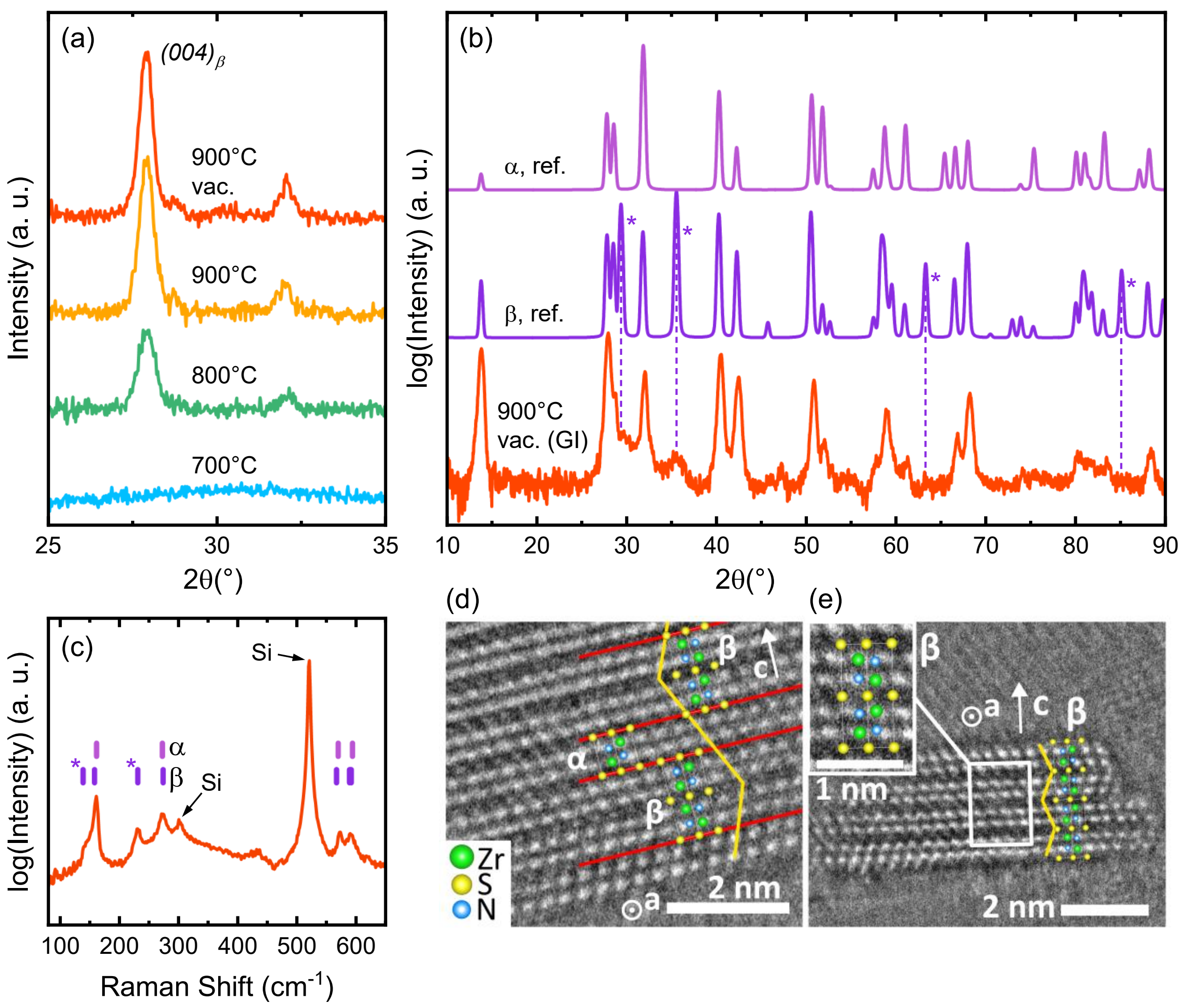

*Figure 3: (a) ϑ-2ϑ XRD pattern of $Zr_2SN_2$ film deposited on Si and annealed in $N_2$ at temperatures ranging from 700 °C to 900 °C at 500 Torr, and at 900 °C in vacuum (0.2 Torr, labelled "vac"). The (004) peak is indexed based on β-$Zr_2SN_2$ Miller indices. (b) GIXRD pattern of the film annealed at 900 °C in vacuum and calculated XRD patterns of α- and β-$Zr_2SN_2$. (c) Raman spectrum of an annealed film on Si, with theoretical peak positions of α- and β-$Zr_2SN_2$ taken from the Computational Raman Database [13] entries mp-11583 and mp-1158, respectively; (b)(c) Peaks unique to β-$Zr_2SN_2$ are marked with a purple star (*). (d)(e) High resolution annular dark field STEM images along the a-axis of the film annealed at 900 °C in vacuum. (d) A crystallite showcasing an epitaxial α/β stacking. (e) A β crystallite. Zr columns appear brightest whilst S and N give less contrast. An orange line following Zr atom columns is drawn as a guide for the eye. Red lines mark the epitaxial boundaries between α and β domains.*

Using grazing incidence (GI) XRD, we study if the film crystallizes in the α- or β-$Zr_2SN_2$ phase. Peaks that are unique to the β phase, marked with * in Figure 3(b), are broad (Figure S18), suggesting that any coherently scattering domains of β are small. However, all the peaks shared by the α- and β-$Zr_2SN_2$ phases are present and comparatively sharper, hinting that the film exhibits long-range ordering corresponding to α-$Zr_2SN_2$, or potentially to a hybrid structure between α and β. We extract a = b = 3.595 Å and c = 12.794 Å, using the Le Bail method (β-$Zr_2SN_2$ as crystal model, Figure S19, Table S10). To elucidate the detailed structure of the films, we employ complementary techniques. The Raman spectrum (Figure 3(c)) clearly exhibit modes that are unique to β-$Zr_2SN_2$ (Figure S23, Table S11), indicating a substantial presence of this phase with at least short-range ordering. High resolution STEM (Figure 3(e)) also reveals crystallites that unambiguously match the β phase. However, other areas of the film (Figure 3(d)) exhibit a composite ordering, where α and β crystal domains are stacked along the c-axis, in a sequence of coherent epitaxial α and β (Figure 3(d)). We conclude that the film crystallizes as a mixture of β-$Zr_2SN_2$ crystallites and of interlaced α/β with long-range ordering.

Despite the clear evolution of the films' crystallinity with annealing, their morphology is not strongly affected. STEM cross-sectional morphology and the atomic force microscopy (AFM) topography images (Figure 4), show that the crystallized films are smooth, compact and continuous, apart from isolated protrusions. We measure a density of 5.23 g/cm$^3$ by X-ray reflectivity (XRR) (Figure S17). The 6% lower density compared to the nominal $Zr_2SN_2$ density can be quantitatively explained by the slight Zr deficiency, indicating a compact, high-quality film with negligible porosity. The AFM average roughness ($R_a$) is 0.7 nm and 2 nm, when excluding and including the protrusions from the roughness analysis, respectively, in accordance with ellipsometry and XRR (Table S9, Figure S27). Sub-nm roughness is a feature of both amorphous and crystalline $Zr_2SN_2$ grown with our process route, and it is essential to minimize light scattering, haze and degradation of optical transparency in optical coatings [14,15].

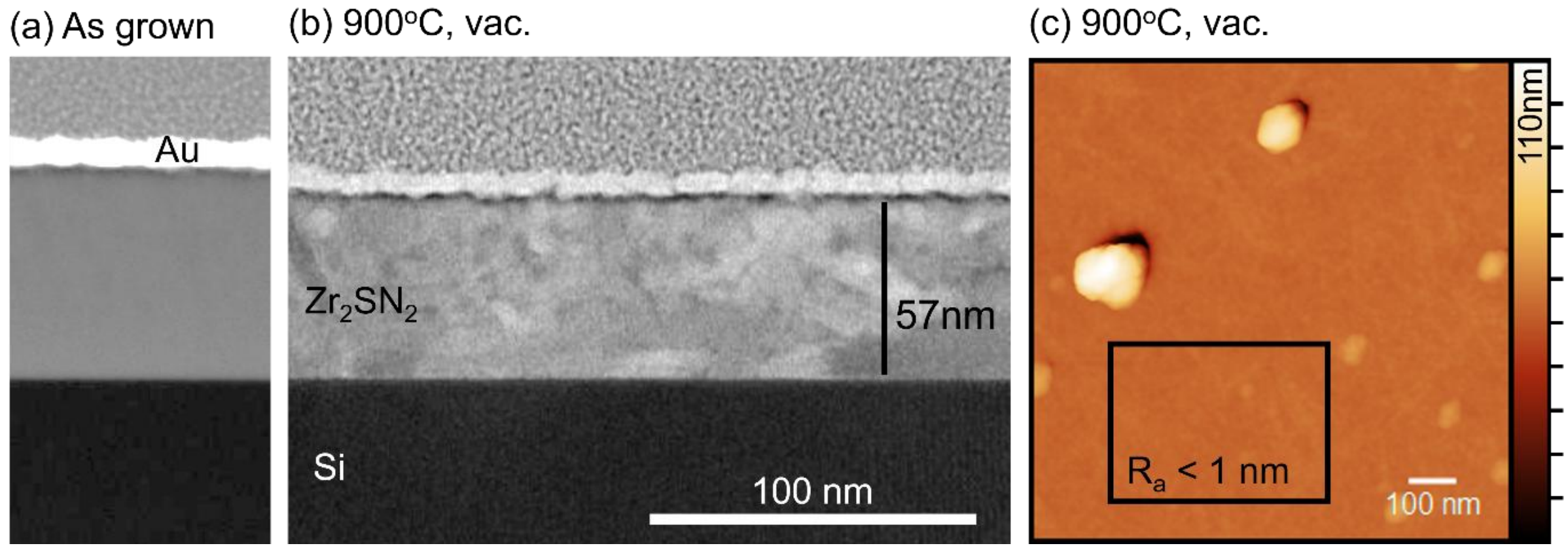

*Figure 4: TEM dark field cross-section images of the (a) as grown film on Si and (b) the film annealed at 900 °C in vacuum, in the same scale. (a,b) The observed height difference is not caused by film shrinkage during annealing but rather arises from sample-to-sample variation. (c) 1 × 1 µm$^2$ AFM image of the film annealed at 900 °C in vacuum.*

## 2.2 Determination of $Zr_2SN_2$ Unique Optoelectronic Properties

The optical properties of $Zr_2SN_2$ in the amorphous and crystalline state are markedly different. Using ellipsometry, we extract the refractive index *n* and the extinction coefficient *k* of the films grown both on fused silica (Figure 5) and on Si (Figure S31-Figure S32), ensuring consistency with reflection and transmission spectroscopy (Figure S30-Figure S36). The optical absorption onset shifts from 2.25 eV in the amorphous films to 2.9 eV in the films crystallized at 900 °C (Figure 5(a)), indicating a drastic change in the nature of optical transitions with crystallization.

To investigate the optical properties of α- and β-$Zr_2SN_2$ individually, we compute their band structures (Figure 5(c,d)) and absorption coefficients (Figure 5(b)) using Density Functional Theory (DFT) within the Heyd-Scuseria-Ernzerhof (HSE06) hybrid functional framework (Figure S3Figure S6). We find an indirect 1.43 eV bandgap from K to Γ for both structures (Table S2) and direct gaps at K of 2.62 and 2.63 eV, respectively. Optical absorption onsets, defined as the energy where the calculated absorption coefficient exceeds $10^3$ $cm^{-1}$, are at 2.80 eV and 2.99 eV, respectively (Figure S2). The absorption coefficient calculation does not include the indirect transitions expected for $Zr_2SN_2$ due to electron-phonon interactions; however, the energy of the main absorption onset is still expected to match the experimental one, due to the typical weakness of indirect transitions [16,17].

In the experimental absorption coefficient, we observe a parasitic absorption tail in the 2.6 - 2.9 eV range. It may be related to indirect transitions or excitonic effects not studied here, to residues of amorphous $Zr_2SN_2$, or to the coexistence of α and β phases, as exposed earlier. Using DFT, we clarify why the transparent regions of α- and β-$Zr_2SN_2$ extend beyond their direct band gaps. Optical absorption at the direct bandgap (K point) is minimal because of a negligible transition dipole moment $|\mu|^2$ (TDM, Figure 5, Figure S2). For the second-lowest direct transition (Γ point), the analysis of the TDM shows that it is two orders of magnitude weaker in β than in α (Figure 5), leading to negligible absorption in β-$Zr_2SN_2$ up to about 3.2 eV.

A more detailed symmetry analysis (Table S3) at the K point reveals that, even though the direct transitions are allowed by symmetry, poor orbital overlap and low joint density of states undermine the optical absorption in both phases. Furthermore, all direct transitions from the valence band maximum to the conduction bands at Γ are symmetry-allowed in the α-phase ("a" in Figure 5(c)). Combined with strong spatial overlap between the initial and final wavefunctions, these features lead to a large transition dipole moment (81.6 $D^2$) and substantial light absorption at Γ. In contrast, most direct transitions at Γ in the β-phase (from valence band maximum to $2^{nd}$, $3^{rd}$ and $4^{th}$ lowest conduction bands) are symmetry-forbidden (Table S4). The only allowed transition is the one to the lowest conduction band, which however produces a negligible $|\mu|^2$ of 1.3 $D^2$. Even though this transition is allowed by symmetry, the transition is so weak that we still label this transition as "f" in Figure 5(d). The β-phase is therefore far less effective than the α-phase at absorbing light at the band edges, a direct consequence of both symmetry constraints and weak wavefunction overlap. Hence, any α-$Zr_2SN_2$ inclusion in the film may contribute to the experimentally observed absorption tail because of its strongly allowed transition at Γ. This provides a pathway to further improving the transparency of $Zr_2SN_2$ by growing pure β-$Zr_2SN_2$ with neither α-$Zr_2SN_2$ nor amorphous inclusions.

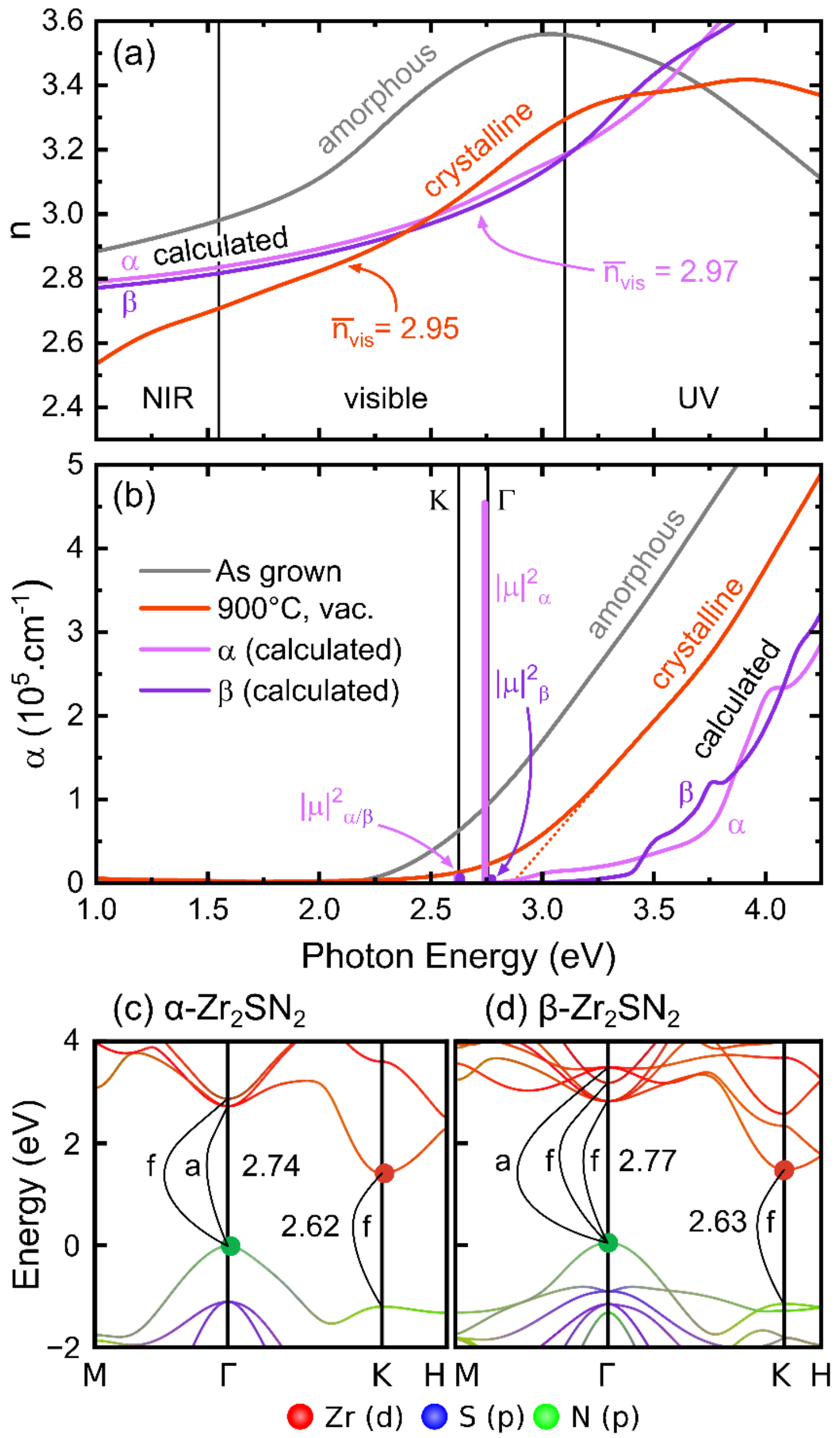

*Figure 5: (a) Refractive index and (b) absorption coefficient extracted from ellipsometry for as-grown (amorphous) and 900 °C-annealed samples (crystalline) on fused silica and calculated for α and β-$Zr_2SN_2$. (a) The average refractive index $\bar{n}_{vis}$ in the visible (1.55 – 3.10 eV photon energy range) is indicated. (b) Absorption edge estimation using the linear extrapolation of the absorption coefficient (dashed line). Transition dipole moments ($|\mu|^2$) for the lowest direct electronic transition in α- and β-$Zr_2SN_2$ (at the K point), and for the second lowest (at the Γ point). The transition strength is similarly low for the two phases at K, but differs significantly at Γ. (c,d) Calculated electronic band structure for (c) α-$Zr_2SN_2$ and (d) β-$Zr_2SN_2$. Bands are colored based on the contributions from Zr d (red), S p (blue) and N p (green) orbitals. Forbidden and weak electronic transitions are labelled f, while allowed and strong transitions are labelled a.*

Transparent materials typically have low refractive indices and are hard to electrically dope. Remarkably, $Zr_2SN_2$ breaks both correlations. The same crystalline film with a wide transparency range (Figure 5(b)) has an average refractive index $\bar{n}_{vis}$ of 2.95 in the visible in good agreement with DFT, which predicts refractive indices of 2.96 and 2.94 for α- and β-$Zr_2SN_2$, respectively. In addition, $Zr_2SN_2$ films exhibit electrical properties approaching those of metals. First, the DFT calculated effective masses are remarkably low for both carrier types (between 0.43 $m_0$ and 0.46 $m_0$ for both electrons and holes, and for both the α and β phases, Table S2). Second, Hall effect measurements on the same $Zr_2SN_2$ film that was optically characterized show degenerate n-type conductivity with carrier concentration of $(2.3 \times 10^{20} \pm 0.2)$ $cm^{-3}$ and mobility of $(0.36 \pm 0.03)$ $cm^2V^{-1}s^{-1}$ (Table S12). Due to the geometry of the Hall measurement, this mobility value includes the effect of scattering from grain boundaries and other extended defects (*intergrain* mobility). To obtain a better estimate of the intrinsic mobility of $Zr_2SN_2$, we fit our ellipsometry spectra with the Drude model in the low-photon-energy range (Figure S39) and extract an *intragrain* mobility of 8.3 $cm^2V^{-1}s^{-1}$, over 20 times higher than the intergrain mobility and in line with typical values for transparent conductive oxides [18]. We also extract a carrier concentration of $3.2 \times 10^{20}$ $cm^{-3}$, in good agreement with the Hall measurement. This suggests that the low Hall mobility is not an intrinsic material property, but is instead limited by extended defects in nanocrystalline $Zr_2SN_2$. Consequently, optimized growth and annealing conditions may substantially enhance the intergrain mobility towards the ellipsometry-measured value.

# 3 Discussion: $Zr_2SN_2$ as an emerging high-refractive-index transparent multifunctional thin film

The refractive index measured in $Zr_2SN_2$ is unusually high for a material with an absorption edge at 2.9 eV. Indeed, the Moss rule heuristically states that the sub-bandgap refractive index $n_0$ is related to the optical absorption edge $E_{edge}$ as $n_0 \propto E_{edge}^{-1/4}$ [19] (Figure 6), meaning that high refractive index and high transparency are generally mutually exclusive in materials.

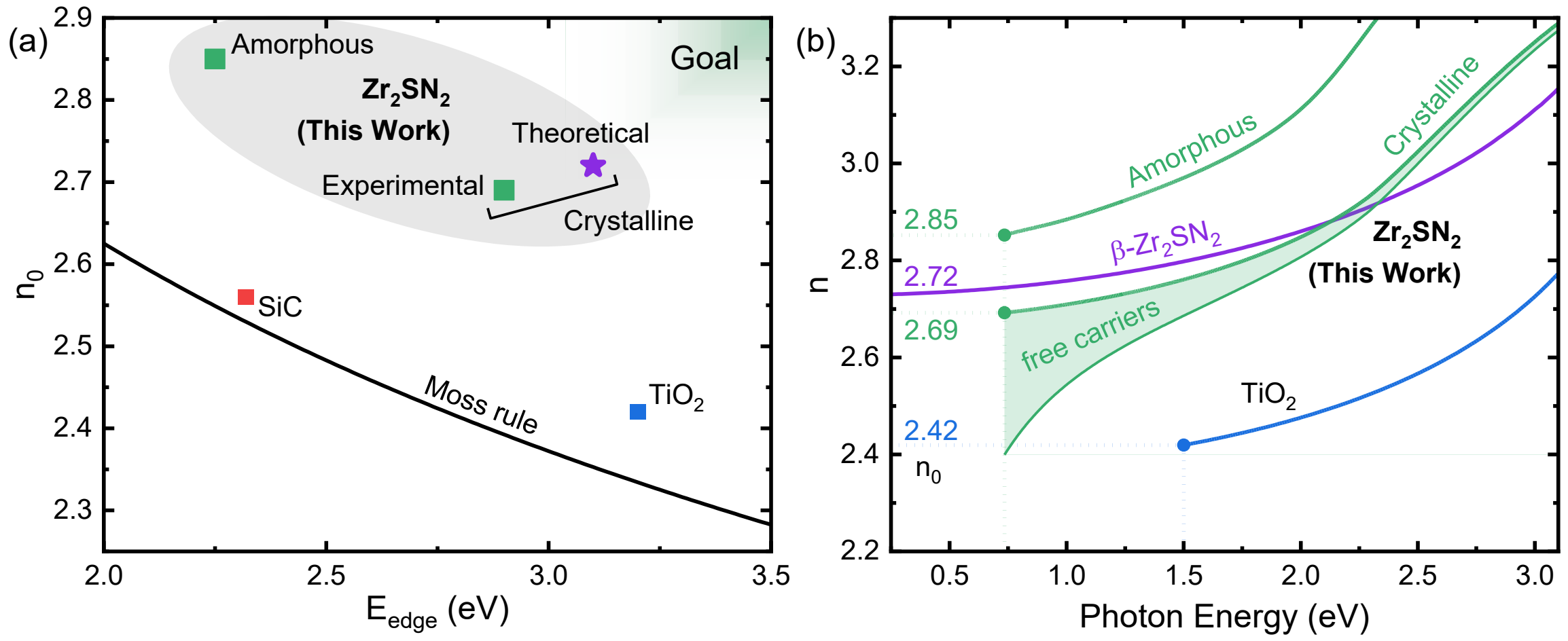

*Figure 6: (a) Moss rule plot: sub-bandgap refractive index $n_0$ as a function of the optical absorption edge $E_{edge}$, for $Zr_2SN_2$ and other high-n transparent thin films (anatase-$TiO_2$ [20], amorphous-SiC [21]). For $Zr_2SN_2$, both the experimental and theoretical (β-$Zr_2SN_2$) $n_0$ /$E_{edge}$ pairs are given. Moss rule $n_0^4 E_{edge} = 95\ eV$. (b) Experimental refractive index spectra in the visible range for crystalline and*

*amorphous $Zr_2SN_2$ and anatase-$TiO_2$ thin films, alongside the theoretical index for β-$Zr_2SN_2$. Vertical drop lines indicate the photon energies used to evaluate the sub-bandgap index $n_0$. For crystalline-$Zr_2SN_2$, we remove the contribution of free carriers to the refractive index (modeled using a Drude model) to evaluate $n_0$.*

We report the sub-bandgap refractive index $n_0$ as a function of absorption edge $E_{edge}$ (Figure 6(a)) and the refractive index spectra (Figure 6(b)) for our $Zr_2SN_2$ films and for state-of-the-art high-refractive-index thin films [21], such as amorphous SiC and anatase $TiO_2$ [20,21]. Crystalline $Zr_2SN_2$ surpasses the refractive index of anatase $TiO_2$ while displaying a relatively high absorption edge of 2.9 eV. The HSE06-calculated absorption coefficient of β-$Zr_2SN_2$ is on par with anatase $TiO_2$ (Figure S2, [22]), suggesting that a pure β-$Zr_2SN_2$ film could reach the transparency of $TiO_2$. We note that the amorphous $Zr_2SN_2$ film also displays super-Mossian behavior (Figure 6(a)), with appealing transparency below its absorption edge of 2.25 eV and significantly higher sub-bandgap refractive index than amorphous SiC thin films.

Interestingly, design criteria for transparent conductors [23] and high-refractive-index materials [24,25] converge on similar chemical strategies½, such as incorporating early transition metals like Zr, yet they impose fundamentally different requirements on the electronic structure. High refractive index in the transparent region requires strong and numerous oscillators above the absorption edge, associated with large joint density of states and large TDMs [21, 27]. These conditions typically emerge from flat, weakly dispersive electronic bands with heavy charge carriers, which are unfavorable for electrical transport. In contrast, transparent conductors benefit from strongly dispersive electronic bands with light charge carriers to enable high mobility [26], which result in lower joint density of states and reduce the refractive index. This creates a trade-off between refractive index and electrical conductivity in transparent materials.

In β-$Zr_2SN_2$, this trade-off is mitigated because optical and electrical properties originate from different regions of the band structure. Optical absorption increases rapidly above 3.4 eV due to large TDMs and high density of states in the valence and conduction band (Figure S3-Figure S6), creating conditions for a high refractive index in the visible. At lower energies, highly dispersive electronic bands host charge carriers with low effective masses, enabling high mobility, while dipole- and momentum-forbidden transitions with weak optical activity suppress most absorption in the visible range. Finally, the narrow fundamental band gap of $Zr_2SN_2$ facilitates ambipolar dopability [4,8]. Together, these features enable the coexistence of high refractive index, optical transparency and electrical conductivity.

$Zr_2SN_2$ combines high transparency, carrier concentration, and carrier mobility, opening opportunities for β-$Zr_2SN_2$ as a next-generation transparent conductor with applications in solar cells and display technologies. Although the present films are n-type, computational work indicates promising prospects for ambipolar dopability in β-$Zr_2SN_2$ [4]. If realized, ambipolar dopability in $Zr_2SN_2$ would represent a significant advance for transparent electronics [26].

Beyond transparent conductors, the simultaneous achievement of low absorption and high refractive index is foundational to modern photonics [27–29]. This positions β-$Zr_2SN_2$ as a compelling candidate for integrated photonic circuits, enabling device miniaturization and improving the efficiency of metasurfaces and advanced display architectures. For applications in red and near-infrared photonics, amorphous $Zr_2SN_2$ may be even more suitable, with a refractive index more than 0.2 higher than its crystalline counterpart and a CMOS-compatible synthesis temperature.

Crucially, the coexistence of high transparency, high refractive index, and high mobility in β-$Zr_2SN_2$ enables functionalities that are not accessible with conventional transparent conductors or photonic materials. For example, it opens the possibility of electrically modulating a high refractive index material through controlled changes in carrier concentration [30]. This could enable tunable optical elements as well as devices where the same material layer provides both optical and electrical functionality, thereby eliminating the need for lossy metal contacts in metasurfaces and integrated photonic devices.

# 4 Conclusions

This work establishes the first synthesis route to sulfonitride thin films, with deposition of $Zr_2SN_2$ via dual-gas reactive sputtering and rapid thermal annealing. Crystallization of amorphous $Zr_2SN_2$ results in an extension of its transparency window up to 2.9 eV photon energy, while maintaining a high refractive index of 2.95 in the visible and optical-grade smoothness. DFT calculations attribute this remarkable property trade-off to an unusually wide range of optical transitions that are either symmetry-forbidden or with low dipole strength. These features render $Zr_2SN_2$ unique among known high-n, low loss material such as anatase $TiO_2$ and amorphous SiC. Beyond their optical properties, $Zr_2SN_2$ films support degenerate n-type doping with carrier densities above $10^{20}$ $cm^{-3}$ and intragrain mobilities above 8 $cm^2V^{-1}s^{-1}$. Together, these properties position $Zr_2SN_2$ as a new multifunctional material for transparent electronics, flat optics, and integrated photonics, and demonstrate that the broader family of metal sulfonitrides is now experimentally accessible in thin-film form.

# 5 Methods

## *5.1 Synthesis: Reactive sputtering step*

Amorphous Zr-S-N films were deposited by reactive magnetron sputtering. The deposition apparatus was a custom sputtering chamber for multi-anion materials (Kurt J. Lesker), described in greater detail in Ref. [31]. A metallic Zr target (Testbourne, 99.95% purity, 2-inch diameter) was sputtered at 70 W radio-frequency power (RF, 13.56 MHz) using an $Ar/N_2/H_2S$ gas mix with relative concentration 96/3/1 %, at a total pressure of 2 mTorr unless stated otherwise. Films were grown on n-type Si and fused silica substrates clamped to a metallic platen maintained at 465 °C, unless otherwise stated. The deposition rate was about 1 nm/min.

## *5.2 Synthesis: Annealing step*

As-deposited, amorphous Zr-S-N films were crystallized by rapid thermal annealing in a MILA-5050 Mini lamp furnace (Advance Riko). The samples were placed on a SiC-coated graphite susceptor and annealed in $N_2$ with nominal ramp rates of 100 °C/min and a hold time of 5 min at the targeted annealing temperature unless otherwise specified. The temperature was measured by a thermocouple placed in a hole inside the susceptor. The ramp rate during cool-down was in practice slower below 500 °C due to natural cooling (Figure S15). Unless otherwise specified, annealing experiments were run at 900 °C at a $N_2$ pressure of 0.2 Torr. They are labelled "vac." (vacuum) in the text and figures. Other annealing experiments were run in 500 Torr of $N_2$ at different temperatures between 700 °C and 900 °C. In all cases, the $N_2$ gas was continuously flown through the furnace at 100 sccm, and the pressure inside the furnace was controlled by a butterfly valve placed between the furnace and the pump. The different samples considered in this study, and the measurements that they undertook, are summarized in Table S8.

### *5.3 Characterisation*

The depth-averaged elemental composition of the thin films was determined by energy dispersive X-ray spectroscopy (EDX) using a FEI Quanta FEG 250 scanning electron microscope (SEM) equipped with an Oxford Instruments EDX detector at a beam voltage of 20kV. The elemental composition was extracted using the LayerProbe software (Oxford Instruments), which extracts a self-consistent film composition and mass thickness by fitting EDX spectra to a model consisting of a flat film on a substrate.

X-ray diffraction (XRD) and reflectivity (XRR) were recorded in parallel beam geometry using a Rigaku SmartLab system equipped with a high-power 9 kW Rotating Anode Cu Kα source and a HyPix-3000 2D detector. For XRD experiment conducted in θ/2θ geometry, a CBO-f focusing element coupled with 0.5 mm length limiting slit were used as incident optics. Grazing incidence XRD was recorded at an incident angle of 1°. XRR was fitted using the GenX software v3.7 [32].

A Thermo Fisher Helios Hydra plasma focused ion beam (FIB) – scanning electron microscope (SEM) instrument was used to prepare a thin lamella of the thin film. Scanning TEM (STEM) images were recorded on this lamella using a Thermo Fisher Spectra Ultra (S)TEM instrument. The microscope is equipped with aberration correction on the probe forming lenses and an Ultra-X energy dispersive x-ray spectroscopy (EDX) detector. STEM images of the sample were recorded at 300 keV electron beam energy and with an electron probe convergence angle of 30 mrad.

Raman spectra were recorded using a Renishaw InVia confocal Raman microscope with a 532 nm laser. No laser-induced damage was observed.

Surface topography was characterized using a Dimension Icon-PT atom force microscope (AFM), from Bruker AXS, and analysed using Gwyddion [33].

Spectroscopic ellipsometry data was recorded using a VASE M2000-XI Ellipsometer from J.A. Woollam. Analysis of the data was performed using the Complete EASE software, by fitting the data with a Kramers-Kronig-compliant B-spline model if not stated otherwise. Surface roughness was modeled using Bruggeman's effective medium approximation. To validate the optical functions derived from ellipsometry data, an Agilent Cary 7000 instrument with a Universal Measurement Accessory was used to measure the transmission T and reflection R of the thin film sample deposited on the transparent fused silica substrate. For samples deposited on Si, an integrating sphere was used to measure total (direct and diffuse) UV-Vis-NIR reflection R spectra.

The long-range (intergrain) electrical properties of the films were characterized in the substrate plane by Hall effect measurements. The samples were measured with the DC Hall method in the van der Pauw geometry using In contacts and spring-loaded probes in a Lake Shore M91 FastHall station equipped with a 0.5 T permanent magnet. Values of carrier concentration, mobility, and conductivity type were confirmed by magnetic field reversal.

The short-range (intragrain) electrical properties of the films were characterized by fitting ellipsometry spectra with a model consisting of a Drude oscillator to describe free carriers, and a single Tauc Lorentz oscillator to describe interband transitions. Carrier concentration and mobility were the fitted parameters of the Drude oscillator, while the effective mass of electrons was fixed to the value calculated by DFT with the HSE06 functional (see below).

All characterization was conducted at room temperature and under ambient conditions, except for the techniques requiring vacuum (EDX and STEM).

### *5.4 Computational details*

Calculations of frequency-dependent linear optical properties for $\alpha$-$Zr_2SN_2$ and $\beta$-$Zr_2SN_2$ phases were performed using density functional theory (DFT) with the projector-augmented wave (PAW) method [34] as implemented in VASP [35,36] and with the aid of the ASE package [37]. The HSE06 [38] hybrid functional (25% Hartree-Fock exchange, screening parameter 0.208 $Å^{-1}$) was employed in all calculations with 550 eV plane wave energy cutoff, Γ-centered k-point meshes, and the tetrahedron method with Blöchl corrections. The structures were relaxed until all forces were below 0.005 eV/Å forces, yielding optimized lattice parameters for the α-phase (space group P-3m1) and the β-phase (space group $P6_3mm$). Band structures were calculated on standard high-symmetry paths with 353 and 308 k-points, respectively, and were plotted with the SUMO package [39]. Linear optical properties were extracted using the LOPTICS framework to compute the imaginary dielectric tensor $\varepsilon_2(\omega)$ from momentum matrix elements on a dense 16×16×8 Γ-centered k-point mesh. The real part of the dielectric tensor $\varepsilon_1(\omega)$ was derived by Kramers-Kronig transformation. Refractive index n, extinction coefficient k, and absorption coefficient $\alpha(\omega)$ were derived by standard relations. Transition dipole moments were derived from the momentum operator matrix elements $\langle c|p|v\rangle$ as implemented in VASPKIT [40]. For the group-theoretical analysis, symmetry eigenvalues and corresponding irreps were computed using the IrRep Python package [41]. Extended computational details available in the SI.

## 6 Acknowledgement

We thank Andriy Zakutayev, Nicolas Stenger, Marc Bernet I Gracia, Antoine Letoublon and Olivier Durand for the insightful discussions. We are grateful to Luc Villibord and Evgeniy Shkondin for assistance with characterization and to Roy Cork for his technical support. This work was funded by the European Union (ERC, IDOL, 101040153). Views and opinions expressed are, however, those of the authors only and do not necessarily reflect those of the European Union or the European Research Council. Neither the European Union nor the granting authority can be held responsible for them. Anat Itzhak acknowledges the support from the European Union's Horizon Europe MSCA grant No 101152844 (CAPSELL). This work was supported in part by a research grant (42140) from VILLUM FONDEN. S. Raza acknowledges support by the Novo Nordisk Foundation (NNF24OC0096142, NNF25OC0095542). J. M. Castillo-Robles, M. Dillenz, J. M. García Lastra, and I. E. Castelli thankfully acknowledge the computer resources at MareNostrum and the technical support provided by Barcelona Supercomputing Center (FI-2025-1-0023, FI-2025-2-0053), as well as the Danish e-Infrastructure Cooperation (DeiC) for awarding this project access to the LUMI supercomputer, owned by the EuroHPC Joint Undertaking, hosted by CSC (Finland) and the LUMI consortium through the project DeiC-DTU-N5-2025140. The authors acknowledge support from the Novo Nordisk Foundation Data Science Research Infrastructure 2022 Grant: A high-performance computing infrastructure for data-driven research on sustainable energy materials, Grant no. NNF22OC0078009.

# 8 Supplementary information

## 8.1 Scarcity of the literature on sulfonitrides

The scarcity of the reports on the chemistry of metal sulfonitrides (SNs), and the limited understanding of their chemical and physical properties is striking. While phosphosulfides and arsenosulfides are relatively well explored [3], a query for experimentally reported inorganic ternary SNs (excluding O, H, C and halogens) only yields 31 entries in Inorganic Crystal Structure Database [42], most of them with rare-earth metals. The same query in the computational database Materials Project [5] only identifies 13 materials. No naturally occurring sulfonitride minerals are known [43].

Early on in the 60s, $Pd(NS_3)_2$, $U_2SN_2$, $Th_2SN_2$ structures have been determined [44]. Later, many other SNs, such as $La_3S_3N$, $Ce_3S_3N$ or $Nd_3S_3N$, were discovered by Lissner et al. in a series of articles [45,46]. More recently, $Hf_2SN_2$ was investigated in 2013 in the search for new superconductors, and some of its properties were computational and experimental determined [2]. $Y_4S_3N_2$ is Materials Project's single ternary sulfonitride predicted stable and not experimentally reported.

## 8.2 Comment on the similarities between α- and β-$Zr_2SN_2$: Computational Study

### 8.2.1 Crystal Structure and Synthesis

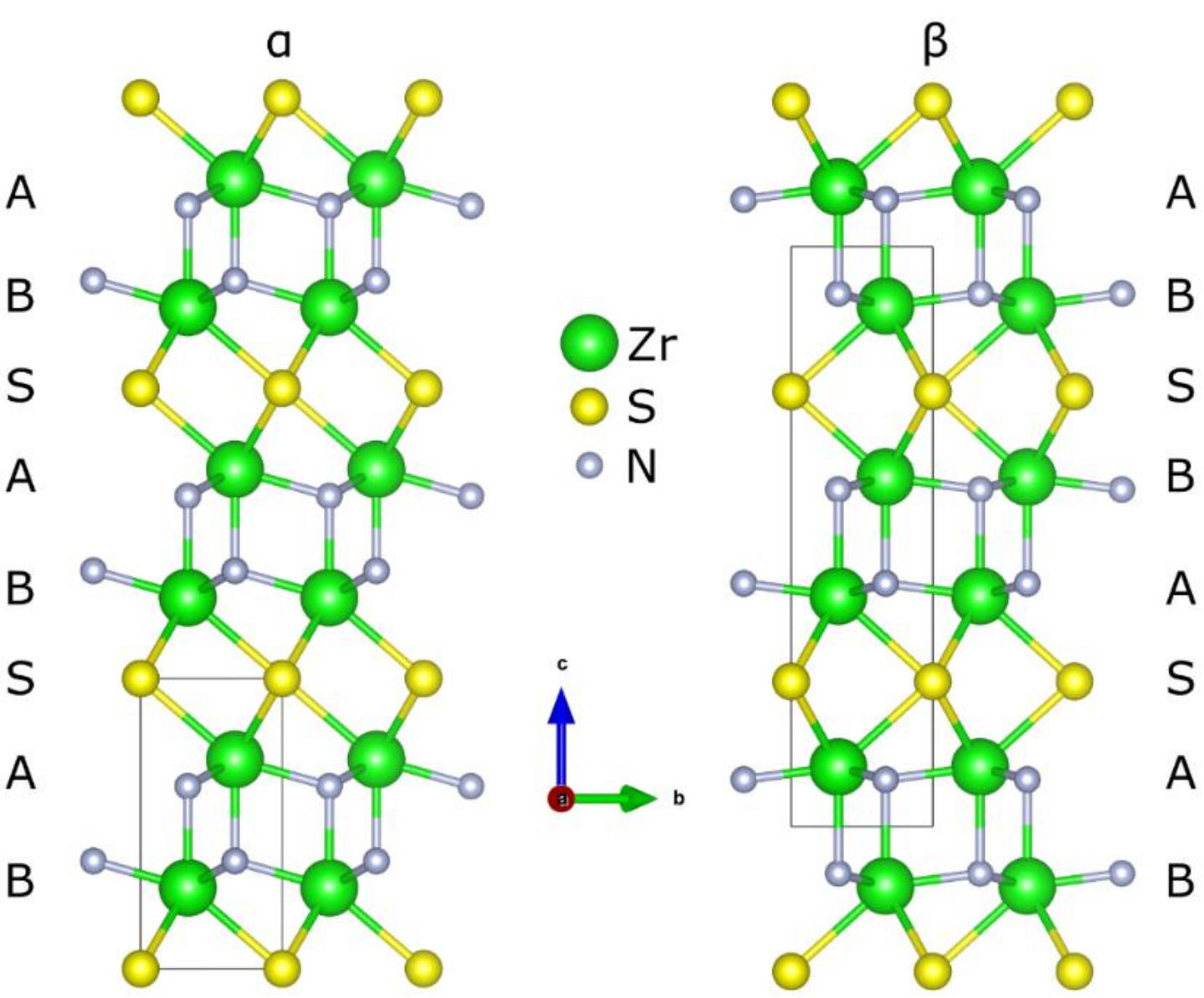

*Figure S1: Crystal structure representation for α (left) and β (right) structures of $Zr_2SN_2$, as viewed in the <100> zone-axis, equivalent to the <110> and <010> directions for the discussed structure.*

β-$Zr_2SN_2$ was first synthesized in 2003 by Stoltz et al. [4] (space group: *P*$6_3$/*mmc*, Inorganic Crystal Structure Database (ICSD) entry: 96971) in the powder form by anion metathesis reaction between ZrNCl and $A_2S$ (A = Na, K, Rb). It was additionally found that, although β-$Zr_2SN_2$ is the preferred structure above 850 °C, the α-$Zr_2SN_2$ phase (space group: P-3m1, ICSD entry: 96970) was found to crystallize instead below 800 °C. α-$Zr_2SN_2$ adopts the layered structure $La_2O_2S$. The two hexagonal structures α- and β-$Zr_2SN_2$ are closely related (Figure S1);

they can both be described as a stack of $Zr_2N_2$ layer separated by S interlayers. They however differ in the stacking sequence in the c-axis direction, with AB-S-AB-S-AB stacking for the α and AB-S-BA-S-AB stacking for the β, where S represent S interlayers. More quantitively, the β phase differs from the α by an a/2 shift of the interleaving $Zr_2N_2$ layers, leading to a doubling of the unit cell in the c-axis direction. $(hk2l)_β$ planes of the β phase are therefore symmetrically equivalent to $(hkl)_α$ of the α phase, where the subscript stands for the reference unit cell to define the Miller indices.

*Table S1: Crystallographic and synthesis parameters of the two hexagonal $Zr_2SN_2$ polymorphs. The α- and β-phases share closely related layered structures but differ in stacking sequence along the c-axis, resulting in a doubling of the unit cell in the c-direction for the β-phase. Both phases were first synthesized in powder form by Stolz et al. (2003) via anion metathesis reactions between ZrNCl and $A_2S$ (A = Na, K, Rb)*

| | Space group | ICSD entry number | Structure type | Symmetry | Synthesis temperature (° C) [9] |
|---|---|---|---|---|---|
| α-$Zr_2SN_2$ | $P\bar{3}m1$ | 96970 | $La_2O_2S$ | Trigonal | < 850 |
| β-$Zr_2SN_2$ | $P6_3/mmc$ | 96971 | $Zr_2SeN_2$ | Hexagonal | > 850 |

### 8.2.2 Extended computational details

Density Functional Theory (DFT) calculations have been used to investigate the frequency-dependent linear optical properties of the α-$Zr_2SN_2$ and β-$Zr_2SN_2$ phases. The calculations have been performed using the Vienna Ab initio Simulation Package (VASP) package [35,36] and the Atomistic Simulation Environment (ASE) [37]. All the computed crystal structures were visualized using VESTA[47]. All electronic and optical structures have been calculated using the HSE06 hybrid functional [38], incorporating 25% exact Hartree-Fock exchange and a screening parameter of 0.208 $Å^{-1}$ to accurately describe the semiconducting band gaps. We employed the recommended Projector Augmented Wave (PAW) [34] pseudopotentials with an energy cutoff of 550 eV. The tetrahedron method with Blöchl corrections was used for all electronic structure and structural relaxation calculations. The structural models for the $\alpha$-and $\beta$-$Zr_2SN_2$ phases were defined using their respective hexagonal lattice vectors. All ionic positions were fully relaxed until the interatomic forces dropped below a convergence threshold of 0.005 eV/Å, using a $\Gamma$-centered $10 \times 10 \times 5$ and $10 \times 10 \times 3$ $k$-point meshes for the $\alpha$ and $\beta$-phase structures. Post-optimization, the $\alpha$-$Zr_2SN_2$ phase (space group $P\bar{3}m1$, No. 164) yielded lattice parameters of $a = b = 3.595$ Å and $c = 6.372$ Å. The $\beta$- $Zr_2SN_2$ phase (space group $P6_3/mmc$, No. 194) yielded lattice parameters of $a = b = 3.596$ Å and $c = 12.721$ Å. The electronic band structure and optical properties post-processing were obtained using SUMO [39] and VASPKIT [40] software packages.

The high-symmetry path in reciprocal space used for the electronic band structure calculation was generated using the SUMO package. Based on the hexagonal crystal symmetry for both $\alpha$ and $\beta$ phases, the Brillouin zone was sampling along the trajectory $\Gamma \rightarrow A \rightarrow L \rightarrow M \rightarrow \Gamma \rightarrow K \rightarrow H \rightarrow A$. The path coordinates are defined as $\Gamma(0,0,0)$, $A(0,0,1/2)$, $L(0,1/2,1/2)$, $M(0,1/2,0)$, $K(-1/3,2/3,0)$, and $H(-1/3,2/3,1/2)$. To sample the paths, a total of 353 and 308 $k$-points were used for the $\alpha$ and $\beta$-phase structures, respectively, and with their generation specifically adapted to hybrid functional calculations.

The linear optical properties were determined via the LOPTICS framework by calculating the frequency-dependent imaginary part of the dielectric tensor, $\varepsilon_{\alpha\beta}^{(2)}(\omega)$, which represents the electronic transitions between occupied and unoccupied states:

$$\varepsilon_{\alpha\beta}^{(2)}(\omega) = \frac{4\pi^2 e^2}{\Omega} \lim_{q\to 0} \frac{1}{q^2} \sum_{c,v,k} 2\, w_k \delta(\epsilon_{ck} - \epsilon_{vk} - \hbar\omega) \times \langle u_{ck+e_\alpha q} | u_{vk} \rangle \langle u_{vk} | u_{ck+e_\beta q} \rangle$$

where $\omega$ is the frequency, $q$ is the volume of the unit cell, $w_k$ is the $k$-point weight, $u_{ck}$ and $u_{vk}$ are periodic component of the electronic wavefunction at $k$, and $c$ and $v$ refer to the conduction and valence band states, respectively. The real part of the dielectric function, $\varepsilon_{\alpha\beta}^{(1)}(\omega)$, was subsequently derived through the Kramers-Kronig transformation:

$$\epsilon_{\alpha\beta}^{(1)}(\omega) = 1 + \frac{2}{\pi} P \int_0^\infty \frac{\epsilon_{\alpha\beta}^{(2)}(\omega')\omega'}{\omega'^2 - \omega^2 + i\eta} d\omega'$$

where $P$ is the principal value operator [48]. A complex shift $\eta$ of $10^{-6}$ eV was applied to the frequency grid. From the complex dielectric function, the macroscopic optical constants, including the refractive index $n(\omega)$, and the absorption coefficient $\alpha(\omega)$, were calculated as follows:

$$n(\omega) = [\frac{\sqrt{\epsilon_{\alpha\beta}^{(1)}(\omega)^2 + \varepsilon_{\alpha\beta}^{(2)}(\omega)^2} + \varepsilon_1(\omega)}{2}]^{1/2},$$

$$\alpha(\omega) = \frac{2\omega k(\omega)}{c},$$

where $k(\omega)$ is the extinction coefficient, and $c$ the speed of light. For the optical calculations, a dense $\Gamma$-centered of $16 \times 16 \times 8$ $k$-point mesh was used to sample the Brillouin zone.

Finally, the transition dipole moment and its square, which dictates the intensity of these optical transitions, is calculated from the matrix elements of the momentum operator $\mu_{cv}$ as:

$\mu_{cv} = \frac{i\hbar e}{m_0(\epsilon_{ck} - \epsilon_{vk})} \langle \psi_{ck} | \hat{p} | \psi_{vk} \rangle$, as implemented in VASPKIT [40].

### *8.2.3 Predicted properties*

Table S2 shows the predicted properties for α- and β-$Zr_2SN_2$, as found in Materials Project (as calculated by Woods-Robinson et al. [4]) and as calculated at the HSE06 level in our work. α- and β-$Zr_2SN_2$ have nearly identical formation enthalpy at 0K for both theory levels, with a small difference in formation energy of 0.013 eV/atom, in favor of β-$Zr_2SN_2$. Effective masses of holes and electrons are nearly identical. One can see in Figure S2 that their absorption coefficients are also similar at both theory levels, the main difference being the absorption onset as discussed below.

*Table S2: Formation enthalpy ($\Delta H_f$), direct and indirect bandgap $E_G$, electron $m_e$* and hole effective mass $m_h$* and average absorption coefficient in the visible (400-800nm) $\alpha_{avg,vis}$ for α- and β-$Zr_2SN_2$ at the PBE level (taken from Materials Projects entries mp-553875 and mp-11583) and at the HSE06 level (our calculation). $\Delta H_f$ is only consistent for the same theory level and cannot be compared across from PBE to HSE.*

| | α-$Zr_2SN_2$ | | β-$Zr_2SN_2$ | |
|---|---|---|---|---|
| | mp-553875 | Our Calc. | mp-11583 | Our Calc. |
| | PBE | HSE | PBE | HSE |
| $\Delta H_f$ (eV/atom) | -1.878 | -10.961 | -1.889 | -10.974 |
| Energy Transitions (eV) | | | | |
| $E_G^i$ (indirect) | 0.55 | 1.43 | 0.56 | 1.43 |
| $E_G^d$ (direct) | 1.61 | 2.62 | 1.62 | 2.63 |
| $m_e$*/$m_h$* | 0.404/0.443 | 0.45/ 0.43* | 0.409/0.442 | 0.46/0.45* |
| $\alpha_{avg,vis}$ ($cm^{-1}$) | 2680 | < 0.1 | 2490 | < 0.1 |

*calculated at the PBE level, using harmonic mean

As displayed in Table S2 and Figure S2, α- and β-$Zr_2SN_2$ are very similar. They exhibit a 1.43 indirect bandgap from Γ to K, as well as a 2.62 eV and 2.63 eV direct bandgap at K, for α- and β-$Zr_2SN_2$, respectively. This direct transition at K yields near zero absorption. Another direct transition Γ can be seen at energies of 2.74 eV and 2.77 eV, for α- and β-$Zr_2SN_2$, respectively. We define the absorption onset as the energy where the absorption reaches $10^3$ $cm^{-1}$ (Figure S2 inset), one finds absorption onset of 2.80 eV and 2.99 eV for α and β, respectively. A similar treatment with a threshold of $2.10^3$ $cm^{-1}$ yields finds absorption onset of 2.83 eV and 3.10 eV for α and β, respectively.

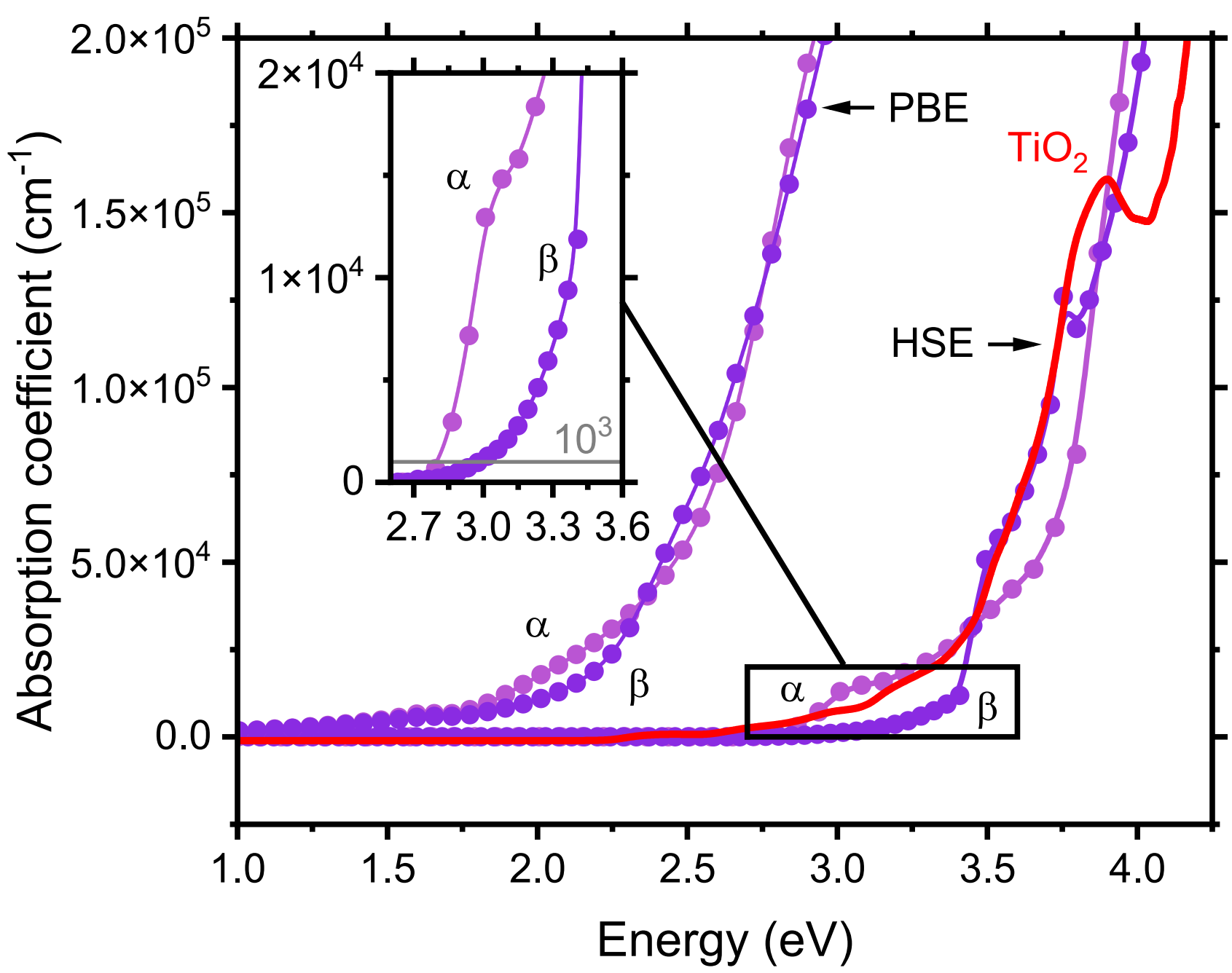

*Figure S2 Calculated absorption coefficient for α and β at the PBE level (α: mp-553875 and β: mp-11583) and at the HSE06 level (our calculation). Inset: zoom of the absorption onset region (2.6 – 3.6 eV) and*

*threshold value of $10^3$ $cm^{-1}$. Overlayed HSE06 calculated absorption coefficient of anatase $TiO_2$, reproduced from Ref. [22].*

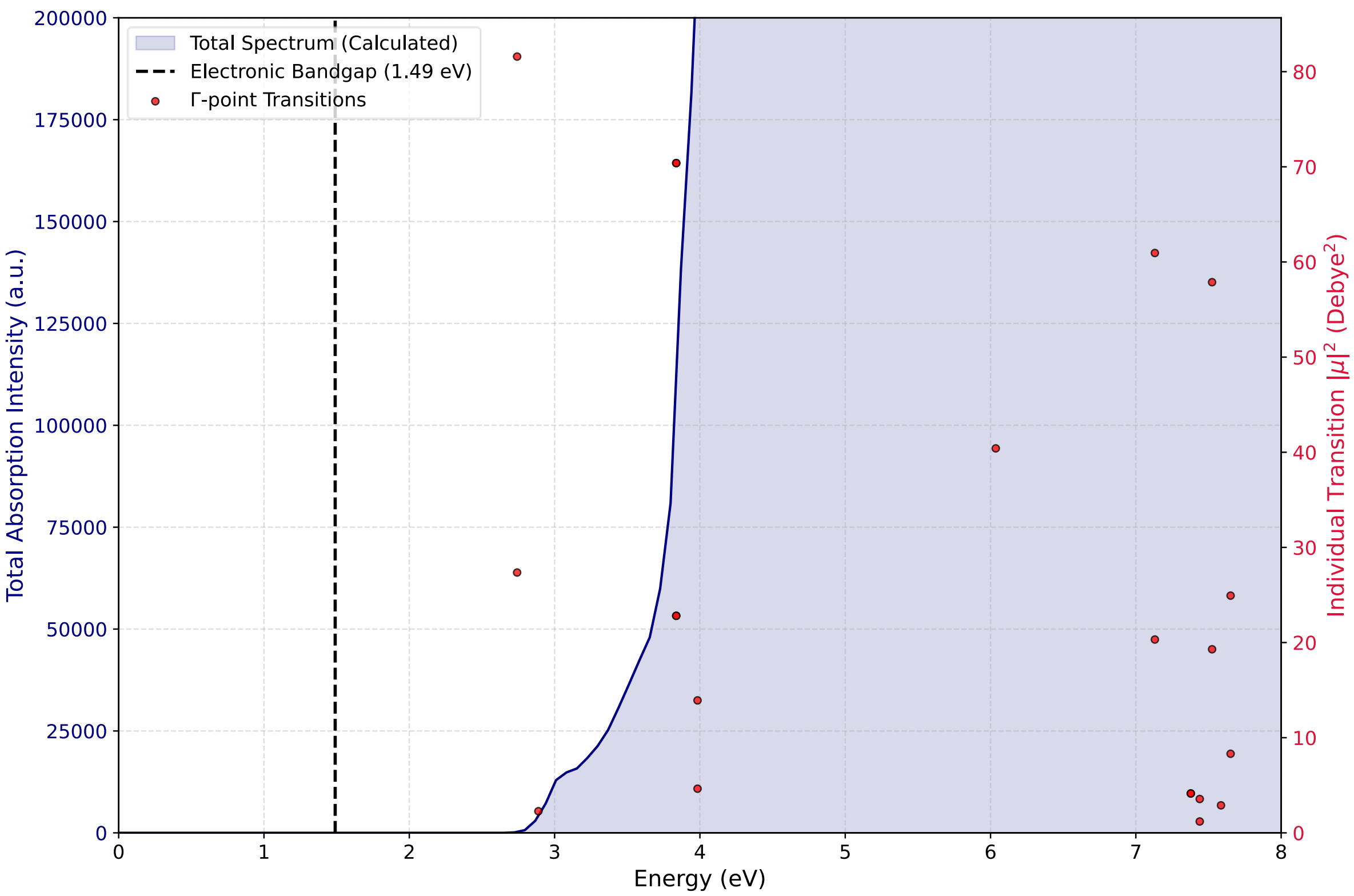

*Figure S3: Calculated absorption spectrum for α-Zr2SN2. The total absorption intensity is shown as the blue shaded curve (left axis). The fundamental electronic band gap (1.49 eV) is indicated by the vertical black dashed line. The red circles represent the square of the transition dipole moments for individual transitions calculated at the Γ-point (right axis)*

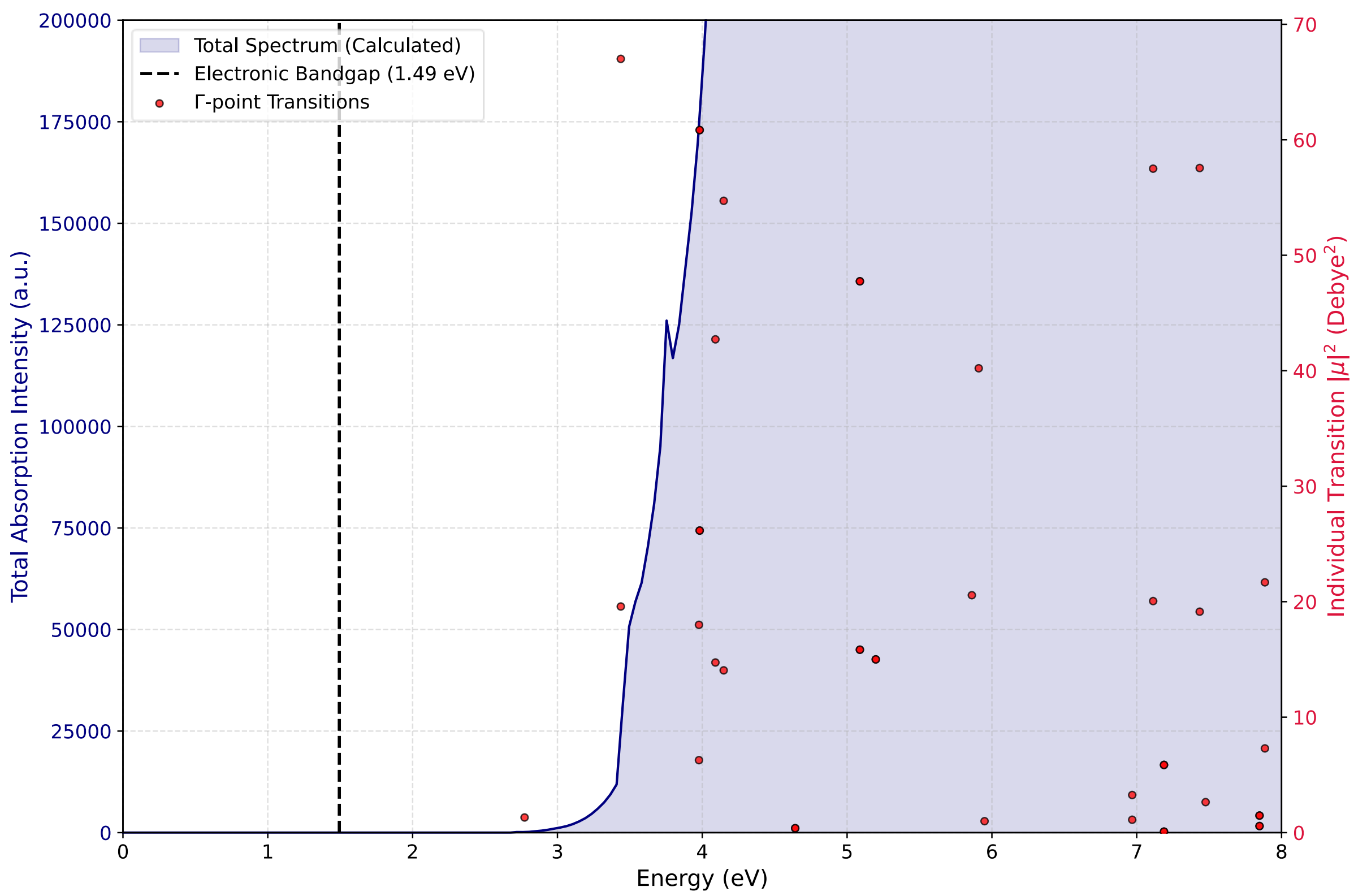

*Figure S4: Calculated absorption spectrum for β-Zr2SN2. The total absorption intensity is shown as the blue shaded curve (left axis). The fundamental electronic band gap (1.49 eV) is indicated by the vertical black dashed line. The red circles represent the square of the transition dipole moments for individual transitions calculated at the Γ -point (right axis)*

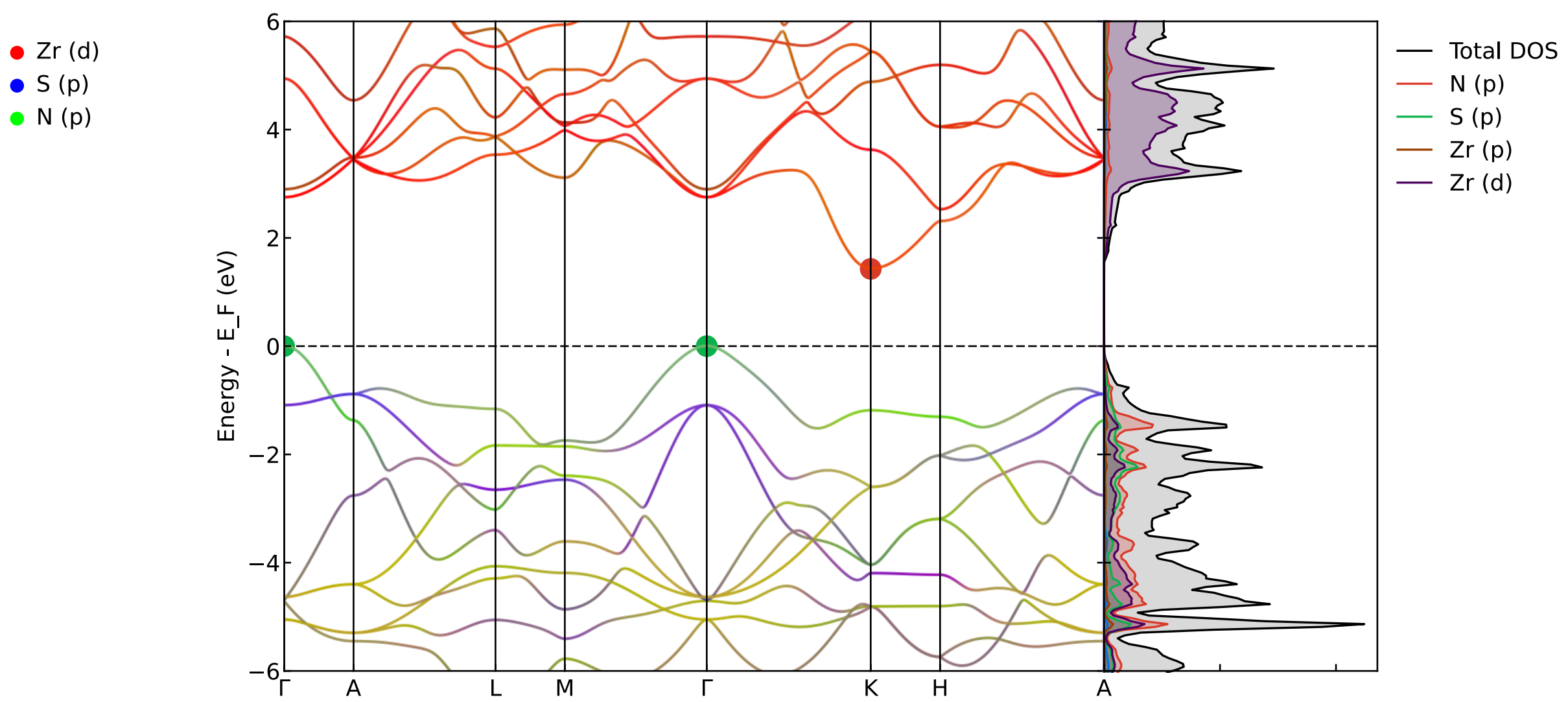

*Figure S5: Calculated band structure (left) and the corresponding projected density of states (pDOS) (right) for α-$Zr_2SN_2$. The Fermi Level ($E_F$) is set to 0 eV, and it is indicated by the horizonal dashed line. In both panels, the different colors indicated the contribution of the orbitals to the band (pDOS). The plot reveals an indirect band gap transition, with the Valence Band Maximum (VBM) located at the Γ-point and the Conduction Band Minimum (CBM) located at the K-point.*

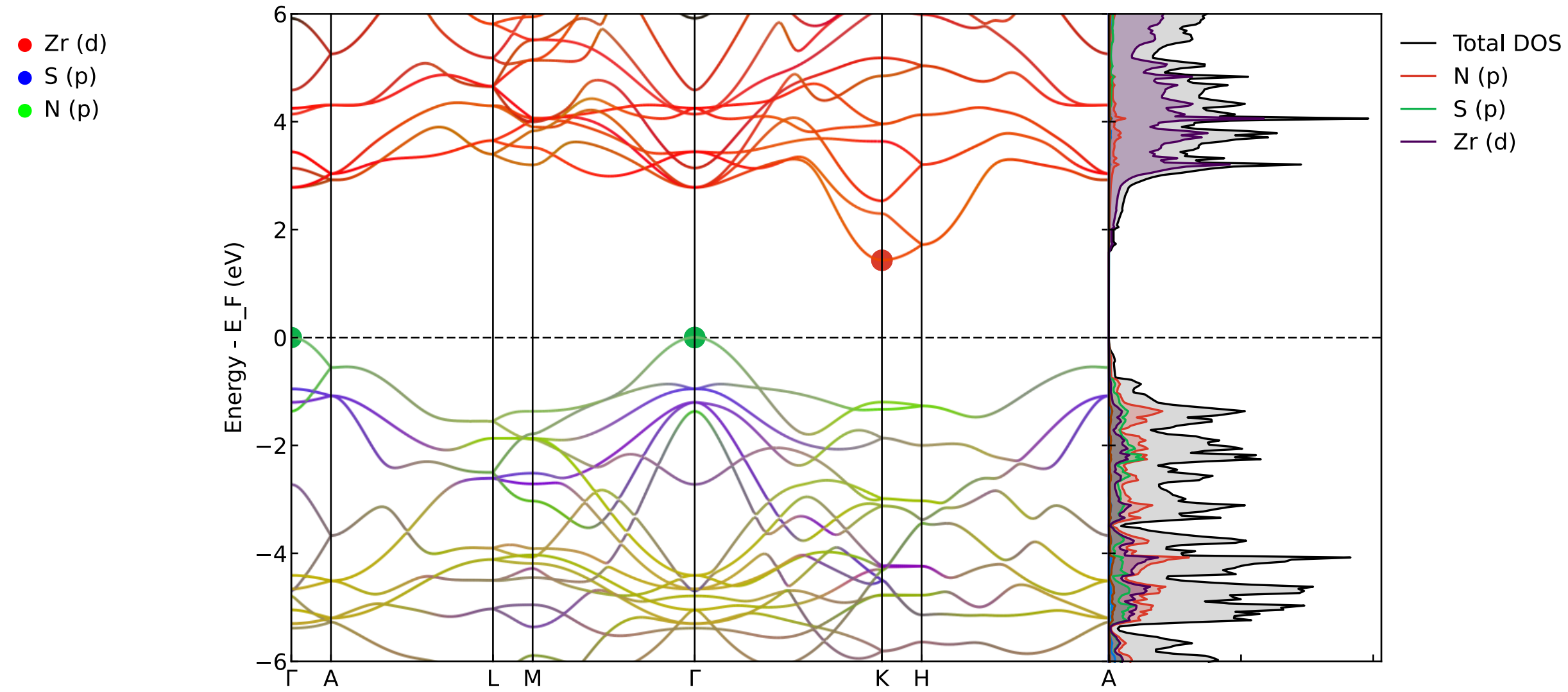

*Figure S6: Calculated band structure (left) and the corresponding projected density of states (pDOS) (right) for $\beta$-$Zr_2SN_2$. The Fermi Level ($E_F$) is set to 0 eV, and it is indicated by the horizonal dashed line. In both panels, the different colors indicated the contribution of the orbitals to the band (pDOS). The plot reveals an indirect band gap transition, with the Valence Band Maximum (VBM) located at the $\Gamma$-point and the Conduction Band Minimum (CBM) located at the K-point.*

To elucidate the origin of the contrasting transition dipole moments (TDMs) at the valence and conduction band edges of both phases, a rigorous group-theoretical symmetry analysis was performed at the $\Gamma$ $k$-point. This approach determines the spatial symmetries of the electronic Bloch states by mapping them to the irreducible representations (irreps) of the crystal's point group. The symmetry eigenvalues and corresponding irreps were computed using the IrRep Python package [41]. By evaluating the direct products of the initial and final state irreps, optical selection rules were established based on the symmetries of the electric dipole operator.

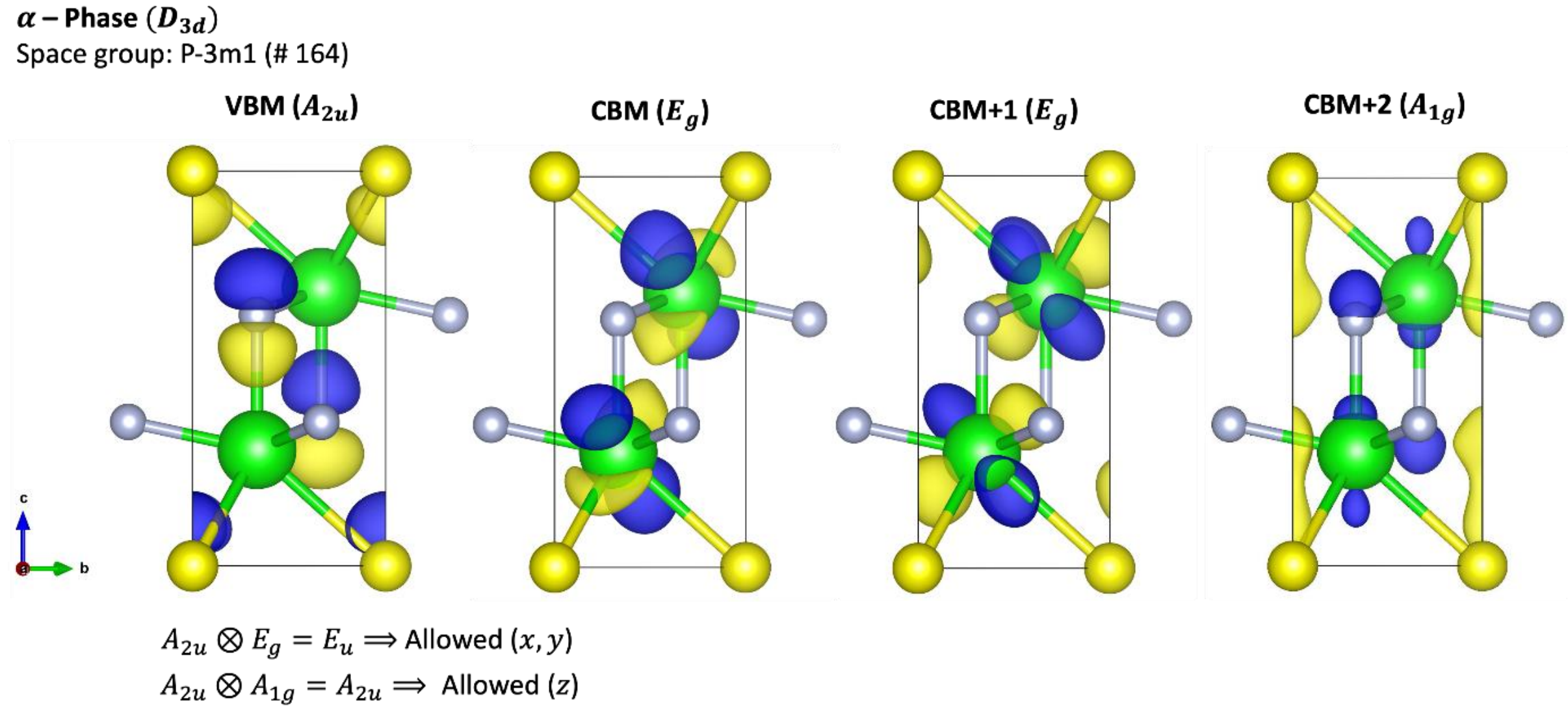

*Figure S7: Wavefunctions and transition selection rules for α-$Zr_2SN_2$ at the Γ special point. The wavefunctions phases are shown in blue (positive) and yellow (negative). The direct product calculations indicate that the optical excitations from the VBM (Valence Band Maximum) are strictly symmetry-allowed to the CB (Conduction Band), CB+1 and CB+2 states.*

*Table S3. Group-theoretical optical selection rules and transition dipole moments for the α-phase at the Γ k-point (point group $D_{3d}$ and $D_{6h}$ for α and β respectively). Direct products of the irreducible representations of the valence band maximum (VBM, $A_{2u}$) with the lowest conduction band states (CB, CB+1, CB+2) determine the symmetry-allowed transitions and their corresponding light polarization vectors under the $D_3$d point group. Transition energies and squared transition dipole moments (TDM) $|\mu|^2$ are obtained from DFT calculations.*

| Transition | k-points | Direct product | Polarization vector | Selection Rule | Energy (eV) | $\|\mu\|^2$ ($D^2$) |
|---|---|---|---|---|---|---|
| VBM – CB | Γ – Γ | $A_{2u} \otimes E_g = E_u$ | (x, y) | Allowed | 2.742 | 81.586 |
| VBM – CB +1 | Γ – Γ | $A_{2u} \otimes E_g = E_u$ | (x, y) | Allowed | 2.742 | 27.360 |
| VBM – CB +2 | Γ – Γ | $A_{2u} \otimes A_{1g} = A_{2u}$ | z | Allowed | 2.888 | 2.273 |

For the α phase (space group $P\overline{3}m1$, point group $D_{3d}$, Table S3, Figure S7) the analysis focused on the valence band maximum (VBM) and Γ's lowest conduction band (CB, CB+1, CB+2), as shown in Table S3. The calculations assign VBM to the $A_{2u}$ irrep, the degenerate CB/CB+1 states to $E_g$, and CB+2 to $A_{1g}$. The direct products evaluate as $A_{2u} \otimes E_g = E_u$ and $A_{2u} \otimes A_{1g} = A_{2u}$. Because $E_u$ and $A_{2u}$ correspond to the in-plane $(x, y)$ and out-of-plane$(z)$ light polarization vectors respectively, these transitions are strictly symmetry-allowed. Consequently, the non-zero TDMs observed for these states are governed entirely by the physical spatial overlap of the initial and final wavefunctions (Figure S7).

**$\beta$-Phase ($D_{6h}$)**
Space group: P6_3/mmc (# 194)

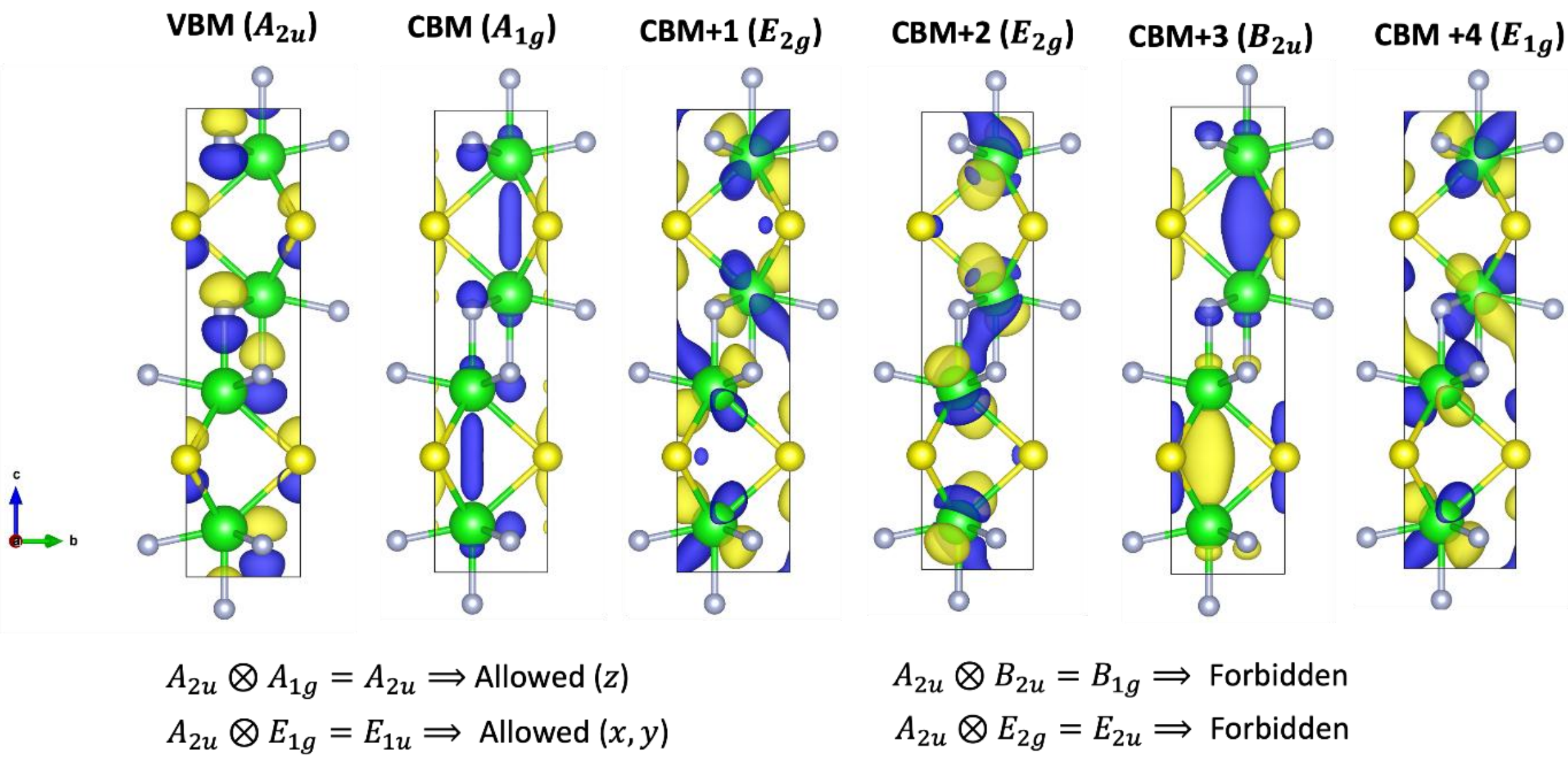

*Figure S8: Wavefunctions and transition selection rules for β-$Zr_2SN_2$ at the Γ k-point. The wavefunctions phases are shown in blue (positive) and yellow (negative). The direct product calculations indicate that the optical excitations from the VBM (Valence Band Maximum) are strictly symmetry-allowed only to the CB (Conduction Band Minimum) and CB+4 states.*

*Table S4: Group-theoretical optical selection rules and transition dipole moments for the β-phase at the Γ k-point (point group $D_{3d}$ and $D_{6h}$ for α and β respectively). Direct products of the irreducible representations of the valence band maximum (VBM, $A_{2u}$) with the lowest conduction band states (CB to CB+5) reveal a combination of symmetry-allowed and symmetry-forbidden transitions under the relevant point group. The forbidden transitions to CB+1, CB+2 and CB+3 arise from direct products that do not correspond to any component of the electric dipole operator, resulting in strictly zero transition dipole moments (TDMs). Transition energies and squared TDMs $|\mu|^2$ are obtained from DFT calculations.*

| Transition | k-points | Direct product | Polarization vector | Selection Rule | Energy (eV) | $\|\mu\|^2$ ($D^2$) |
|---|---|---|---|---|---|---|
| VBM – CB | Γ – Γ | $A_{2u} \otimes A_{1g} = A_{2u}$ | z | Allowed | 2.773 | 1.315 |
| VBM – CB +1 | Γ – Γ | $A_{2u} \otimes E_{2g} = E_{2u}$ | | Forbidden | 2.778 | 0 |
| VBM – CB +2 | Γ – Γ | $A_{2u} \otimes E_{2g} = E_{2u}$ | | Forbidden | 2.778 | 0 |
| VBM – CB +3 | Γ – Γ | $A_{2u} \otimes B_{2u} = B_{1g}$ | | Forbidden | 3.136 | 0 |
| VBM – CB +4 | Γ – Γ | $A_{2u} \otimes E_{1g} = E_{1u}$ | (x, y) | Allowed | 3.437 | 19.583 |
| VBM – CB +5 | Γ – Γ | $A_{2u} \otimes E_{1g} = E_{1u}$ | (x, y) | Allowed | 3.437 | 67.005 |

In the $\beta$ phase (space group $P6_3/mmc$, point group $D_{6h}$, Table S4, Figure S8), the structural transformation imposes stricter symmetry constraints, as exposed in Table S4. The band-edge states are identified as follows: VBM is $A_{2u}$, CB is $A_{1g}$, the degenerate CB+1/CB+2 states are $E_{2g}$, CB+3 is $B_{2u}$, and the degenerate CB+4/CB+5 states are $E_{1g}$. Direct product analysis reveals that transitions from VBM to CB+1/CB+2 ($A_{2u} \otimes E_{2g} = E_{2u}$) and to CB+3 ($A_{2u} \otimes B_{2u} = B_{1g}$) are strictly forbidden. The former yields an $E_{2u}$ symmetry that lacks a corresponding dipole vector, while the latter yields an even-parity ($B_{1g}$) state, which violates the Laporte rule. Conversely, transitions to CB ($A_{2u} \otimes A_{1g} = A_{2u}$) and CB+4/CB+5 ($A_{2u} \otimes E_{1g} = E_{1u}$) remain symmetry allowed. The drastic variance in their calculated TDMs is therefore attributed to the degree of spatial overlap between the highly localized and interstitial wavefunctions, as depicted in Figure S8.

## 8.3 Extended Microscopy Methods

A Thermo Fisher Helios Hydra plasma focused ion beam (FIB) – scanning electron microscope (SEM) instrument was used to prepare thin lamella of the sample for transmission electron microscopy (TEM) analysis. The preparation involved using a 30 keV $Xe^+$ ion beam to cut a thin cross-section of the sample (approximately 20 µm in length, 2 µm in thickness and 5 µm in depth) and transfer it to a TEM compatible grid. The cross-section was then thinned to electron transparency (approximately 100 nm thick) through further $Xe^+$ ion beam milling, followed by final polishing using an 8 – 2 keV $Ar^+$ ion beam at low currents (60 – 20 pA). Scanning TEM (STEM) images were recorded using a Thermo Fisher Spectra Ultra (S)TEM instrument. The microscope is equipped with aberration correction on the probe forming lenses and an Ultra-X energy dispersive X-ray spectroscopy (EDX) detector. STEM images of the sample were recorded at 300 keV electron beam energy and with an electron probe convergence angle of 30 mrad. The detector inner collection angles for the annular dark-field (ADF) and high-angle annular dark-field (HAADF) STEM images were 30 mrad and 56 mrad, respectively. The approximate current and size of the electron probe were 97 pA and 1 Å, respectively. EDX maps of the sample were acquired using the Ultra-X detector and analyzed using the Velox software.

## 8.4 O impurity and chemical composition control during sputtering

### 8.4.1 Influence of Deposition Temperature on O content and chemical composition

Zr-S-N films have been sputtered at two different temperatures, 25 and 300 °C. All other parameters have been kept identical. Therefore, the same gas mix and the same deposition pressure of 5 mTorr have been used. Higher temperature has the following effects on the composition: enrichment in N and loss in S and O. Higher deposition temperature is therefore preferred for lower O impurity content.

*Table S5: Film composition determined by EDX (LayerProbe) for films deposited at 5 mTorr, at two different substrate temperatures during deposition.*

| Substrate Temperature (°C) | Zr (%at) | S (%at) | N (%at) | O (%at) | Zr/N | S/N | O/N |
|---|---|---|---|---|---|---|---|
| 25 (Room Temperature) | 33 | 18 | 23 | 26 | 1.43 | 0.78 | 1.13 |

| 300 | 39 | 13 | 35 | 12 | 1.11 | 0.37 | 0.34 |
|---|---|---|---|---|---|---|---|

### *8.4.2 Influence of Sputter Pressure at on O content and chemical composition*

Zr-S-N films have been sputtered at two different pressures, 5 and 2 mTorr. All other parameters have been kept identical. Therefore, the same gas mix and the same deposition temperature (room temperature) have been used. One can see that a lower deposition pressure has the following effects on the composition: enrichment in N and S and loss in O. The effect on the O content is drastic. Indeed, it should be noted that the composition, and therefore the O content, has been extracted using LayerProbe (see Methods section in main article), assuming a simple geometrical model of a Zr-S-N-O film on Si. Any O present in the Si native oxide, or in the Zr-S-N films surface oxid (most likely $ZrO_2$ as explained below), is interpreted as belonging to the film, artificially raising the film's O content. This effect is strongest in thinner films. Based on this argument, the O content of 2%at is overestimated, showing that the O content is greatly minimized with lower pressure. Lower deposition pressure is therefore preferred for lower O impurity content.

*Table S6: Film composition determined by EDX (Layer Probe) for films deposited at 25°C (Room Temperature), at two different pressures.*

| Deposition Pressure (mTorr) | Zr (%at) | S (%at) | N (%at) | O (%at) | Zr/N | S/N | O/N |
|---|---|---|---|---|---|---|---|
| 5 | 36 | 17 | 30 | 16 | 1.20 | 0.57 | 0.53 |
| 2 | 36 | 21 | 41 | 2 | 0.88 | 0.51 | 0.05 |

## *8.5 Sputtering of $Zr_2SN_2$*

A Zr-N-S film was sputtered on a 4 × 4 cm$^2$ n-type Si wafer. The sputter system base pressure was 1.4 × 10$^{-7}$ Torr. The substrate temperature was programmed to reach 465 °C at a rate of 20 °C/min in 15 mTorr of Ar. The Zr target was ignited with 12 W RF power at 30 mTorr and then brought to 75 W for a 6 min pre-sputtering cleaning in the pressure range 5 to 2 mTorr. The deposition was carried in a 96/3/1% Ar/$N_2$/$H_2S$ gas mix, achieved by flowing 56 sccm, 2 sccm and 7.1 sccm of Ar, $N_2$ and a 10% $H_2S$ in Ar gas mixture, respectively. At the end of the deposition, the substrate temperature was decreased from 465 °C to about 200 °C in the same gas mix. $N_2$ and $H_2S$ flow was stopped at 200 °C to let the substrates cool down to room temperature in Ar.

The color of the Zr RF plasma in the Ar/$N_2$/$H_2S$ gas mixture was pink/orange.

Optical emission spectroscopy (Gencoa OPTIX) has been used to follow the composition of the gas permeating the chamber during the process. A flange allows us to sample the chamber gas into a smaller chamber where a secondary DC plasma is generated. The peaks from this secondary plasma emission spectra can be tracked to quantitively extract the composition of the gas mixture. No S-species emission peaks could be detected in the secondary plasma despite the presence of $H_2S$ in chamber; therefore, S was not included in the following analysis. Figure S9 displays the partial pressure of $N_2$ and Ar as a function of time during the process. The $N_2$ partial pressure is stable during the deposition process. $N_2$ partial pressure increases at the very end of the deposition, at the instant of the Zr target shutdown.

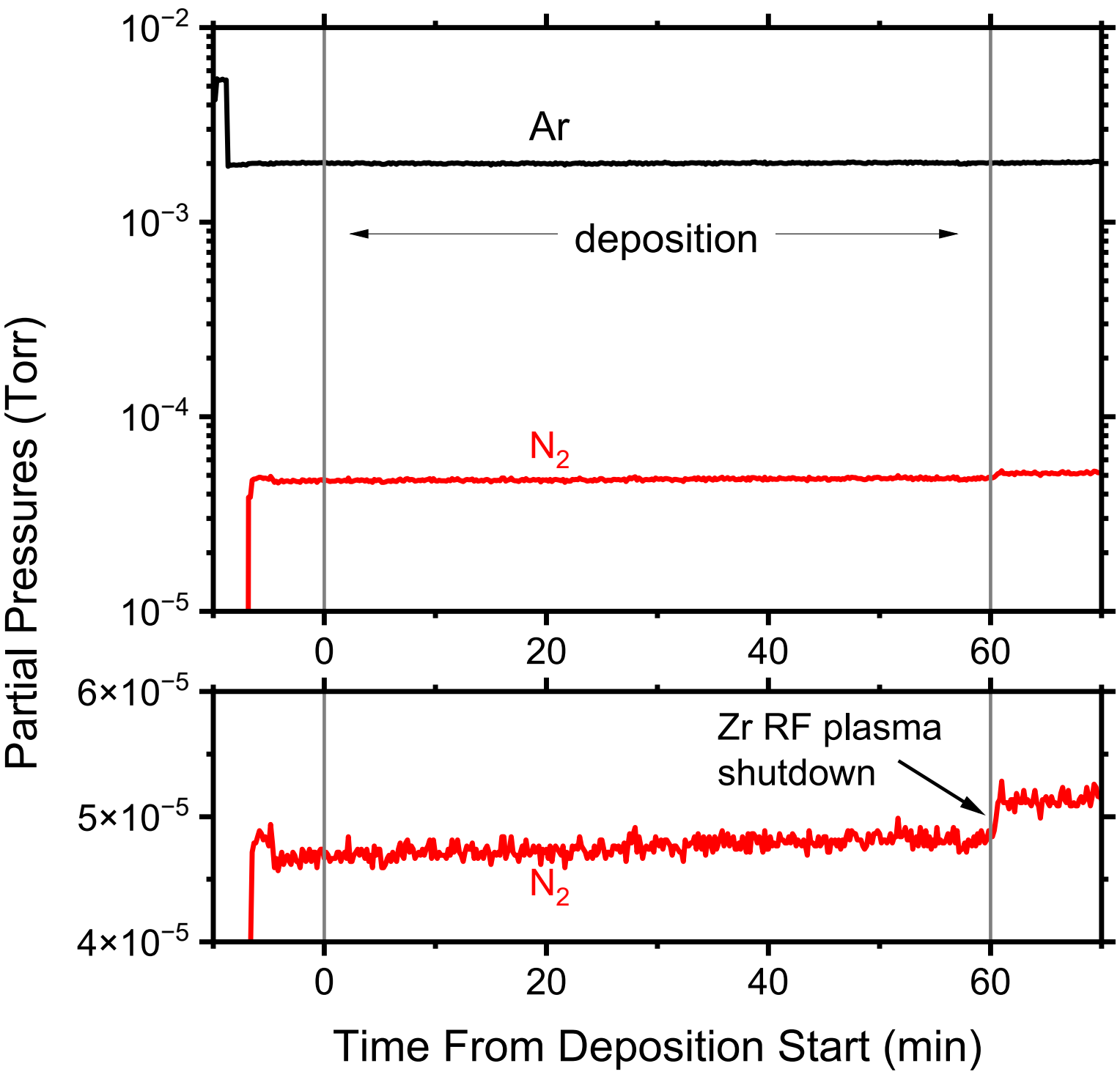

*Figure S9: Ar and $N_2$ gas partial pressure in the chamber during deposition. Partial pressure is measured by calibrated optical emission spectroscopy of a secondary plasma probing a gas sample away from the main deposition chamber.*

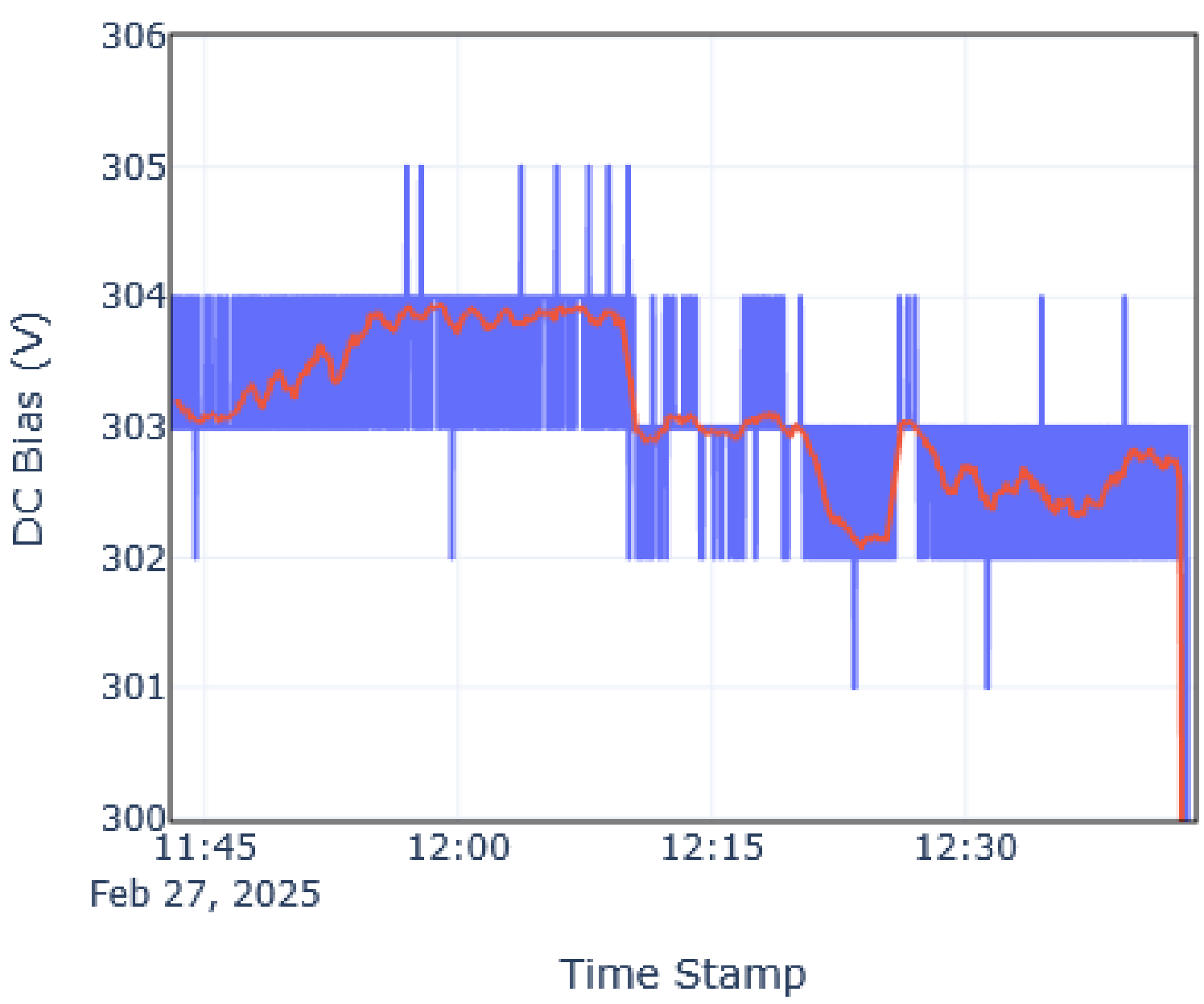

*Figure S10: DC bias of the Zr RF plasma as function a time during deposition. Raw signal (blue curve, 1 V interval) and smoothed signal (red) are plotted.*

Figure S10 displays the DC bias of the Zr RF plasma during the deposition. Both the raw signal and smoothed signal are plotted. The DC bias is stable during the deposition process, in the range 302 to 304 V, demonstrating the stability of the Zr target during deposition

Figure S11 displays different signals of the chamber, giving insight into the effect of the introduction of reactive gases into chamber into the Zr target RF plasma. At the instant of the introduction of $N_2$ gas ($N_2$ flow increases from 0 to 2 sccm), the DC bias responds with a small increase from 302 V to 304 V while a drastic 2-fold decrease in the deposition rate can be seen. Introduction of $H_2S$ gas in the mix only yields a small increase in deposition rate and a small decrease in DC bias.

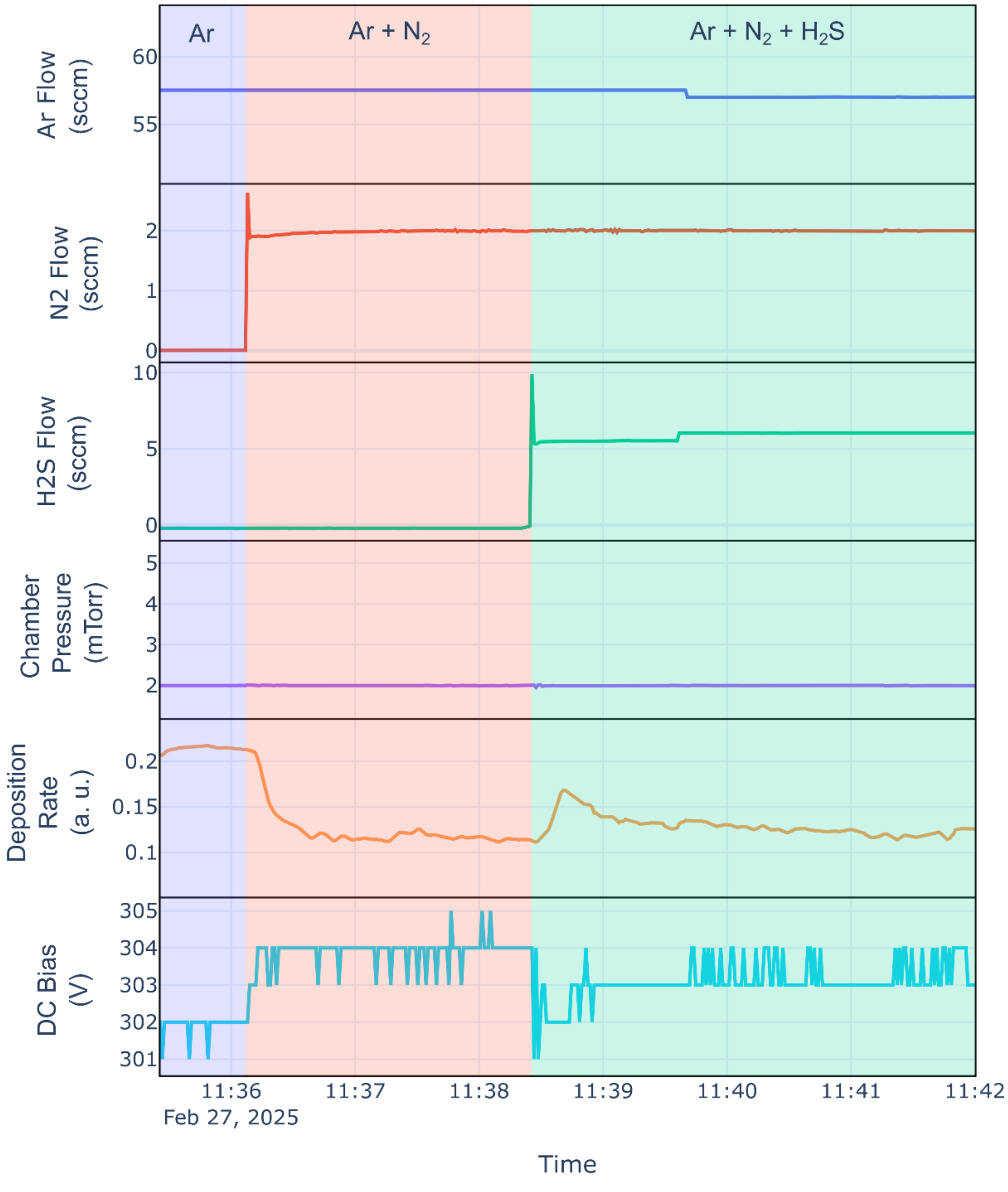

*Figure S11: Chamber signals at the instants of the introduction of the reactive gases in the Zr RF plasma.*

Figure S12 shows the S/N and Zr/N ratio map as extracted from EDX, on the film deposited on a 4 × 4 $cm^2$ Si substrate. The S/N ratio varies between 0.60 and 0.85, above the stoichiometric S/N ratio of 0.5 in $Zr_2SN_2$. The Zr/N ratio varies between 1 and 1.3, above the stoichiometric Zr/N ratio of 1 in $Zr_2SN_2$.

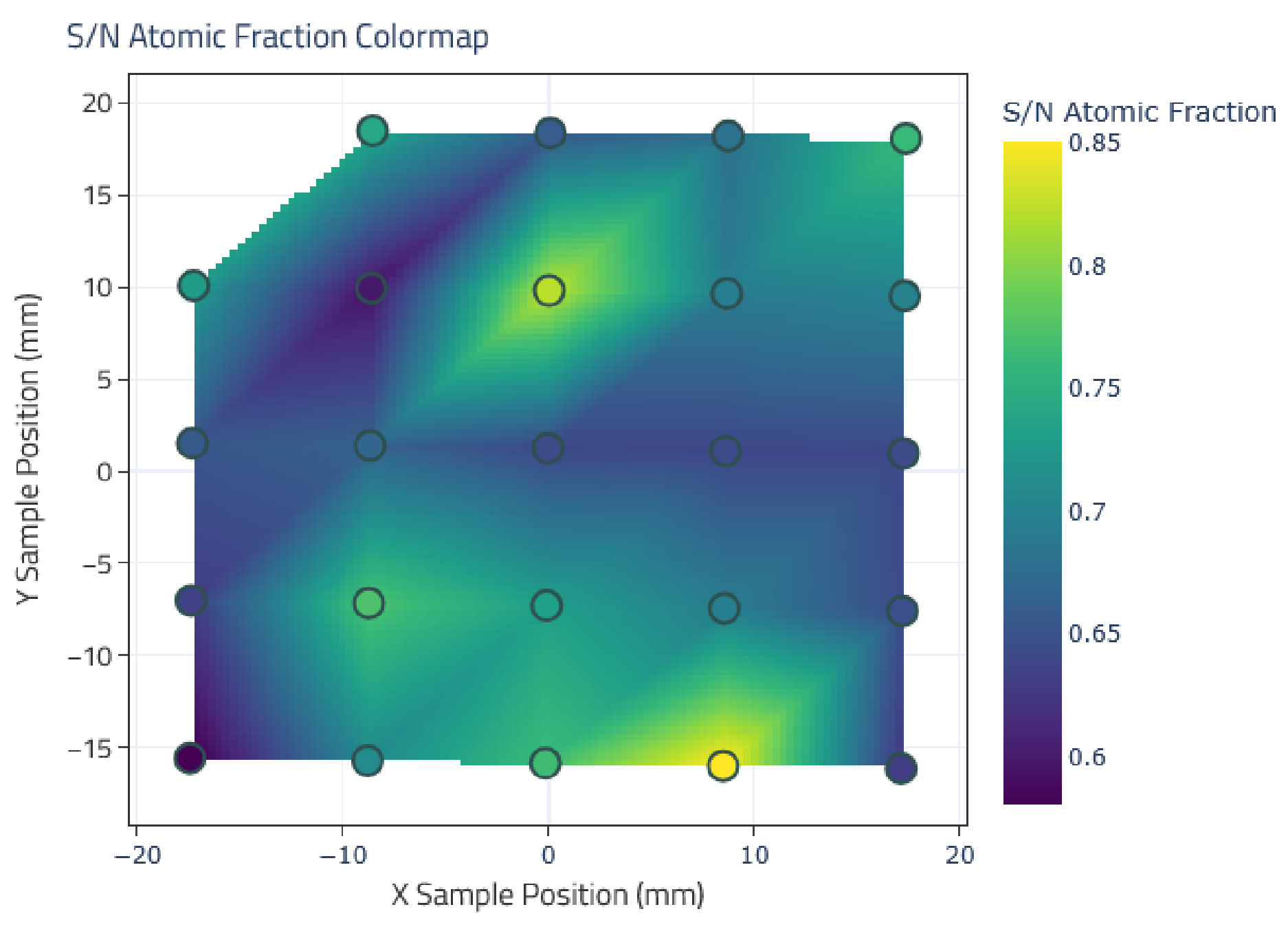

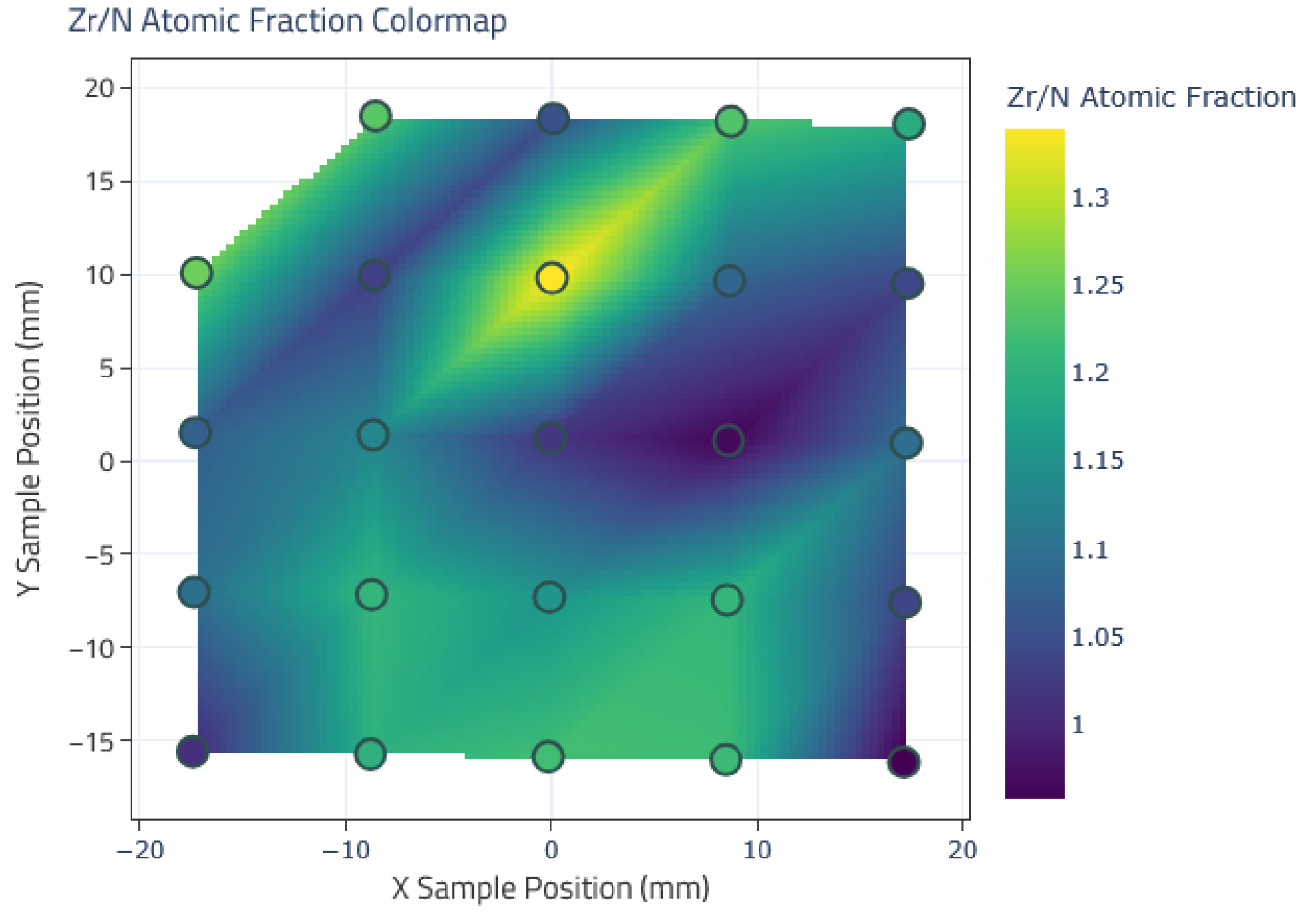

*Figure S12: S/N and Zr/N atomic percent ratios as measured by EDX on 4x4$cm^2$ films deposited on Si.*

Composition spread of the films deposited on Si and (as exposed in Table S7). Thickness extracted using layer probe and assuming a density of 5.55 $g.cm^{-3}$ are also reported.

*Table S7 Composition and thickness t (masses density ρt, assuming a density ρ of 5.55 $g.cm^{-3}$) of the as grown films deposited on Si, as determined by EDX LayerProbe.*

| | t (nm) | Zr (%at) | S (%at) | N (%at) | O (%at) | S/N |
|---|---|---|---|---|---|---|
| Meas. | 40- 65 | 34 - 40 | 22 - 25 | 30 - 37 | 4 -10 | 0.6 - 0.8 |
| Ideal | N.A. | 40 | 20 | 40 | 0 | 0.5 |

XRD and TEM measurements were carried out on the as grown samples deposited on Si. As displayed in Figure S13, XRD does not reveal film-related peaks and only show the Si substrate (004) peaks. TEM dark field cross section images displayed in Figure S14 do not reveal any crystalline ordering, confirming the amorphous nature of the film.

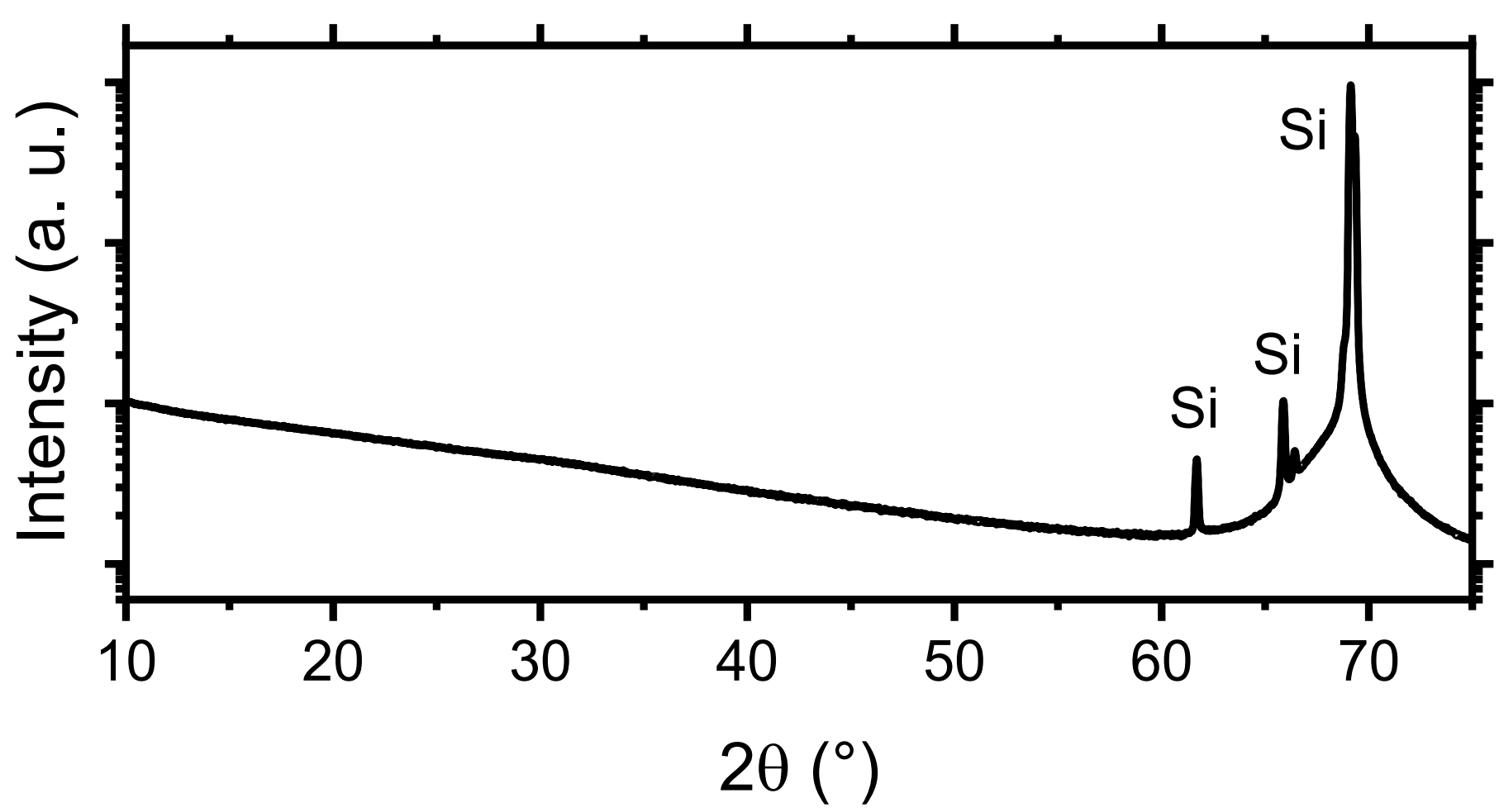

*Figure S13: XRD pattern in ϑ/2ϑ geometry of the as grown film on Si. No peaks other than the substrate peaks are detected.*

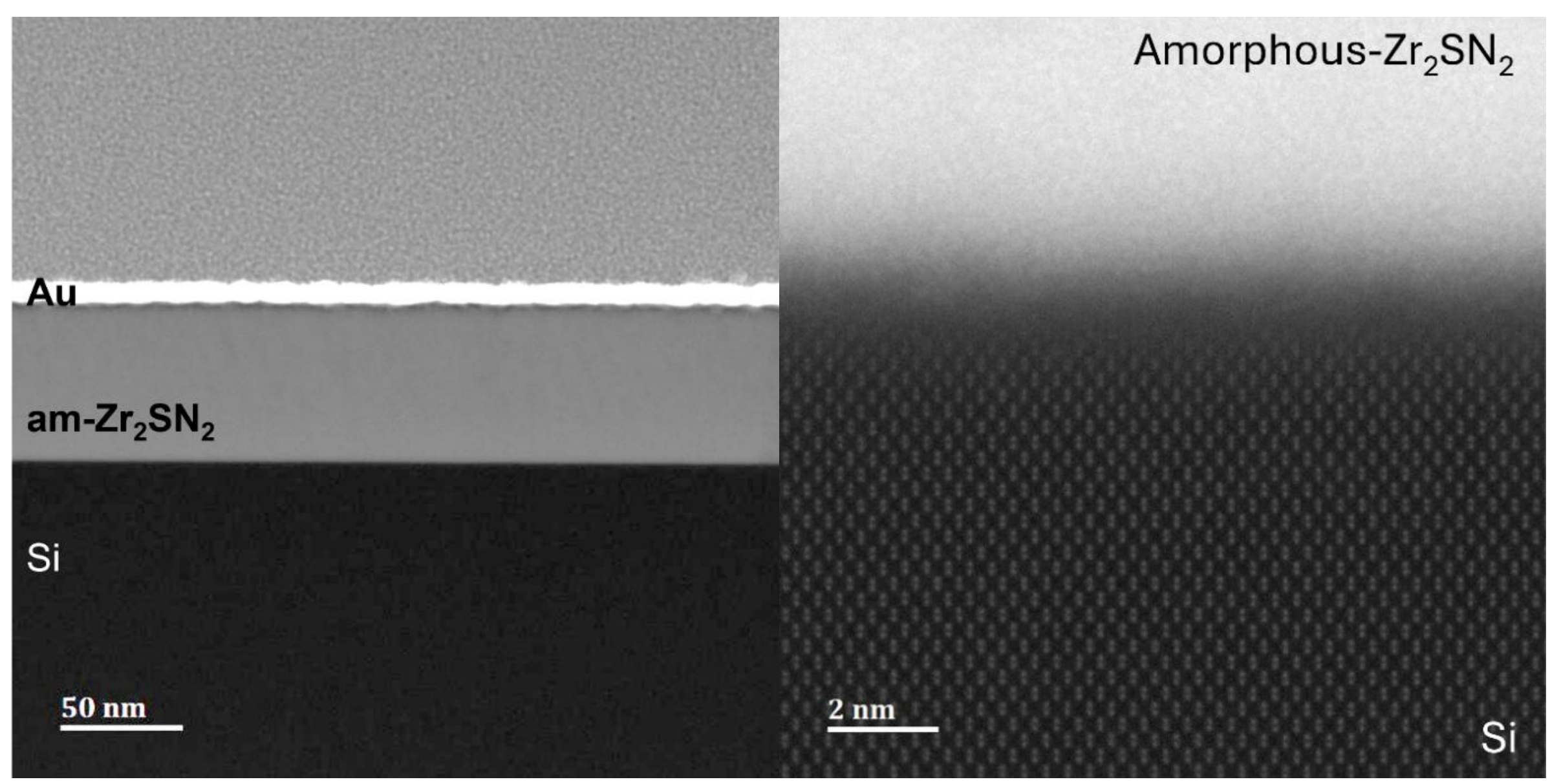

*Figure S14: Dark field TEM cross section images of the as grown film on Si, a different magnification levels. am-$Zr_2SN_2$ stands for amorphous-$Zr_2SN_2$.*

### *8.6 Annealing of as Grown films*

The (4.0 × 4.0) $cm^2$ film deposited on Si was cleaved into smaller pieces of size (1.3 × 1.3) $cm^2$ or (4.0 × 1.3) $cm^2$ and annealed using different pressures and max temperatures as listed in Table S7. A typical rapid thermal process (RTP) annealing process can be visualized in Figure S15, for a maximum annealing temperature T° max of 900 °C, and an $N_2$ pressure of 0.2 Torr (vacuum). As detailed in the Methods section, all annealing processes were carried out with ramps of 100 °C/min.

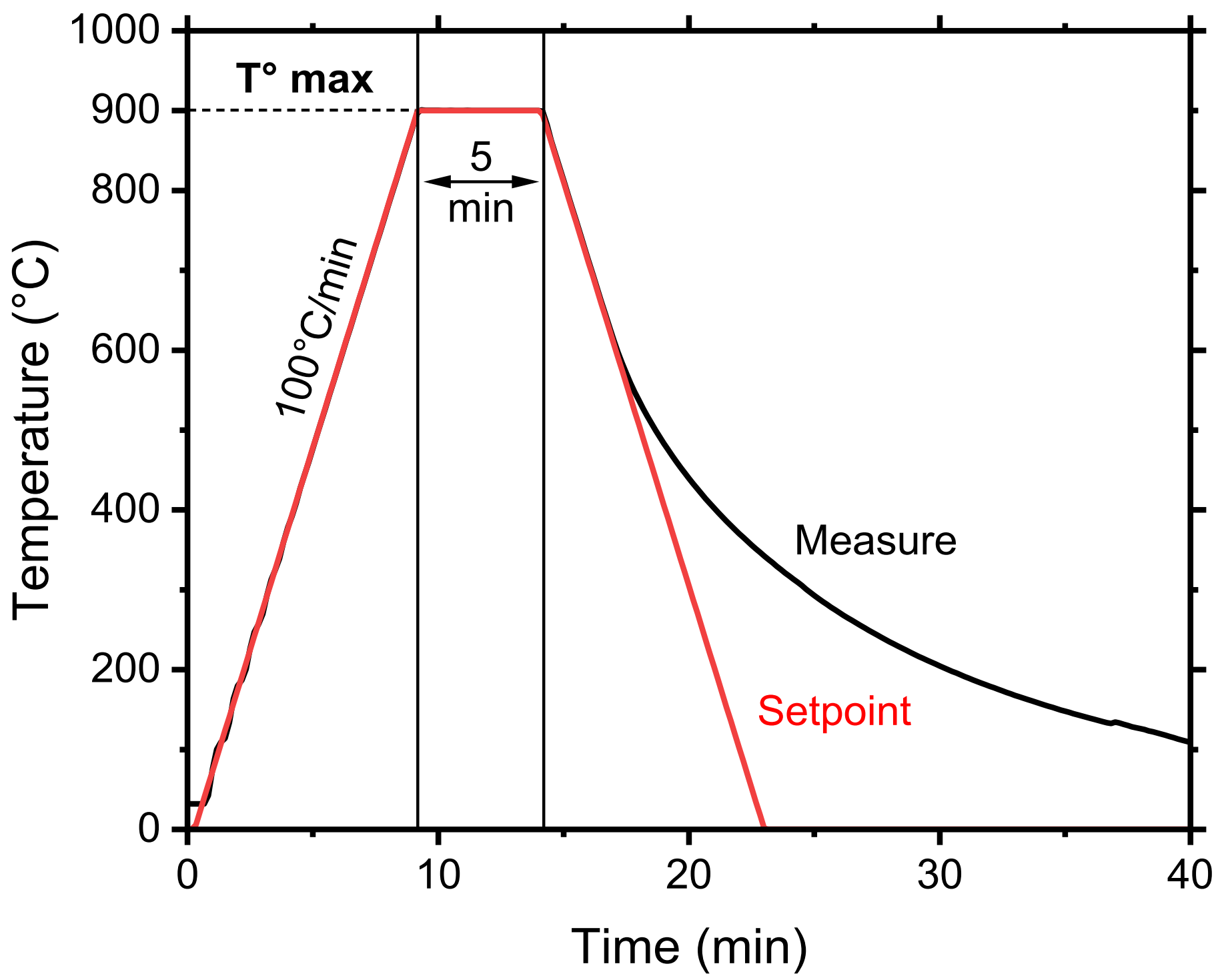

*Figure S15: Temperature setpoint and measured temperature as a function of time for the film annealed at 900 °C at a $N_2$ pressure of 0.2 Torr (vacuum).*

The following table summarizes the different samples produced in this study. The Raman spectrum has been recorded of the film on Si annealed for 30 min at 0.2 Torr. TEM measurements (STEM-EDX, high resolution (HR)-STEM) were conducted on the as grown (sputtered) film and the 900 °C vac. annealed film, both deposited on Si. The optical characterization and Hall effect measurement presented in the main article have been performed on the film deposited on fused silica and annealed 30 min in vacuum.

Table S8: Summary of annealing parameters used in this study. *The as grown sputtered sample is also given here for reference.

| Label | Substrate | Temp (°C) | Pressure (Torr) | Plateau Time (min) | Comment |
|---|---|---|---|---|---|
| as grown* | Si | | | | Sputtered, Amorphous, Measured in TEM |
| 700 °C | Si | 700 | 500 | 5 | Amorphous |
| 800 °C | Si | 800 | 500 | 5 | Lowest crystallization temperature |
| 900 °C | Si | 900 | 500 | 5 | Higher O content than 900 °C, vac. sample |
| 900 °C, vac. | Si | 900 | 0.2 | 5 | Measured in TEM |
| 900 °C, vac. /$SiO_2$ | $SiO_2$ (fused silica) | 900 | 0.2 | 30 | Crystalline on transparent fused silica, Highest n, Hall effect, Main paper optical characterization |
| 900 °C, vac. 30 min | Si | 900 | 0.2 | 30 | Raman spectra |

## 8.7 Additional chemical composition, XRD and thickness data

The following table shows the difference in thickness and chemical composition before and after annealing, for thin films grown on Si. Additionally, the thickness as extracted from ellipsometry, XRR and TEM are shown if available. Consistent thicknesses are found across different techniques.

*Table S9: Thickness, chemical composition and elemental ratios for samples before and after annealing for films on Si annealed at different temperatures and pressures. Thickness t measured by different methods (Transmission Electron Microscopy: $t_{TEM}$, X-ray Reflectivity: $t_{XRR}$, Spectroscopic Ellipsometry: $t_{SE}$ and Electron Dispersive X-ray Spectroscopy LayerProbe: $t_{EDX}$) is reported when available. Since EDX LayerProbe in fact measures mass thickness ρt, with ρ the density, we report the deduced the derived thickness for two different densities.*

| Temp (°C) | Press. (Torr) | Before/ After RTP | $t_{TEM}$ (nm) | $t_{XRR}$ (nm) | $t_{SE}$ (nm) | $t_{EDX}$ (nm) assuming ρ (g.cm$^{-3}$) = 5.55 | $t_{EDX}$ (nm) assuming ρ (g.cm$^{-3}$) = 4.95 | Zr (%at) | S (%at) | N (%at) | O (%at) | Zr/N | S/N | O/N |
|---|---|---|---|---|---|---|---|---|---|---|---|---|---|---|
| 700 | 500 | Before | | | | 56 | 63 | 38 | 24 | 32 | 6 | 1.19 | 0.73 | 0.19 |
| | | After | | | 89±4 | 66 | 74 | 33 | 17 | 34 | 16 | 0.99 | 0.50 | 0.50 |
| 800 | 500 | Before | | | | 49 | 55 | 37 | 23 | 32 | 8 | 1.13 | 0.71 | 0.25 |
| | | After | | | 67±4 | 56 | 63 | 33 | 16 | 33 | 17 | 0.98 | 0.48 | 0.52 |
| 900 | 500 | Before | | | | 64 | 72 | 39 | 23 | 33 | 5 | 1.17 | 0.68 | 0.15 |
| | | After | | | 81±4 | 72 | 81 | 35 | 17 | 38 | 11 | 0.92 | 0.46 | 0.29 |
| | | Before | | | | 51 | 58 | 36 | 23 | 36 | 6 | 1.02 | 0.64 | 0.16 |

| 900, vac.* | 0.2 (vac.) | After | 57±2 | 58±2 | 65±4 | 57 | 64 | 35 [37] | 18 [14] | 38 [36] | 9 [12] | 0.91 [1.03] | 0.46 [0.38] | 0.23 [0.50]* |
|---|---|---|---|---|---|---|---|---|---|---|---|---|---|---|

* for the 900 °C vac. sample, composition determined by STEM-EDX is given in brackets

After annealing, we measure a higher O impurity content, as displayed in Table S9. However, when comparing the two samples on Si annealed at 900 °C, we achieve a lower O content in the vacuum process (0.2 Torr). We confirm the stronger presence of an annealing-induced surface-oxide by the presence of an extra contribution that we interpret as $ZrO_2$ in GIXRD pattern of the 500 Torr sample, which is absent in the 0.2 Torr sample (Figure S16). This shows that lower annealing pressure yields lower contamination of the film by O.

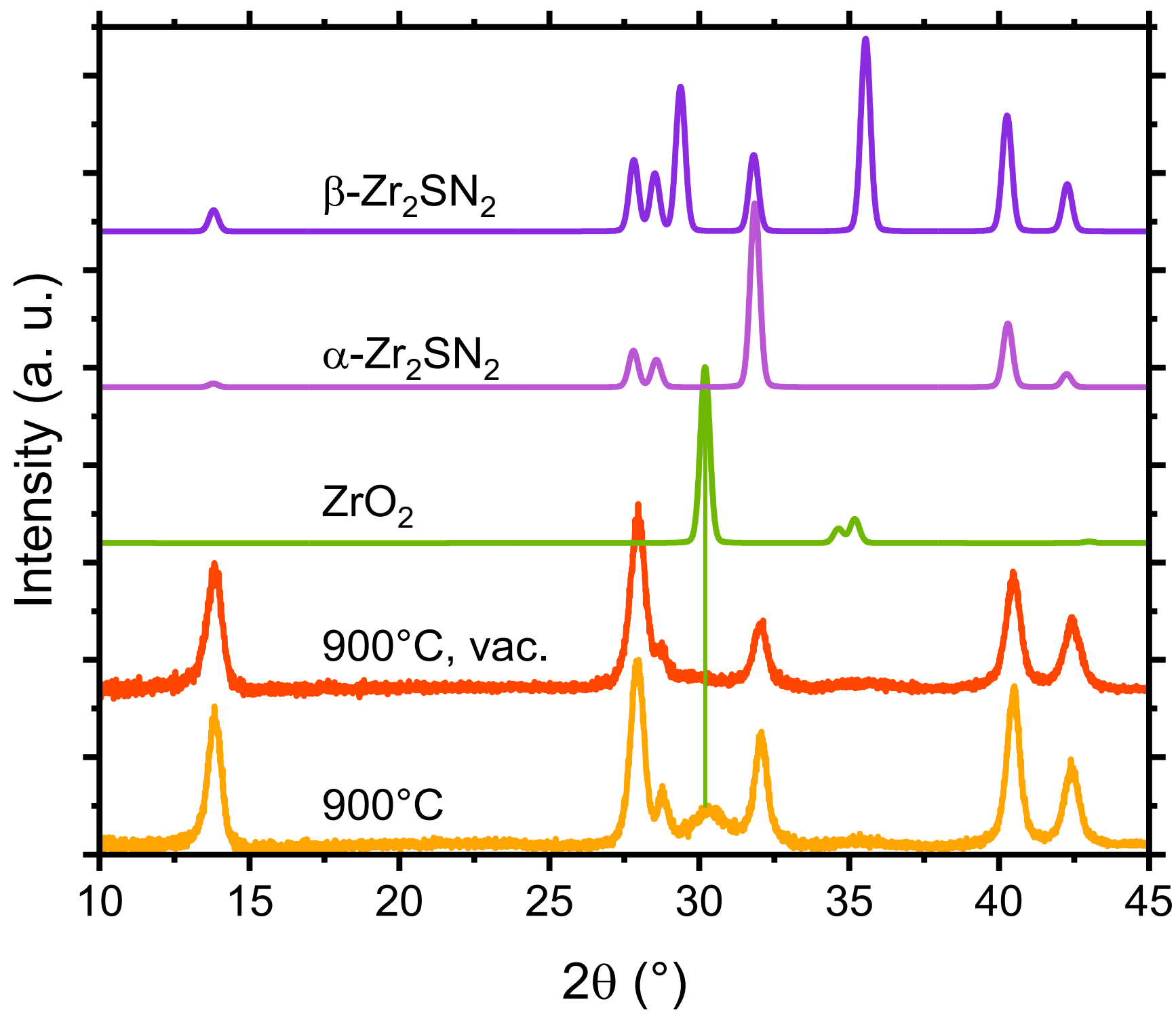

*Figure S16: Grazing incidence GIXRD pattern of the two films deposited on Si and annealed at 900 °C at different $N_2$ pressures 0.2 Torr (labelled 900 °C, vac.) and 500 Torr (labelled 900 °C), alongside theoretical patterns for α- and β-$Zr_2SN_2$ as well as $ZrO_2$.*

### 8.8 All XRD patterns

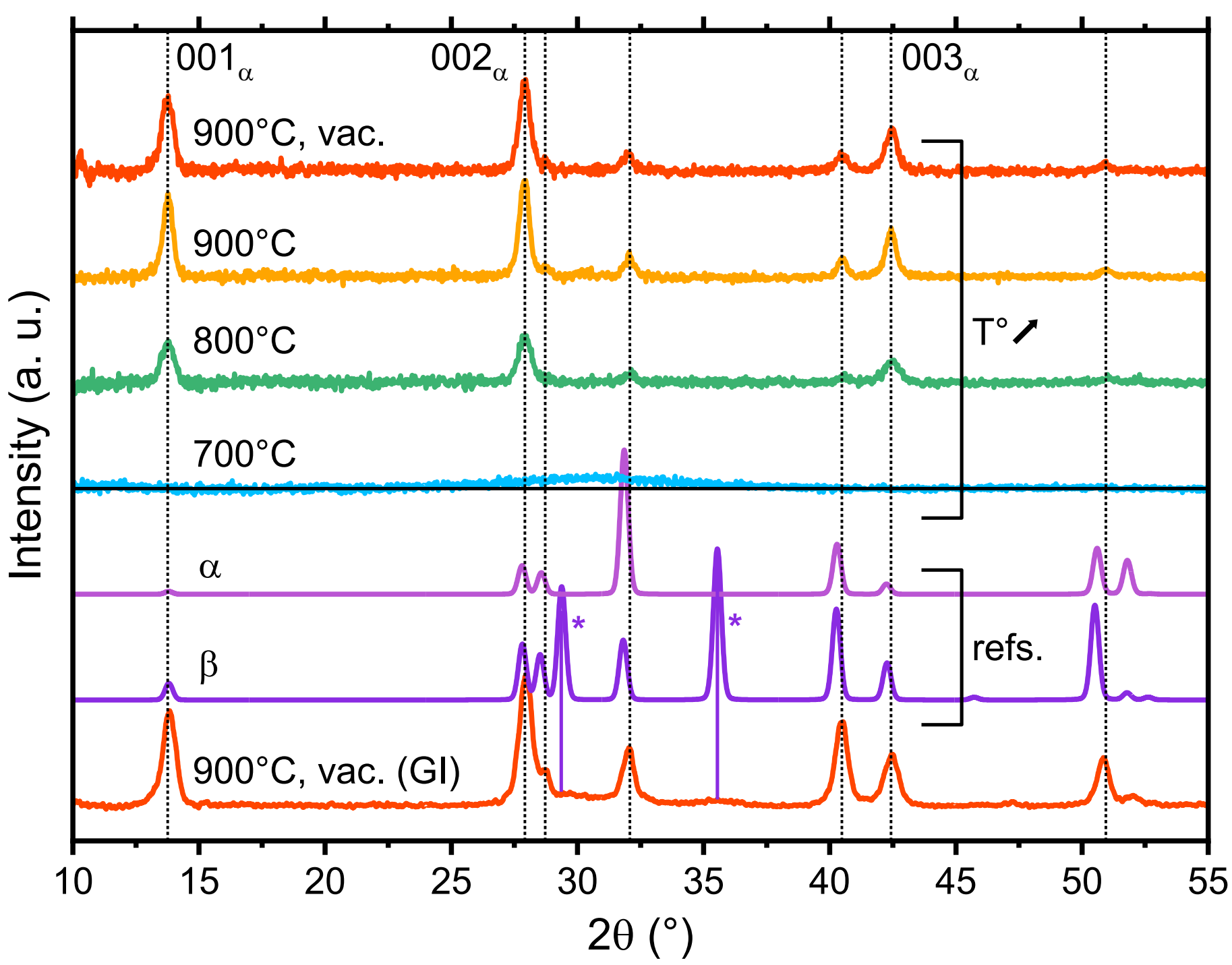

*Figure S17: XRD pattern in the ϑ/2ϑ geometry for films deposited on Si and annealed at different temperatures and pressures (top). Calculated XRD pattern for α- and β-$Zr_2SN_2$ (refs). GIXRD pattern for the film annealed at 900 °C in vacuum (0.2 Torr) (bottom). Planes of the 00l family are labelled using the α lattice.*

Figure S17 shows the XRD patterns of the films deposited on Si and annealed at different temperatures and pressures, together with the reference XRD patterns for α- and β-$Zr_2SN_2$. At 700 °C, one can see a weak and broad contribution centered at 31°, likely related to an amorphous $Zr_2SN_2$ film. Diffraction peaks at 13.68, 27.92 and 42.47° appear at 800 °C annealing temperature and grow more intense at 900 °C. These specific peaks can be indexed at planes of the (00l) family of α- or β-$Zr_2SN_2$, suggesting a strong (00l) texture. More generally all diffraction peaks can be indexed using the reference crystal data for α- or β-$Zr_2SN_2$. Interestingly, no difference can be seen between the 0.2 and 500 Torr annealing process at 900 °C. We observe that the lowest annealing temperature enabling crystallization in 800 °C, and that annealing pressure plays a marginal role in the crystallization process.

### 8.9 Structural analysis on the film annealed at 900 °C in vacuum

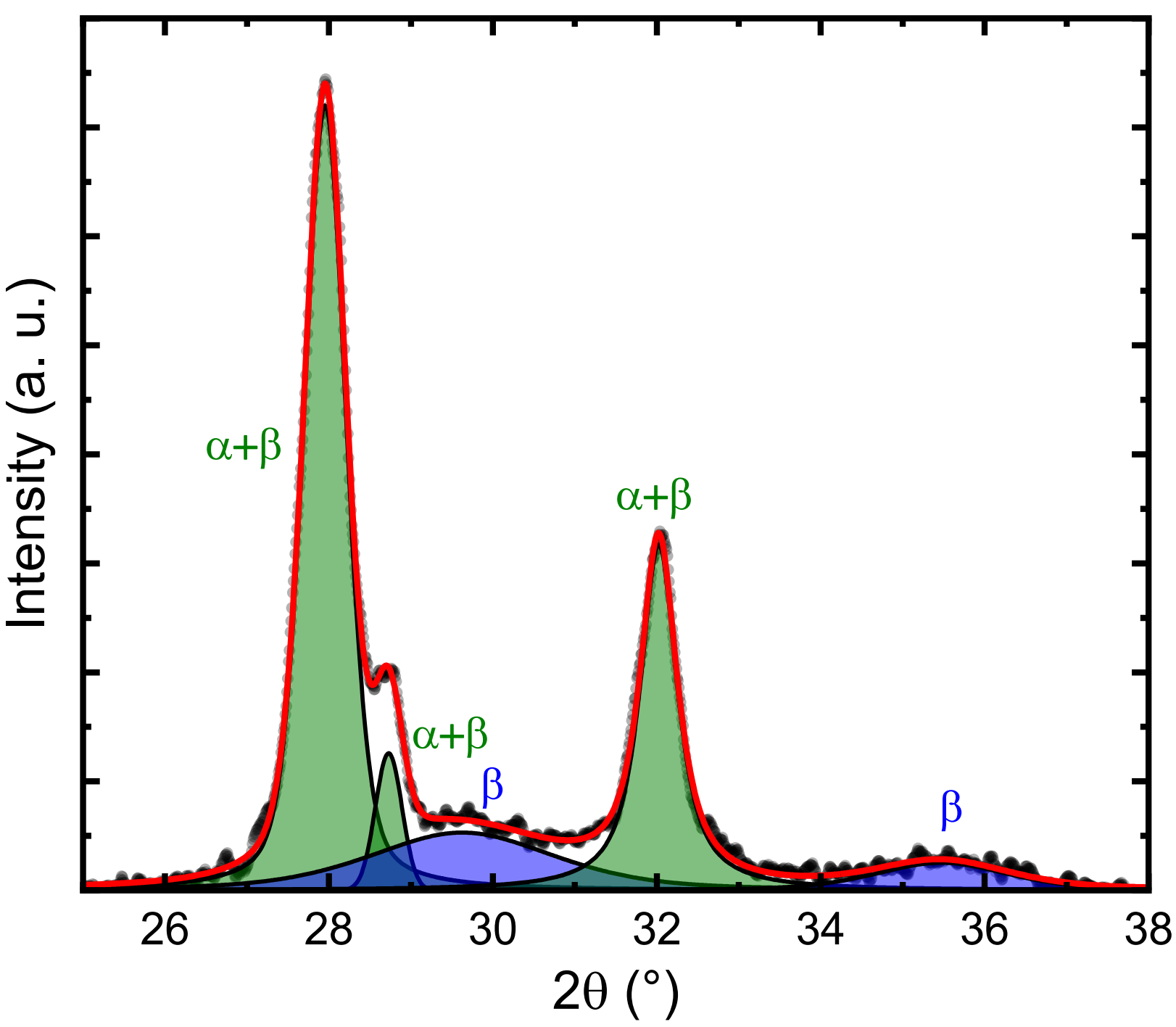

*Figure S18: Fit of the GIXRD pattern of the sample annealed at 900 °C in vac. on Si (red trace of Figure S17) using split Kα₁ + Kα₂ pseudo-Voigt peaks with constrained intensities, width, position and Gaussian-Lorentzian mixing.*

Figure S18 displays the fit of the GIXRD pattern of the sample annealed at 900 °C in vac. on Si. We fit 5 contributions from the (004), (100), (101), (102), (103) reflections, as expressed in the β-phase Miller indices. It is important to note that the (101) and (103) reflections are exclusive to the β phase while all the others have equivalent planes diffracting at nearly the same Bragg angle (for example β's (004) is equivalent to α's (002)).

In the experimental GIXRD pattern of Figure S17, β-exclusive peaks marked with * are broad and not easily detected (Figure S18), suggesting that coherently scattering domains of β are small (approximately 3-5 nm using Scherrer's equation). However, all the peaks shared by the α- and β-$Zr_2SN_2$ phases are present and comparatively sharper, with coherent domains 15 - 20 nm in size. This hints that the film may exhibit a long-range ordering corresponding to a mixture of α- and β-$Zr_2SN_2$. We extract lattice parameters of a = 3.595 Å and c = 12.794 Å, using the Le Bail method with β-$Zr_2SN_2$ as crystal model (as shown later in Figure S19, Table S10).

As detailed later in Figure S23, Raman spectroscopy reveals the characteristic modes of β-$Zr_2SN_2$ (taken from the Computational Raman Database [13], entries mp-11583 and mp-1158, for α and β respectively; Figure S23, Table S11). Therefore, Raman spectroscopy supports the conclusion from XRD that a fraction of the film exhibits β ordering at the local scale.

The coexistence of α- and β-$Zr_2SN_2$ can be understood based on the following argument. We calculate that α- and β-$Zr_2SN_2$ have nearly identical formation enthalpy $\Delta H_{f,\alpha}$ and $\Delta H_{f,\beta}$, at 0K, respectively, with a difference of $\Delta H_{f,\beta} - \Delta H_{f,\alpha} = -0.013$ eV/atom, in favor of β. Furthermore, the annealing temperatures of 800 - 900 °C are near the reported phase transition temperature (850 °C) between the two polymorphs [9]. Thus, there is most likely no strong thermodynamical drive to towards one or the other, making it likely that the two phases coexist in the system.

### 8.10 Le Bail refinement

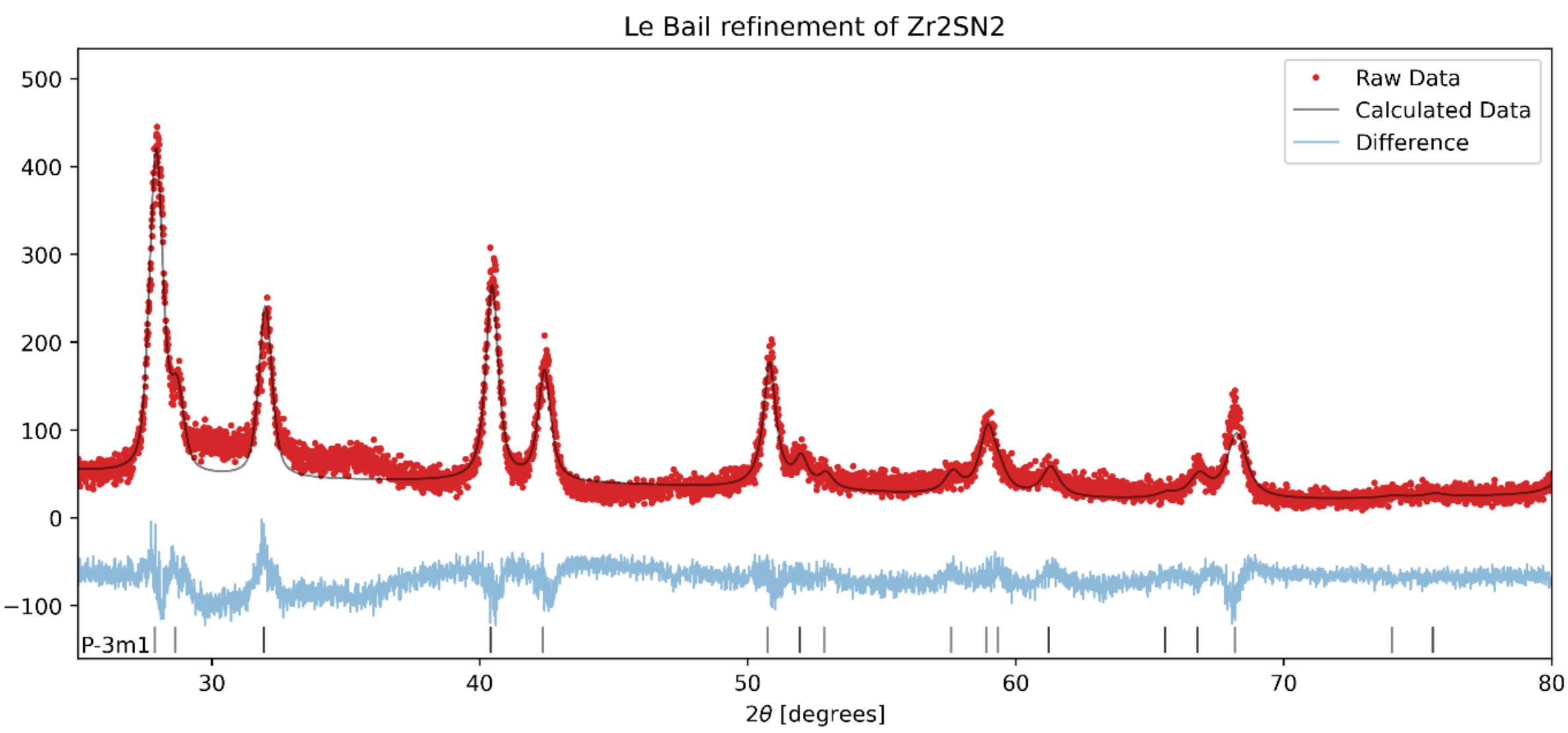

*Figure S19: Le Bail refinement of the GIXRD pattern of the film deposited on Si and annealed at 900 °C in vac (0.2 Torr), using the structure α-$Zr_2SN_2$, with the following constraints a = b, α = β, γ = 120°. The refinement parameters are a = b = 3.5946 ± 0.0003 Å, c = 6.3968 ± 0.0006 Å.*

*Table S10: Results of the Le Bail refinement using either the structure α- or β-$Zr_2SN_2$. Goodness of fit (GOF) is also reported. The obtained parameters are virtually identical.*

| Refinement using | a = b (Å) | c (Å) | GOF |
|---|---|---|---|
| α-$Zr_2SN_2$ | 3.5946 ± 0.0003 | 6.3968 ± 0.0006 | 1.52 |
| β-$Zr_2SN_2$ | 3.5949 ± 0.0003 | 12.7939 ± 0.0011 (= 2 × 6.3969) | 1.42 |

Le Bail refinement was performed on the GIXRD pattern of the film deposited on Si and annealed at 900 °C in vac (0.2 Torr). Virtually identical lattice parameters are extracted using α and β as input structure, since the lattice parameter c of β corresponds to 2 times the parameter c of α, because of a symmetry breaking in the c direction. Fitting with both α and β at the same time leads to unphysical overfitting.

### 8.11 Extra HR-STEM images

Different zone-axis were probed using TEM. In the <1$\bar{1}$0> zone axis, the α and β are virtually indistinguishable, as shown in Figure *S21*. Both structures can explain the observed atomic ordering. However, in the <110> zone-axis (Figure *S20*), specific features of α and β can be observed. Namely, in the 6 × 3 $nm^2$ crystallite, a β-specific stacking can be seen, where each $Zr_2N_2$ layer switches orientation in-between each S interlayer. The size of the investigated β crystallite is good accordance with broad β XRD peaks (Figure S18) and with the Raman spectra (Figure S23). Other crystallites do not exclusively fit one of the two phases. As displayed in the main article's Figure 3(d), some areas of the image better match the atomic ordered of the β phase, as opposed to the α, and vice versa. The two crystal domains are stacked along the c-axis, in a sequence of epitaxial α and β. They share planes marked by red line on Figure 3(d). An orange line following Zr atom columns is drawn as a guide for the eye, and each kink in this orange line makes the presence of a β unit cell. This epitaxial stacking allows long range ordering for diffracting planes shared by α and β, leading to extended coherent domains and explaining the sharpness of XRD peaks shared by α and β (Figure S18). XRD, Raman and STEM combined shows that the film crystallizes as a mixture of β-$Zr_2SN_2$ crystallites and of interlaced α/β with long-range ordering.

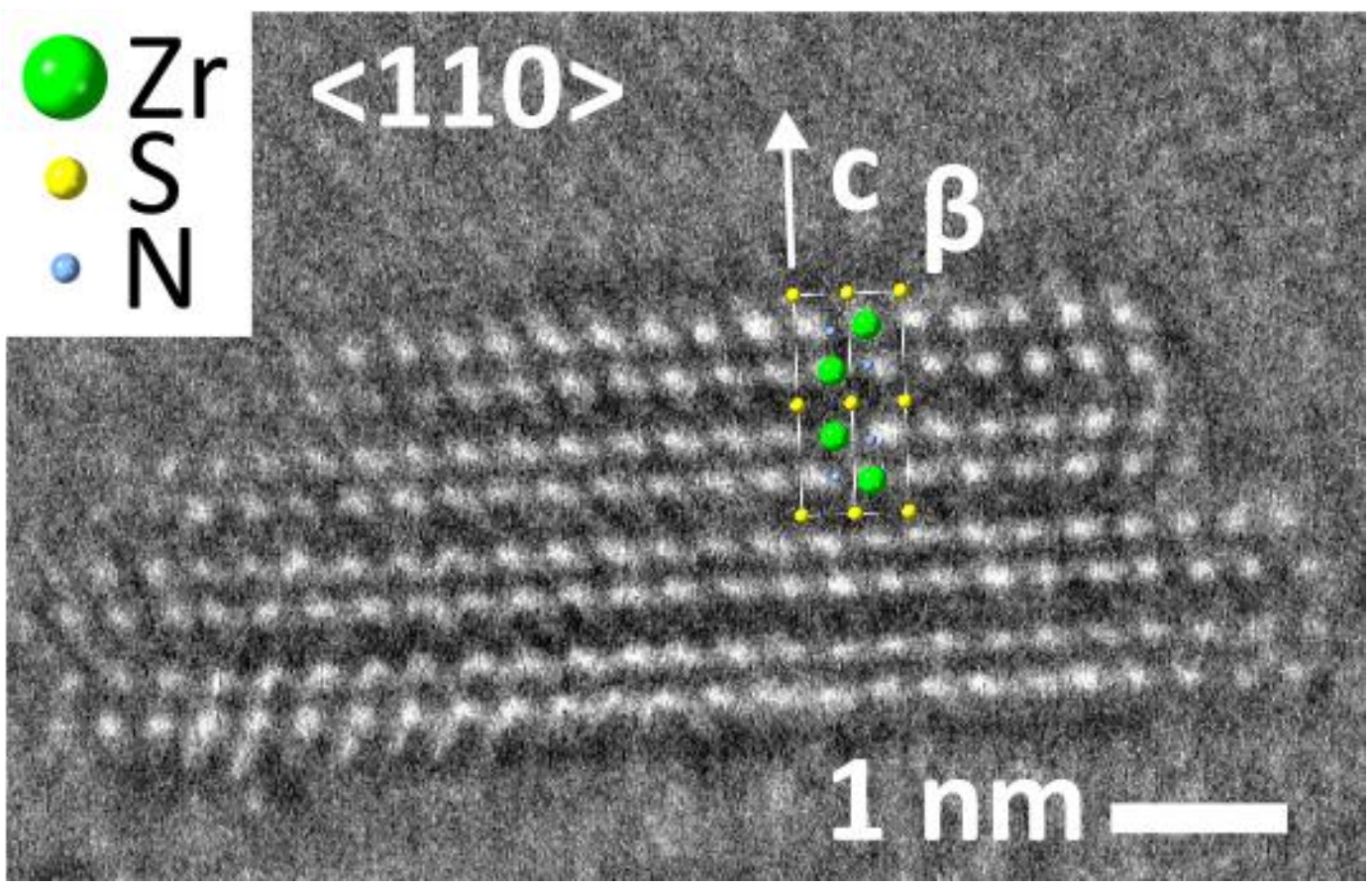

Figure S20: HR-STEM picture recorded on the film deposited on Si and annealed at 900 °C in vac. showing a *6 × 3 nm²* crystallite displaying a β ordering in the <110> zone axis. In this zone axis, β can be distinguished from α.

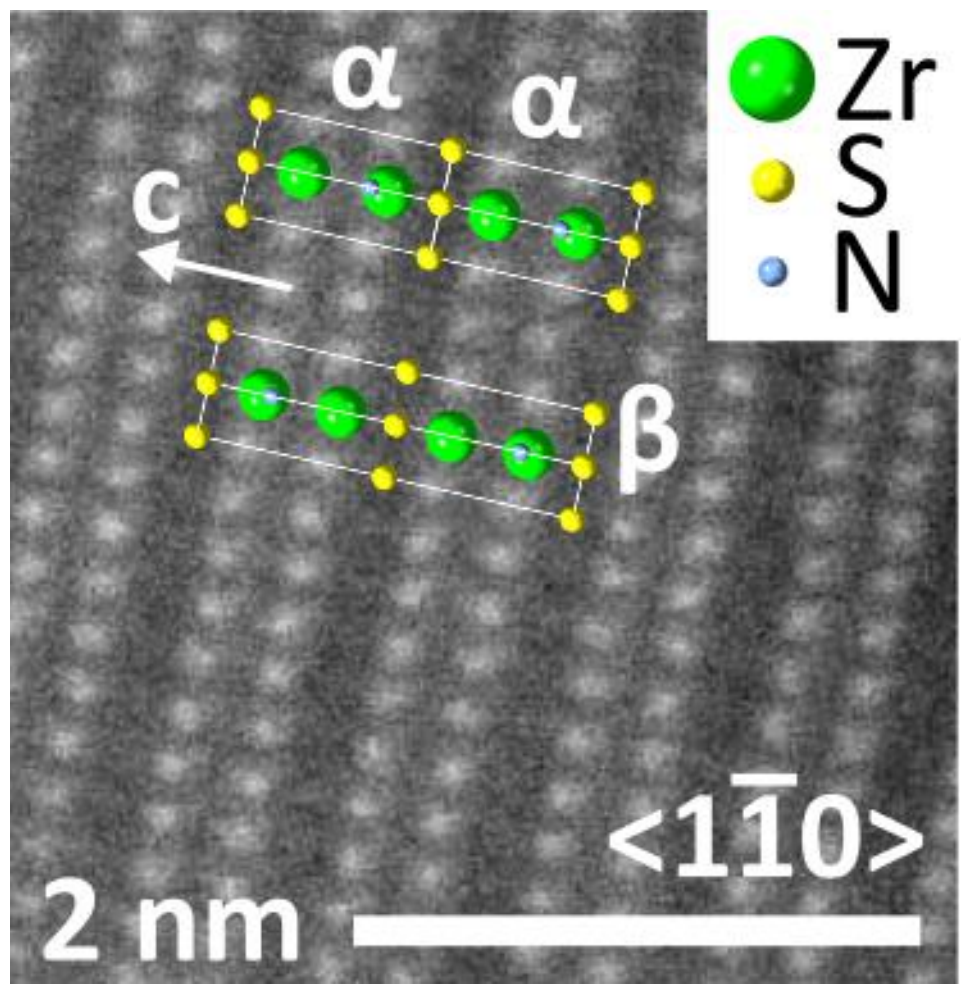

Figure S21: HR-STEM picture recorded on the film deposited on Si and annealed at 900 °C in vac. showing a crystallite displaying a α or β ordering in the <1$\bar{1}$0> zone axis. In this zone axis, α is virtually indistinguishable from β.

### *8.12 Raman spectrum analysis*

To enhance the Raman signal of the film, another film has been produced by annealing for a longer time of 30 min, as opposed to 5 min for all the other films deposited on Si in this work (Table S8). We first verified by means of GIXRD that the two films exhibited the same structure, as shown in Figure S22. Figure S23 shows the Raman spectra of the film annealed for 30 min, alongside the reference Raman mode position and labels as taken from the Computational Raman Database [21], entries mp-11583 and mp-1158, for α and β respectively (see Table S11). Except for the peak originating from the Si substrates, all the observed peaks correspond to the modes of β-$Zr_2SN_2$. The splitting of the $E_g$ mode of higher symmetry α, into three $E_{g1}$ + $2E_{g2}$ for the lower symmetry β, as show in Figure S23. This demonstrates the presence of β in the film, without excluding α.

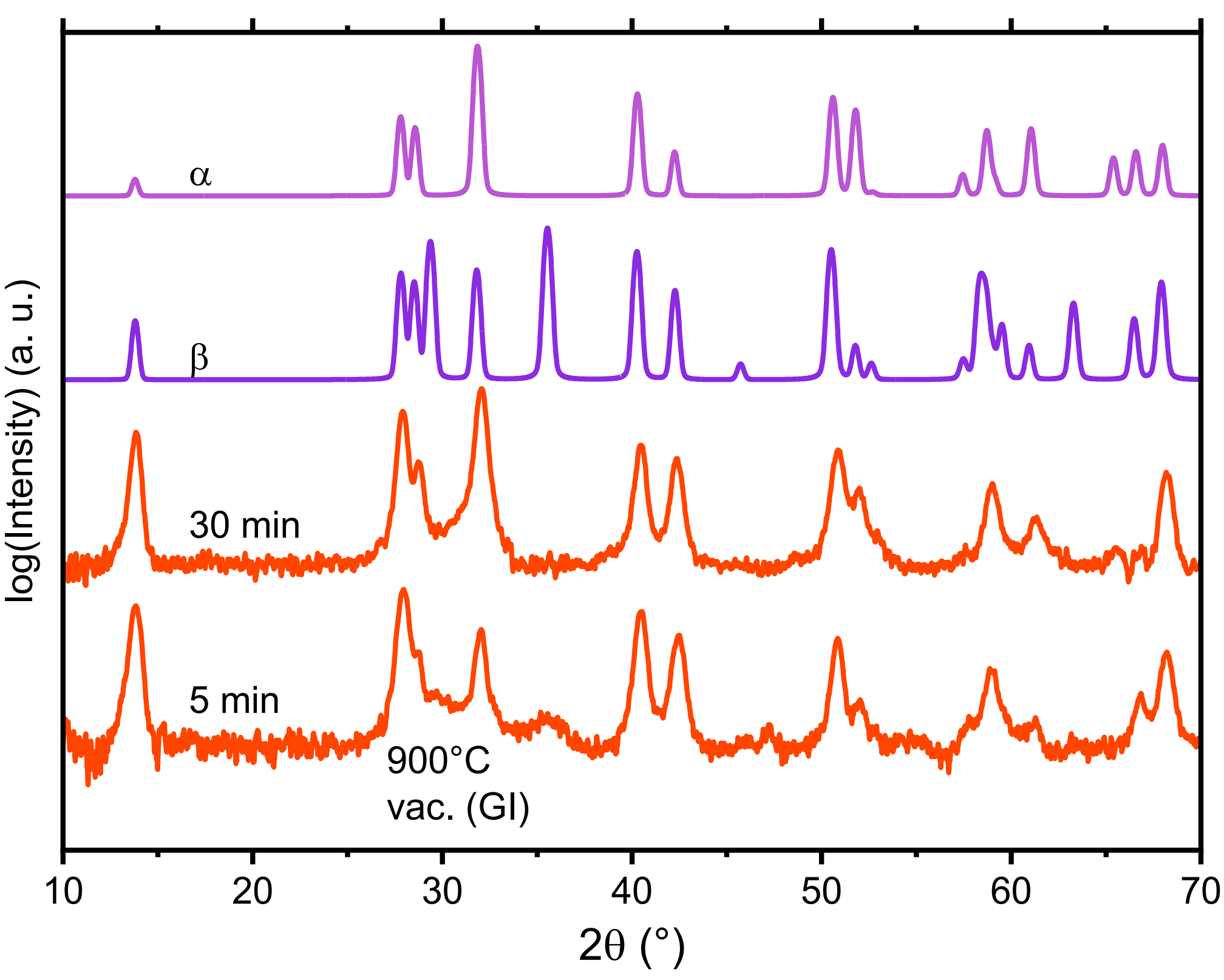

*Figure S22: GIXRD patterns of the films deposited on Si and annealed for 5min and 30min at 900 °C in vacuum, alongside theoretical patterns of α and β-$Zr_2SN_2$.*

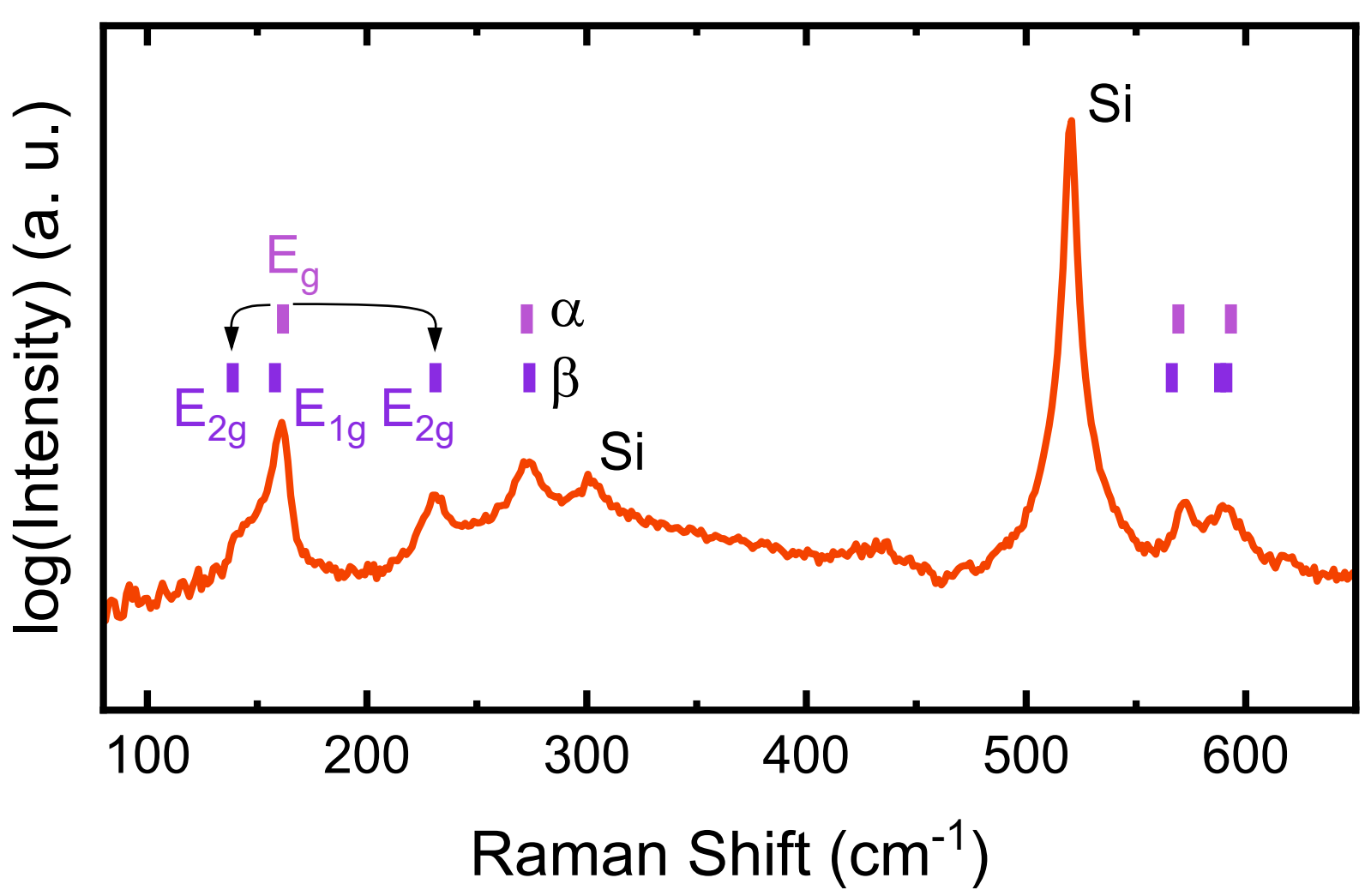

*Figure S23: Raman spectra of the sample deposited on Si and annealed at 900 °C for 30 min in vac. Reference Raman mode position as taken from the Computational Raman Database [21], entries mp-11583 and mp-1158, for α and β respectively. Refer to Table S11 for Raman modes' calculated position and labels.*

*Table S11: Reference Raman mode position and labels as taken from the Computational Raman Database [21], entries mp-11583 and mp-1158, for α and β respectively.*

| Experimental | β-$Zr_2SN_2$ | | α-$Zr_2SN_2$ | |
|---|---|---|---|---|
| Raman Shift ($cm^{-1}$) | Raman Shift ($cm^{-1}$) | Label | Raman Shift ($cm^{-1}$) | Label |
| 149 | 138.878 | $E_{2g}$ | | |
| 160 | 158.071 | $E_{1g}$ | 161.686 | $E_g$ |
| 231 | 231.124 | $E_{2g}$ | | |
| 273 | 273.872 | $A_{1g}$ | 272.675 | $A_{1g}$ |
| 572 | 566.37 | $A_{1g}$ | 569.263 | $A_{1g}$ |
| 590 | 588.238<br>591.078 | $E_{1g}$<br>$E_{2g}$ | 593.262 | $E_g$ |

## 8.13 Comments on the morphology

SEM and AFM top-view images of the films deposited on Si annealed at different temperatures as compared to the as grown sample. Apart from protrusions 10 to 50 nm in height, the as grown film is smooth and continuous with very small grains 10 nm in size, despite the absence of a diffraction signal from the film. Since the protrusions appear sporadically across samples (amorphous or crystalline) and measurement positions, we attribute them to non-ideal substrate preparation rather than to an intrinsic film growth/crystallization mechanism (Figure S26). The Ra (arithmetic average) and Rq (root mean squared) roughness are 0.5 nm and 0.7 nm, respectively when excluding the large protrusions from the roughness analysis, and 2 nm and 6 nm when including them. The roughness quantification using ellipsometry and XRR agrees well the AFM-deduced roughness in the case where the protrusions are included (Table S9, Figure S27). With the increasing annealing temperature, contrast becomes more apparent in the film, which correlates with the increasing crystallinity. At 900 °C annealing temperature; the contrast is maximum, and some “marbling” can be observed. We do not observe a drastic change in morphology with crystallization, and we measure the same Ra and Rq roughness of the underlying film (excluding the protrusions) for the 900 °C samples as for the as grown, 0.5 nm and 0.7 nm, respectively). This is confirmed by the absence of changes in ellipsometry-deduced roughness with annealing temperatures, as displayed in Table S9. We demonstrate that the crystallization of the film does not correlate with important morphological changes, yielding an extremely smooth film of sub-nm roughness.

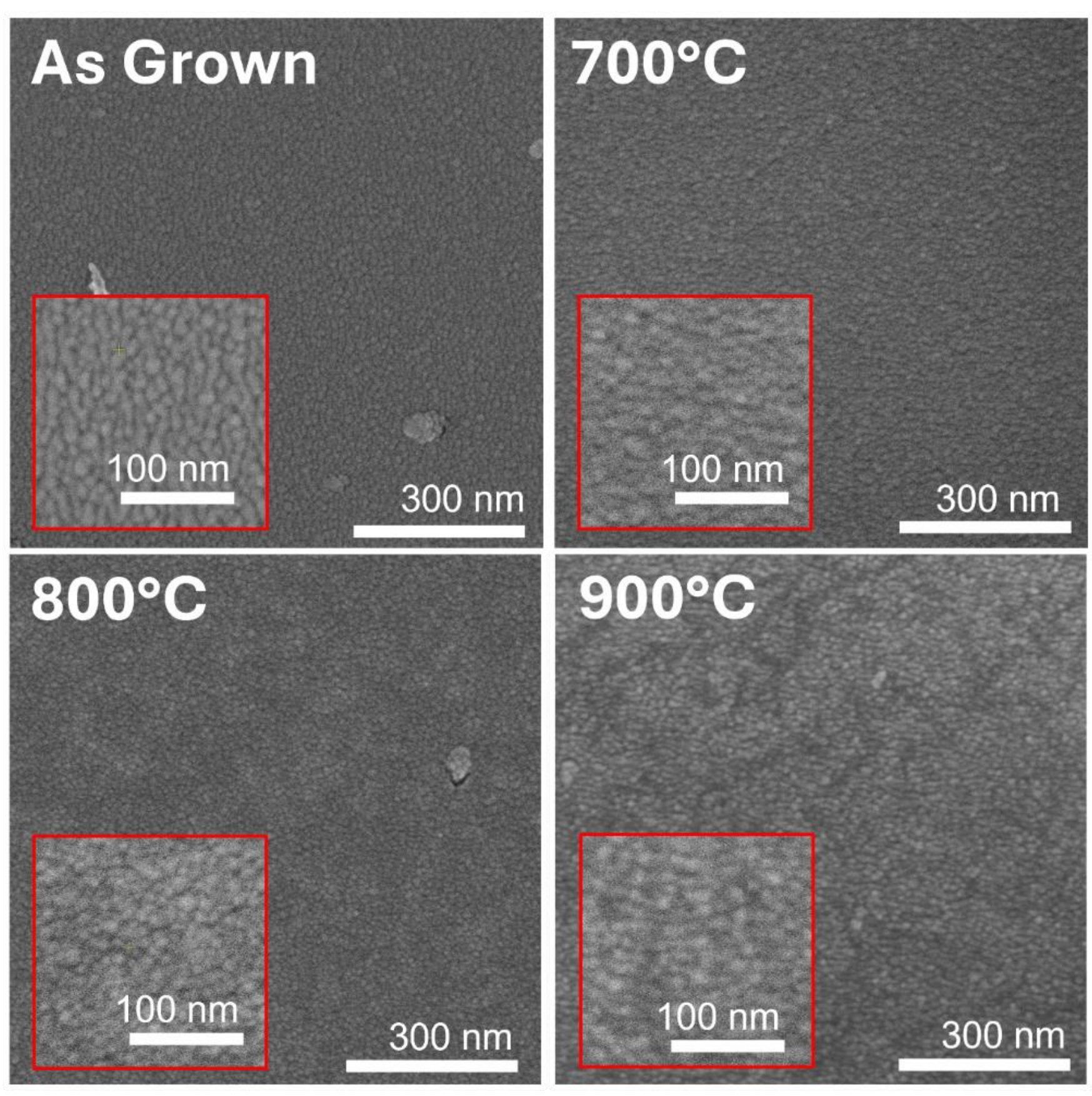

*Figure S24. SEM top view images of the as grown film deposited on Si as well as the films annealed from 700 °C to 900 °C at 500 Torr. Surface morphology was imaged using the Verious 5 UC scanning electron microscope from Thermofisher, at a working distance of 4mm, and acceleration voltage of 5 to 10 kV.*

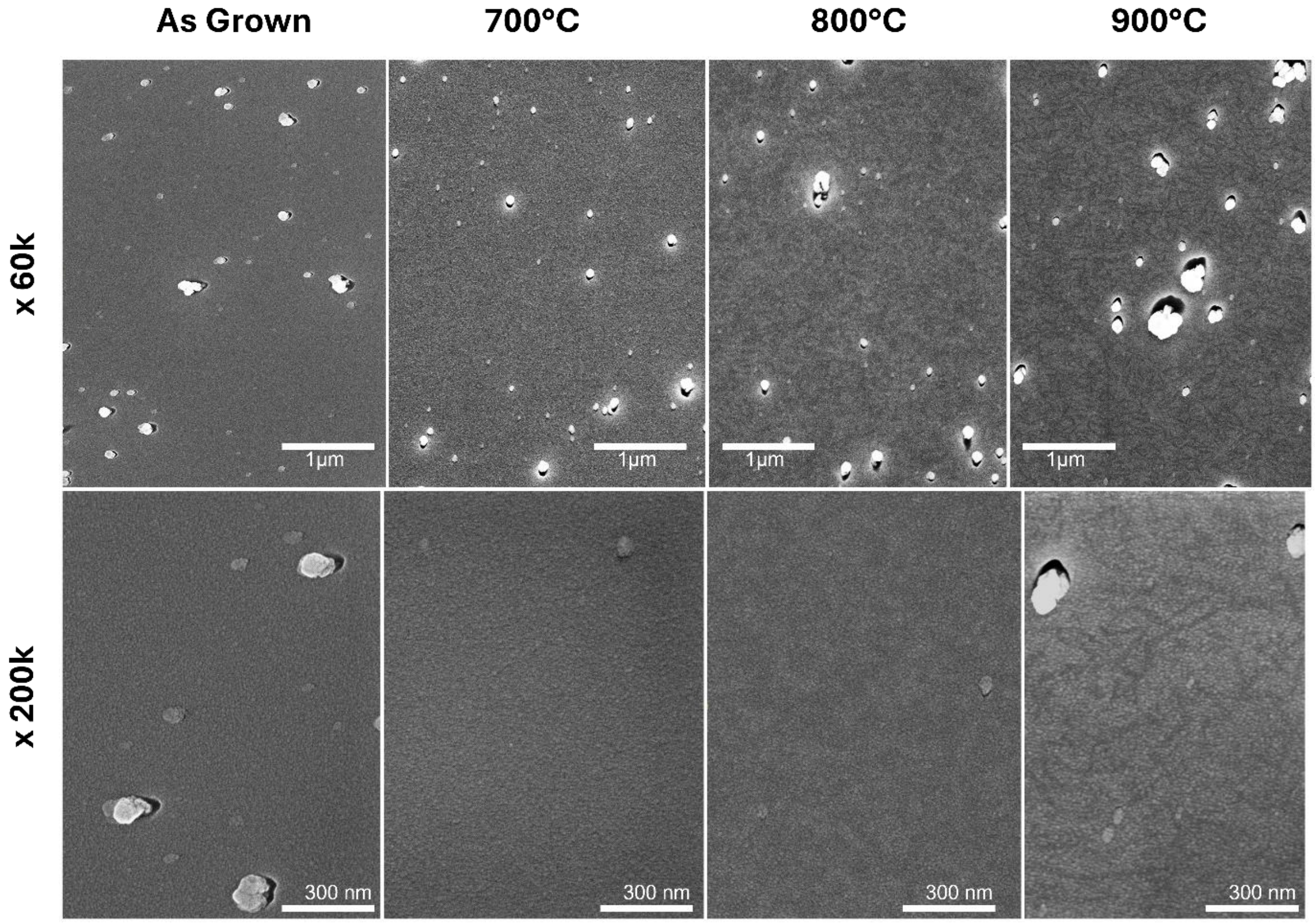

*Figure S25: Overview SEM top view images of the as grown film deposited on Si as well as the films annealed from 700 °C to 900 °C at 500 Torr. Surface morphology was imaged using the Verious 5 UC scanning electron microscope from Thermofisher, at a working distance of 4mm, and acceleration voltage of 5 to 10 kV.*

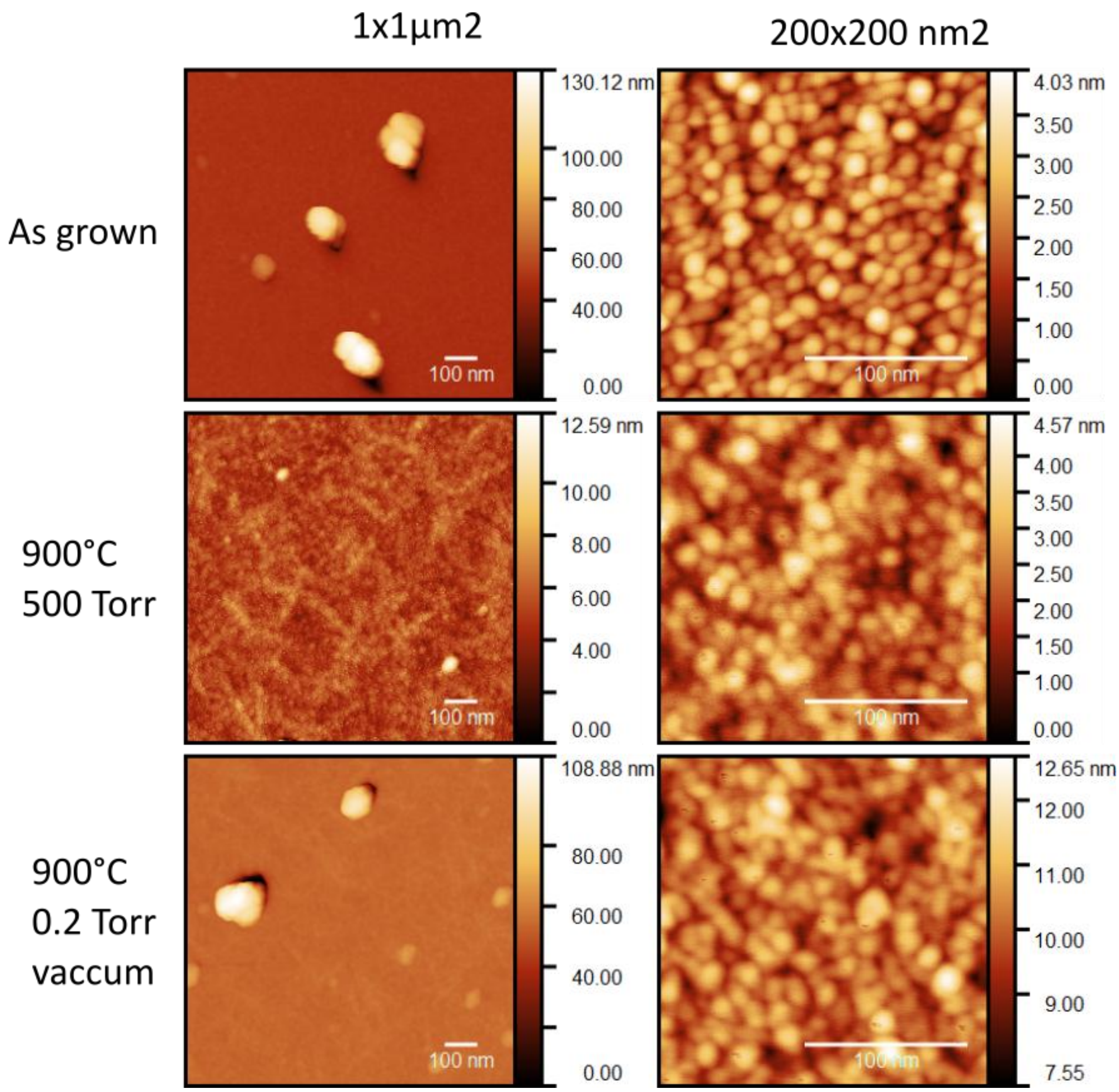

*Figure S26: (1x1) µm$^2$ and (0.2x0.2) µm$^2$ AFM scans of the as grown films on Si as well as the samples annealed at 900 °C in vacuum (0.2 Torr) or at 500 Torr.*

*8.14 XRR Study on the 900 °C, vac. sample*

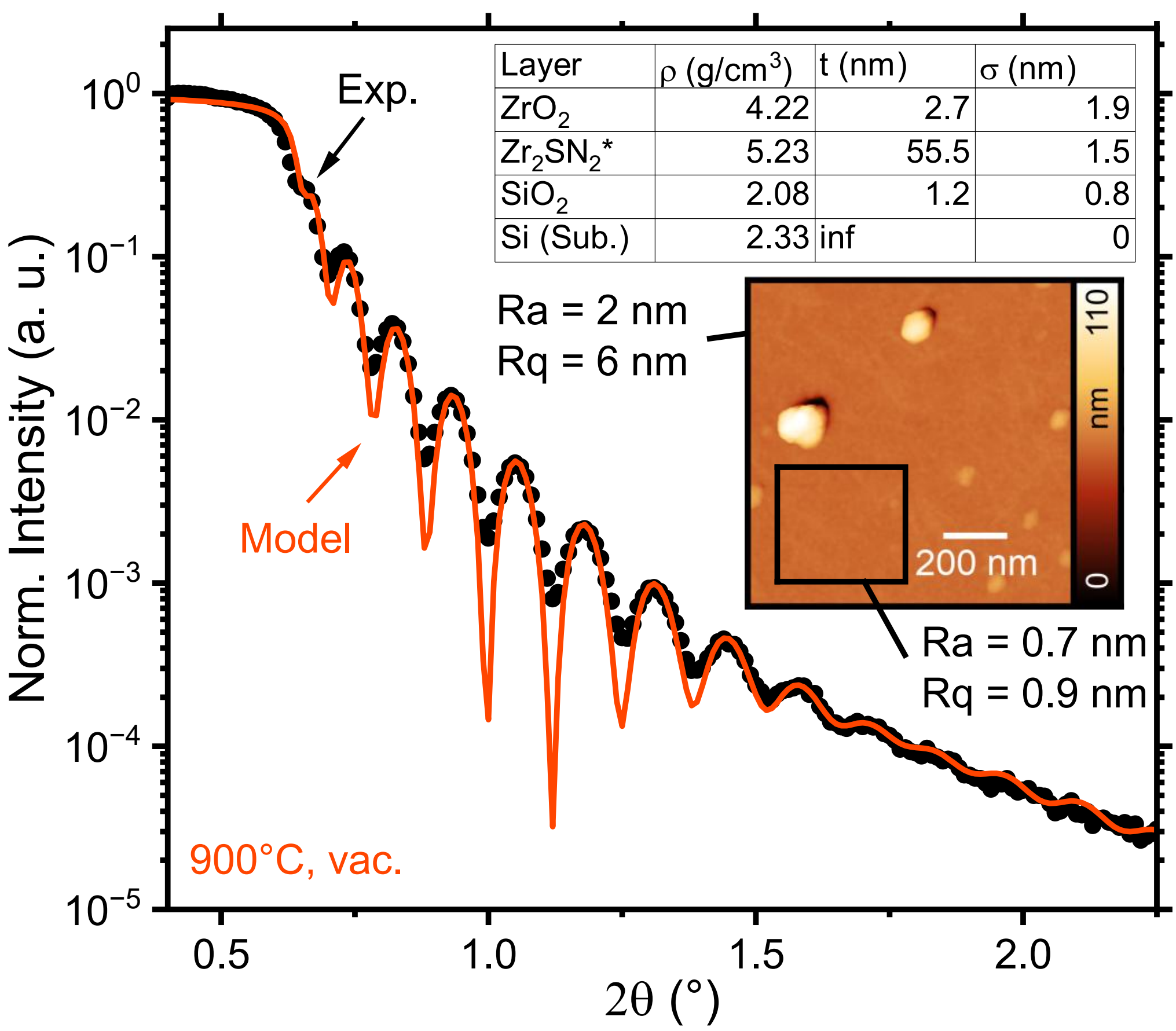

*Figure S27: Experimental X-ray reflectivity profile of the film deposited on Si and annealed at 900 °C in vac. and corresponding fit using Gen X3. Table Inset: Description of the 3-layer model and quantitative fit results with the density ρ, the thickness t and the roughness σ. *the Zr-S-N was assumed to have the composition extracted from SEM $Zr_{35}S_{18}N_{38}O_9$, although the label $Zr_2SN_2$ is chosen here for clarity. Picture Inset: (1 × 1) $\mu m^2$ AFM picture of the same sample. Extracted Ra and Rq roughness on the whole picture or on the specific area marked by a black rectangle.*

X-ray reflectivity was applied to the film deposited on Si and annealed at 900 °C at 0.2 Torr, and the reflectivity profile was fitted using the GenX software. The following assumptions were made:

1. The system can be modelled using a 3-layers system on a substrate (surface oxide / film / interface oxide / substrate)
2. The surface oxide is Zr oxide with the $ZrO_2$ composition. This is justified by the evidence for the surface oxide in the STEM-EDX data displayed in Figure S28 presence of a contribution in the XRD patterns 900 °C at 0.2 Torr
3. The interface oxide or substrate native oxide is a Si oxide with the $SiO_2$ composition
4. The film was assumed to have the composition extracted from LayerProbe EDX namely $Zr_{35}S_{18}N_{38}O_9$, as indicated by a * in the inset Table of Figure S27.
5. The substrate is perfectly flat with a density of 2.33 g/cm³

We measure a film density slightly lower than the theoretical density for α- and β-$Zr_2SN_2$, of 5.58, and 5.56 g/cm$^3$, respectively. Indeed for 0.2 Torr 900 °C-annealed, using XRR, we extract a density of 5.23 g/cm$^3$. This discrepancy may be explained by the overall Zr-poor stoichiometry for this film, Zr being by far the heaviest element in the system. Indeed, we calculate a Zr-to-anion ratio Zr/(S+N+O) of 0.54 instead of 0.66 for O-free stoichiometric $Zr_2SN_2$. The lower density would also be related to porosity in the film. We conclude that the measured density is compatible with the theoretical density of both α- and β-$Zr_2SN_2$.

### *8.15 Chemical composition maps in cross-section*

Spatially resolved composition mapping using STEM-EDX reveal that the film grown on Si and annealed at 900 °C at 0.2 Torr displays a high chemical homogeneity. Figure S28 and Figure S29 displays the STEM-EDX chemical profile extracted from the maps, for the as grown film and the film annealed at 900 °C in vacuum, respectively. While the relative changes in the elemental composition across the profile are robust against quantification errors, the absolute elemental compositions should be taken semi-quantitatively. Indeed, the thin lamella most likely oxidized during transfer from the FIB to the TEM, explaining the fact that more O is measured in STEM-EDX than in SEM-EDX using LayerProbe (see Table S7). As a result of the oxidation of the TEM lamella, the O content is very likely overestimated.

The O map reveals the thin Si native oxide and an oxide on the surface of the $Zr_2SN_2$ film, in accordance with GIXRD and XRR (Figure S16, Figure S27). We cannot rule out the presence of O in the bulk (Figure S28 and Figure S29); however, since the 9 at.% O content was determined assuming homogeneous distribution throughout the film (Table S7, LayerProbe), and interface oxides are present, the bulk O content in the annealed $Zr_2SN_2$ layer must be below 9 at.%. The film is homogenous in the substrate plane. Along the growth direction, we find that Zr, S and N are homogeneously distributed in the film, with no clear signs of phase segregation.

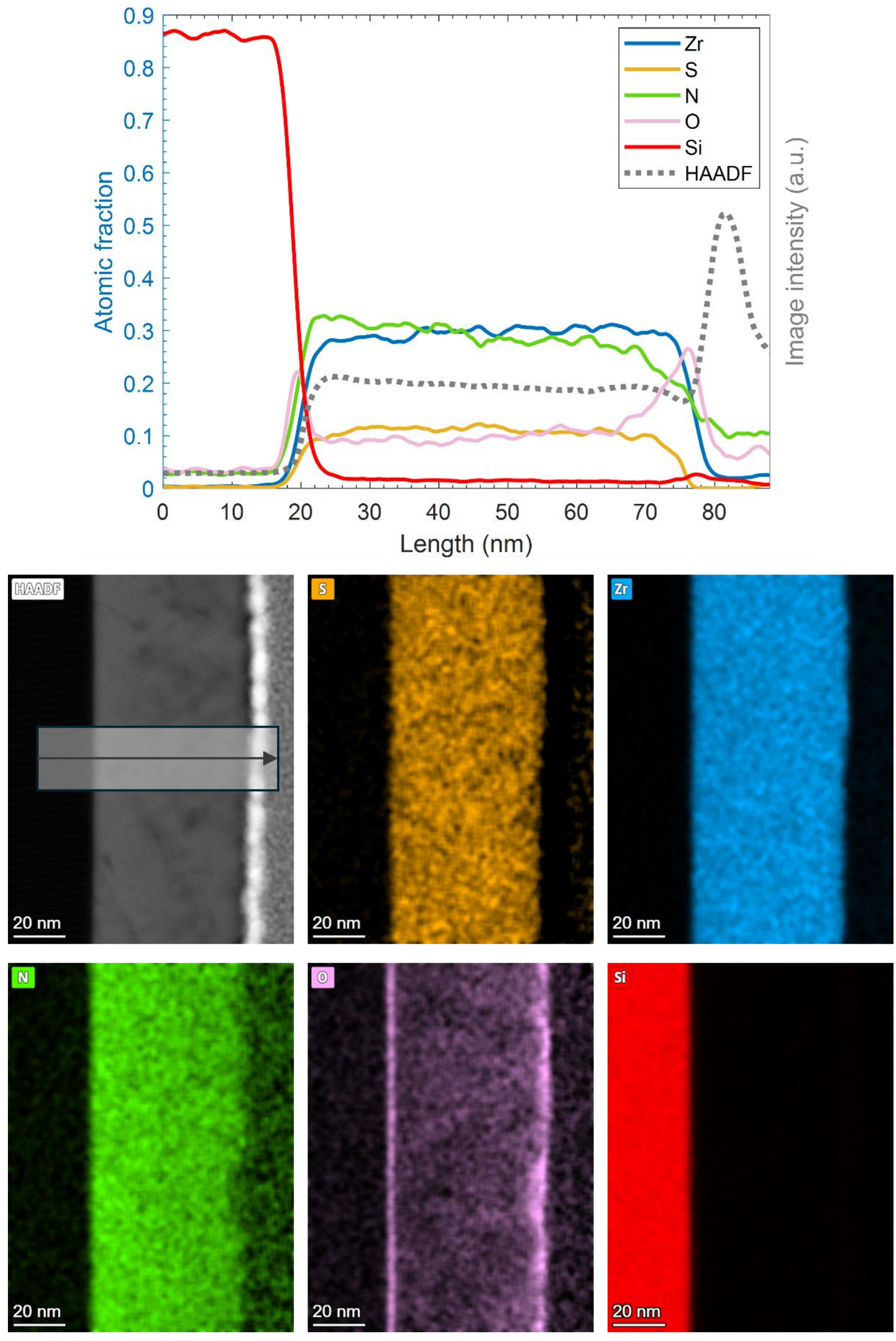

*Figure S28: STEM-EDX maps for the 900 °C vac. sample. Chemical profile along the growth direction, at the location marked by an arrow on the HAADF map.*

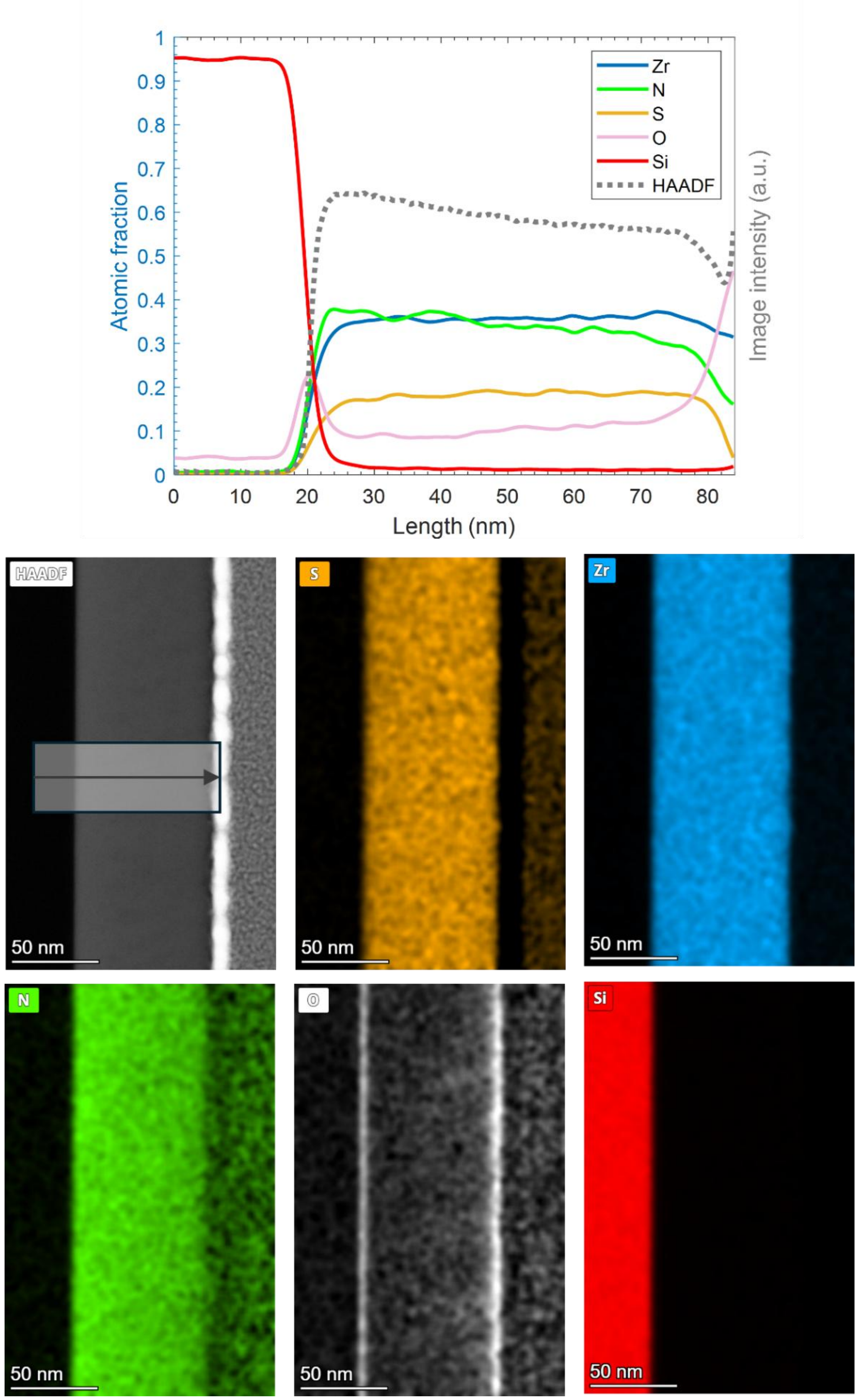

*Figure S29: STEM-EDX maps for the as grown sample on Si. Chemical profile along the growth direction, at the location marked by an arrow on the HAADF map.*

### *8.16 Validation of the ellipsometry analysis on Zr-S-N films grown on Si*

We fit the optical constants of the as grown and annealed samples deposited on Si using spectroscopic ellipsometry measurements. The data has been fitted with an agnostic Kramer-Kronig compliant B-spline model, with surface roughness (more details can be found in the Methods section).

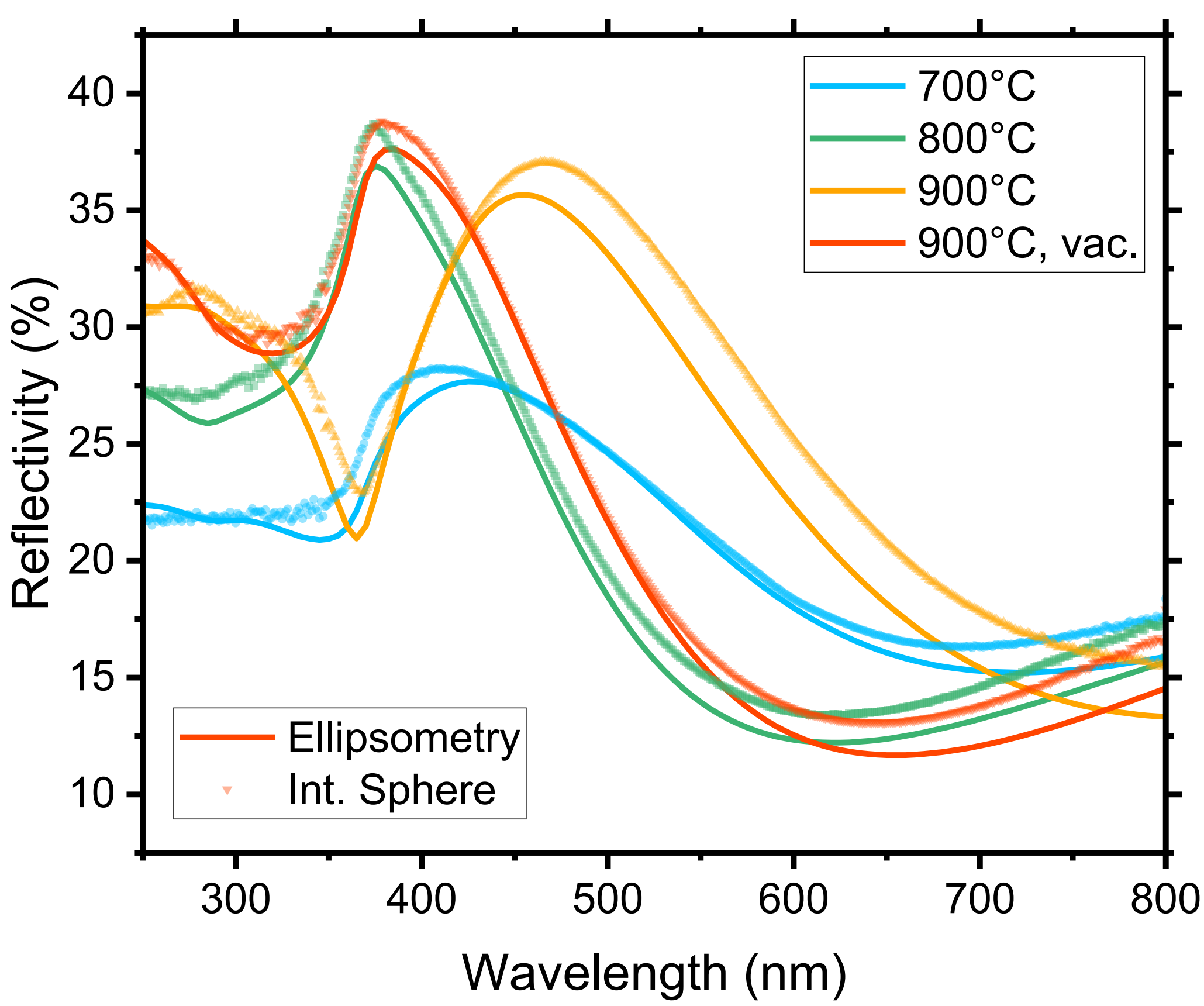

*Figure S30: Overlay of (i) measured reflectivity spectra using an integrating sphere and (ii) simulated reflectivity spectra using thickness and optical constants n and k extracted from ellipsometry measurements.*

Firstly, we find that the model accurately fits the ellipsometry data for all samples, as displayed in Figure S32. Secondly, using CompleteEASE, one can calculate the reflectivity profiles that the films should exhibit, based on the ellipsometry-fitted optical constants, thickness and roughness; we find a good match between the calculated reflectivity profile and the measured reflectivity using an integrating sphere, as shown in Figure S30. Thirdly, the thickness extracted from the model matches the thickness measured by TEM and XRR (Table S9,Figure S27). All three arguments above demonstrate that the optical constants extracted from ellipsometry are accurate. Thickness and roughness deduced from ellipsometry are reported in Table S7.

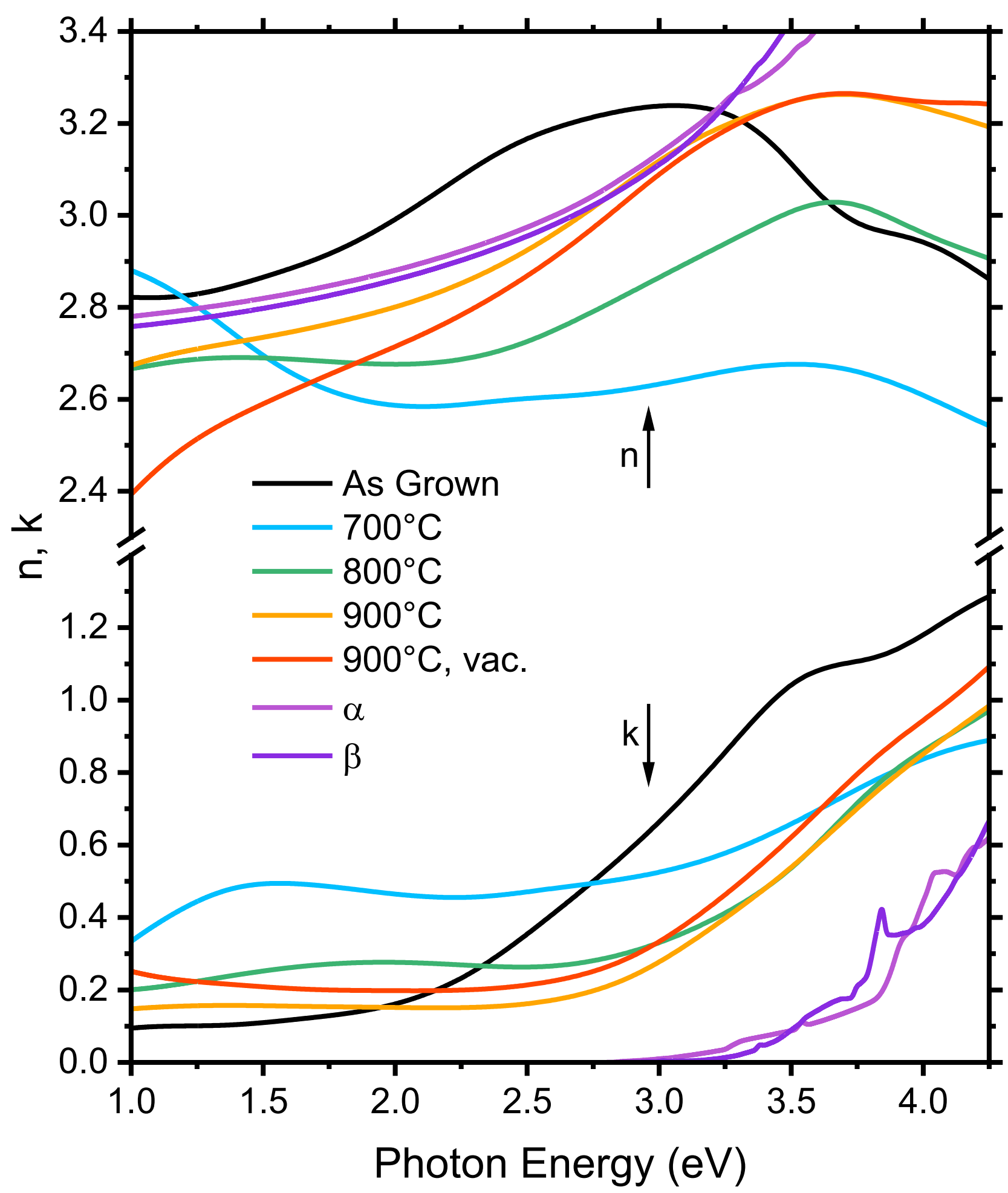

*Figure S31: Experimental n and k spectra of all samples grown on Si as extracted from ellipsometry. Calculated n and k spectra for α and β-$Zr_2SN_2$ as calculated by HSE06.*

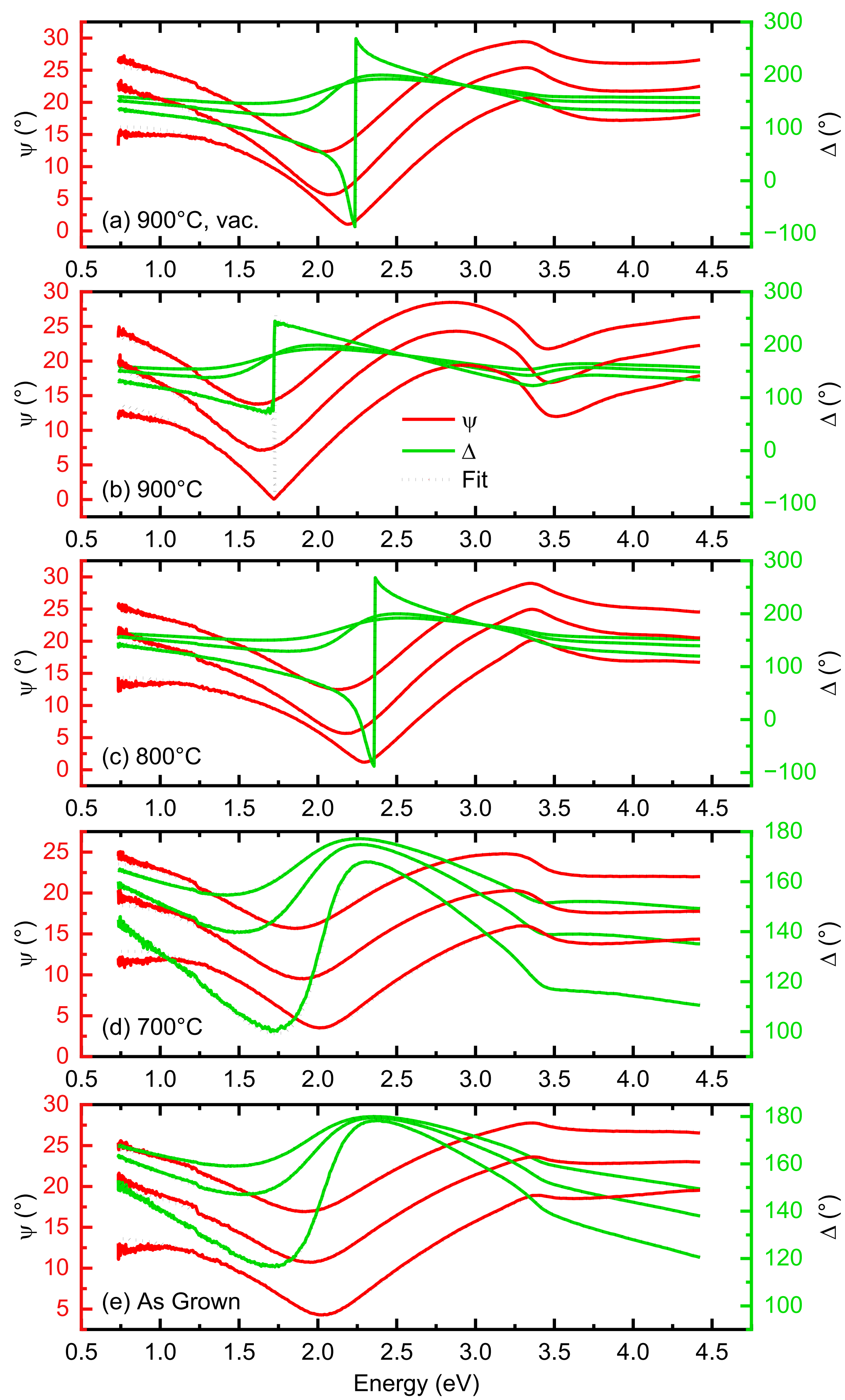

*Figure S32: Measured Δ and ψ ellipsometric quantities at 55, 60 and 65° and corresponding fit using the CompleteEASE software, for all samples grown on Si.*

The annealing drastically modifies the optical properties of the films grown on Si. As display in Figure S31, comparing the as grown sample and 700 °C-annealed sample, one can see a drastic change in both the absorption coefficient spectra and the index of refraction. Firstly, the crystallinity does not change between the two samples as stated earlier. Secondly, the composition is strongly modified by the annealing at 700 °C, namely strong N incorporation and S loss. Therefore, we argue that loss/gain in gaseous species without crystallization and reorganization of the lattice has created many defects, which are detected here. With increasing temperature from 700 °C to 900 °C, one can see that the absorption coefficient morphs relatively smoothly into a more defined absorption profile with an approximate absorption onset around 2.75-3.25 eV. A similar observation is made of the 900 °C vacuum process. Comparing the onset of crystalline samples (800-900 °C) in range 2.75-3.25 eV to the absorption onset of the as grown sample in the range 2.0-2.5, one can conclude that the optical absorption onset is shifted to higher energy with the crystallization of the samples.

## 8.17 Bandgap estimation using Tauc Plots

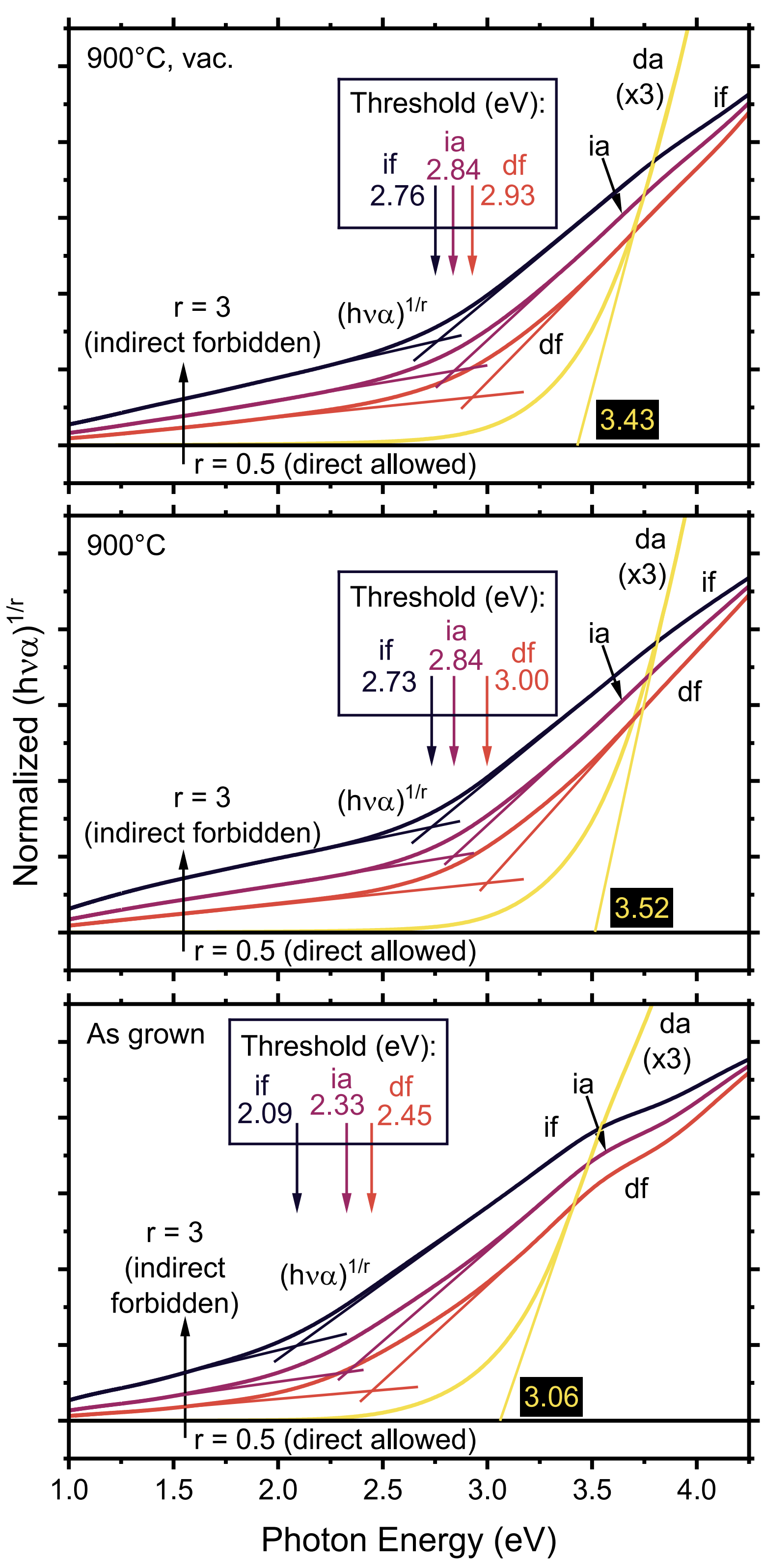

*Figure S33: Tauc plot quantities $(h\nu E)^{1/r}$ for different values of r: direct allowed da r = 1/2, direct forbidden df r = 2/3, indirect allowed ia r = 2, indirect forbidden if r = 3.*

We find that the Tauc method fails in determining a precise onset for the absorption. On three different films grown on Si, we find that the absorption edge extracted using different exponent 1/r for the tauc quantity $(h\nu E)^{1/r}$, is different, demonstrating the limitation of this analysis. This may stem from the fact that both α and β have $Zr_2SN_2$ rather complicated optical properties with dipole forbidden and/or indirect transitions. Furthermore, we have shown (Figure 3, Figure S*20*, Figure S*21*) that some fractions of the films consists composite structure with a stacked sequence of α and β at the atomic scale and that there might be some leftover, uncrystallized amorphous film, which may also contribute in contradicting the base assumptions of absorption edge via the Tauc method.

### *8.18 Additional Data on the sample on fused silica annealed in vac. at 900 °C*

The film deposited on fused silica and annealed in vac. (0.2Torr) at 900 °C for 30min (see Table S8) was characterized by ellipsometry, XRR, EDX and Reflection/Transmission spectroscopy, and Hall effect. The composition determined by EDX is Zr:S:N, 40:22:38, very close to the theoretical 40:20:40 stoichiometry. We do not report the O content for this sample as this information cannot be accessed easily due to the presence of O in the $SiO_2$ substrate.

Figure S34 shows the ellipsometry fit for this sample. A good fit is achieved and the index of refraction as well as the absorption coefficient are reported in Figure 5 of the main article. A thickness of 99nm and a roughness of 3nm are extracted.

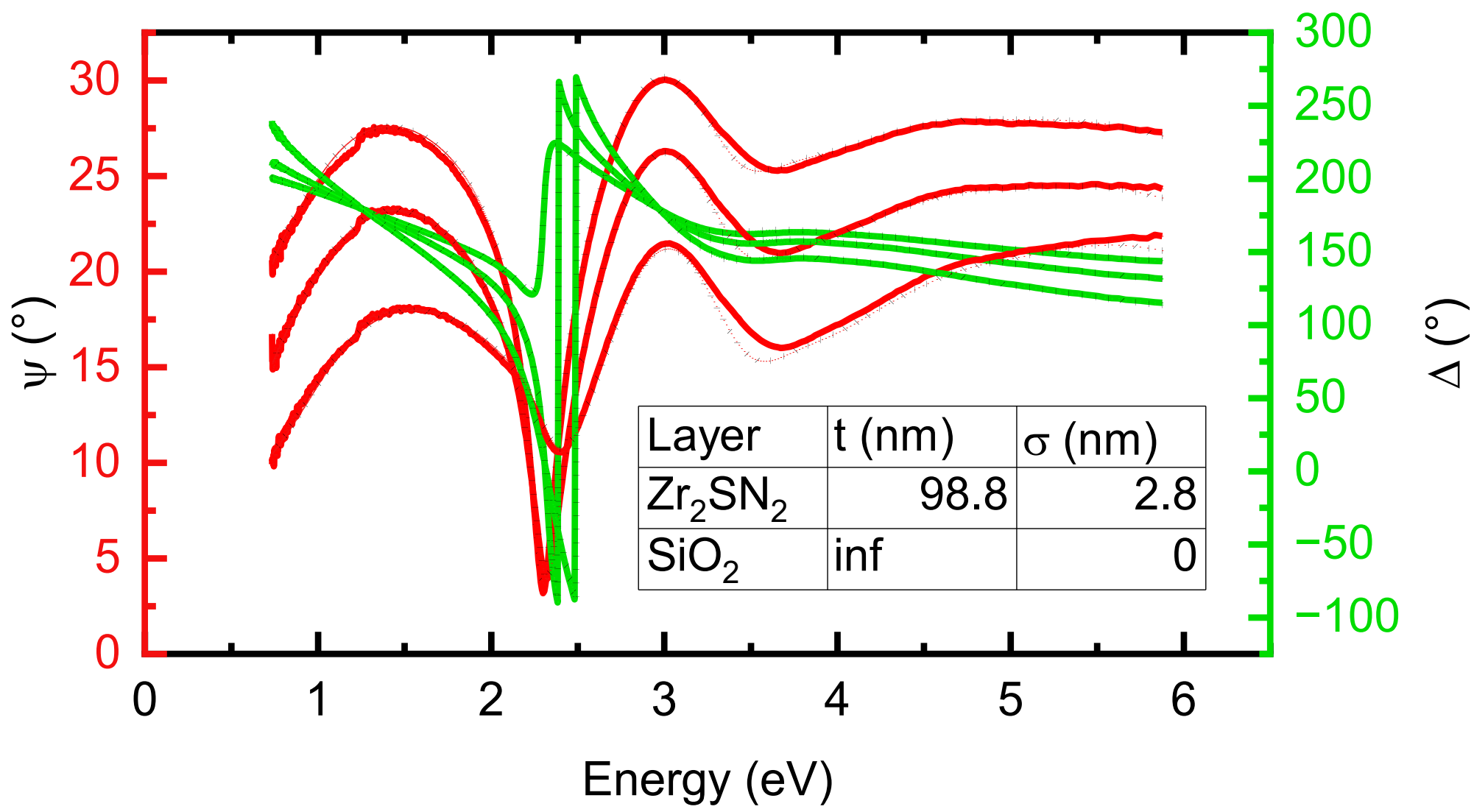

*Figure S34: Raw data and fit using a Kramers-Kronig compliant B-spline model of Δ and ψ for the sample on fused silica annealed 30min in vac.*

An XRR profile was measured and fitted, as shown in Figure S35. The fit is suboptimal for 2θ angles from 0.5 to 1°. On the contrary the fit is accurate above 1°. We explain this by the presence of a thickness gradient for this sample. Indeed, for different 2θ angles, the beam print is different, and therefore different sample locations with different thicknesses, density and roughness may simultaneously contribute to the XRR profile, which contradicts the geometrical model used, i.e. single layer homogenous layer on a substrate. At higher angle, the beam print is smaller, and the region probed by the X-rays is more homogeneous, giving a better fit. We argue that fit accuracy at higher 2θ is preferred in this case to extract accurate information. We extract a film thickness of 92nm and roughness of 2.2, in good accordance with the ellipsometry.

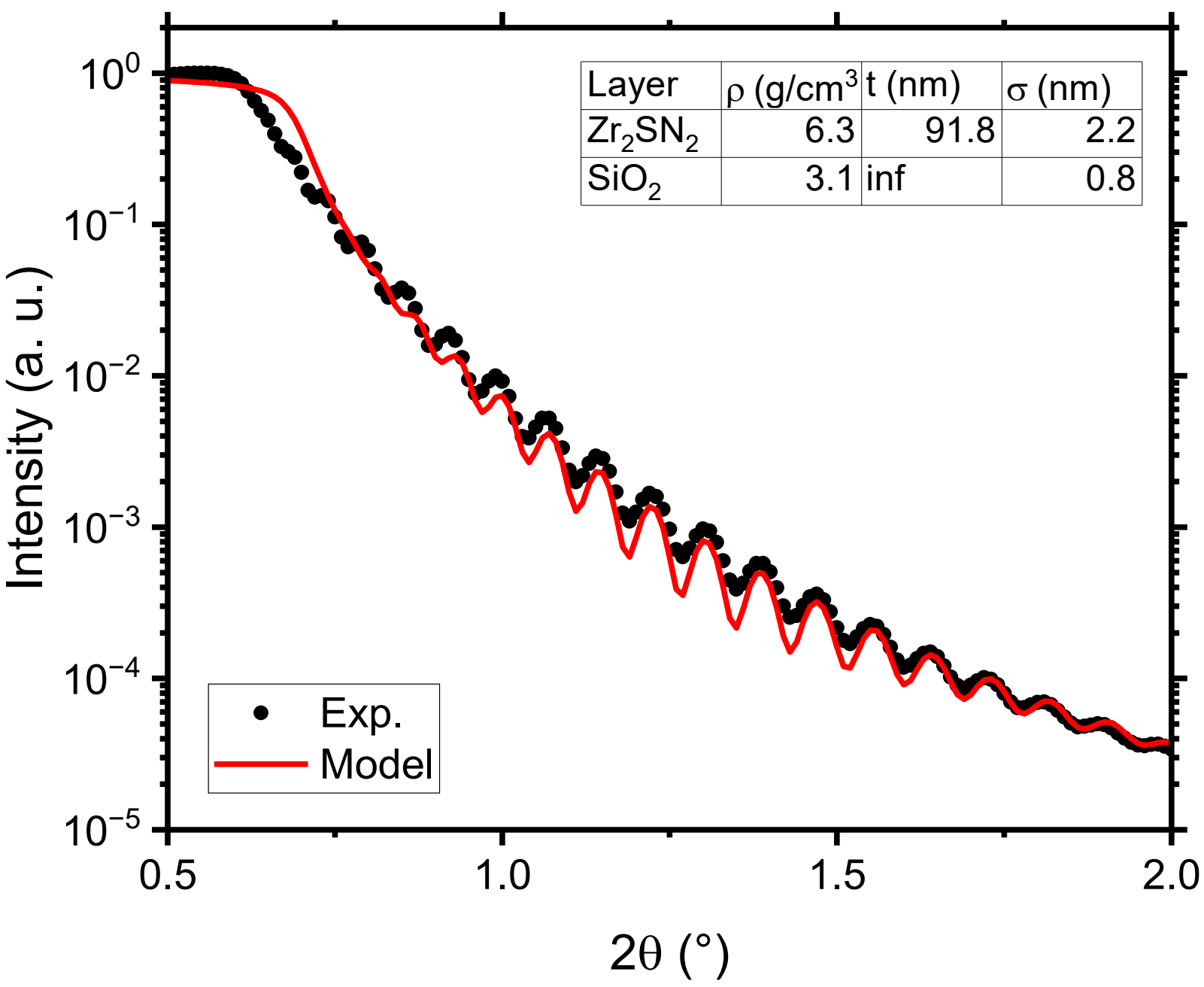

*Figure S35: XRR profile for the film grown on fused silica and annealed 30min in vac.*

Figure S36 display the comparison between the measurements of Reflection and Transmission spectra (UV-vis-NIR spectroscopy) and the corresponding simulated spectra using CompleteEASE using the optical constants and thickness deduced from ellipsometry. The Absorption A is calculated as $A = 1 - R - T$. A good match is achieved between the two techniques, demonstrating the validity of the ellipsometry analysis for the sample annealed at 900 °C for 43min, on fused silica.

To model the dynamics of free carriers in the film grown on fused silica and annealed for 30min, we construct a model consisting of a Drude oscillator (intraband transitions by free carriers) and a Tauc-Lorentz oscillator (interband transitions). We ensure that the model is consistent with a Kramers-Kronig B-spline analysis of the same data (Figure S38). We extract the specific contribution of free carriers (Drude) to n and k (Figure S39).

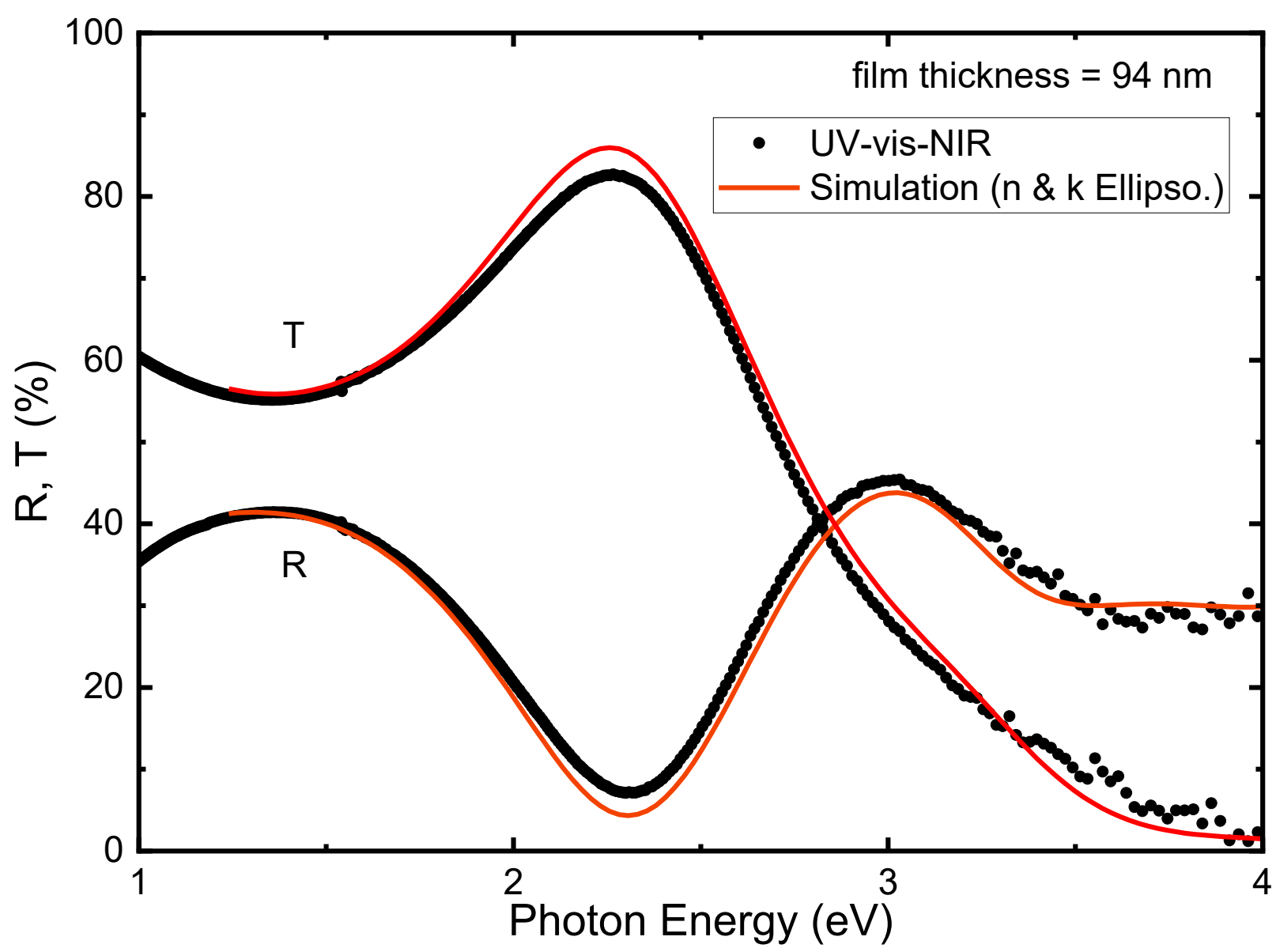

*Figure S36: Overlay of (i) measured Reflection R and Transmission T spectra using UV-vis-NIR spectroscopy and (ii) simulated Reflection R and Transmission T spectra using thickness and optical constants n and k extracted from ellipsometry measurements, for the sample deposited on fused silica and annealed at 900 °C for 30min.*

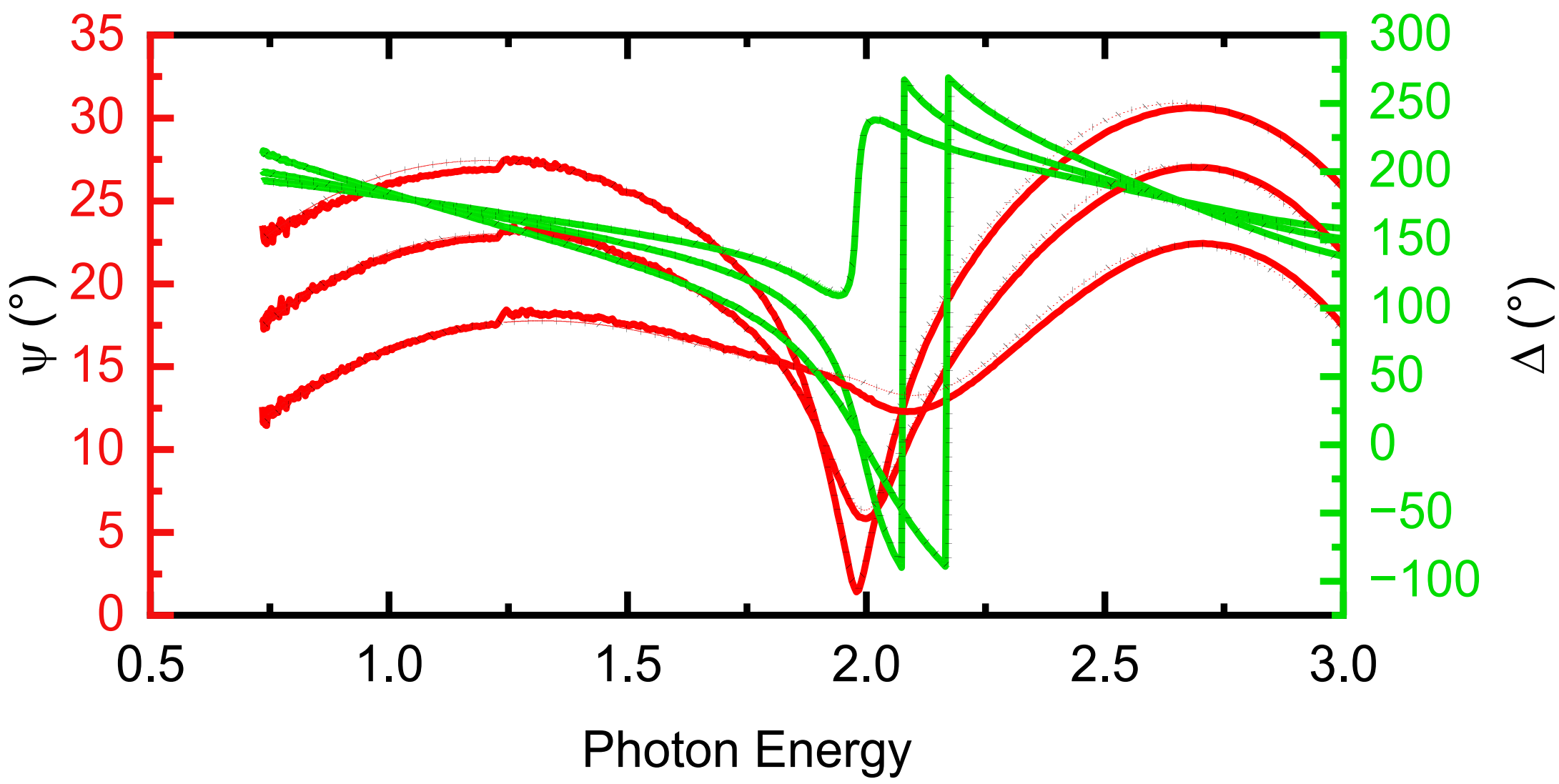

*Figure S37: Fit of the ellipsometric response of the film grown on $SiO_2$ and annealed at 900 °C for 30min at the location of the Hall effect measurement. The model consists in one Drude oscillator and one Tauc-Lorentz oscillator.*

*Table S12: Comparison table of electrical properties of the film grown on $SiO_2$ and annealed at 900 °C for 30min. The film thickness for Hall measurement was extracted from ellipsometry, and the charge carrier effective mass used as input for the ellipsometry Drude model was calculated by DFT-HSE06. We report the resistivity and Hall carrier concentration and mobility based on DC field Hall and on corrected*

*FastHall™ using the two different polarities of the permanent magnet (North and South). The different values are within reasonable agreement.*

| | Ellipsometry Drude (AC) | Hall (DC field) | FastHall™ (corrected) |
|---|---|---|---|
| Thickness (nm) | 115 | 115 (input) | 115 (input) |
| Carrier density ($10^{20}$ $cm^{-3}$) | 3.18 | 2.3 ± 0.2 | 3.4± 0.1 |
| Mobility ($cm^2V^{-1}s^{-1}$) | 8.3 | 0.36 ± 0.03 | 0.33 ± 0.01 |
| Resistivity (mΩ.cm) | 2.4 | 75.8 | 75.8 |
| Scat. Time (fs) | 2.2 | | |
| Electron effective mass | 0.46 (input) | | |

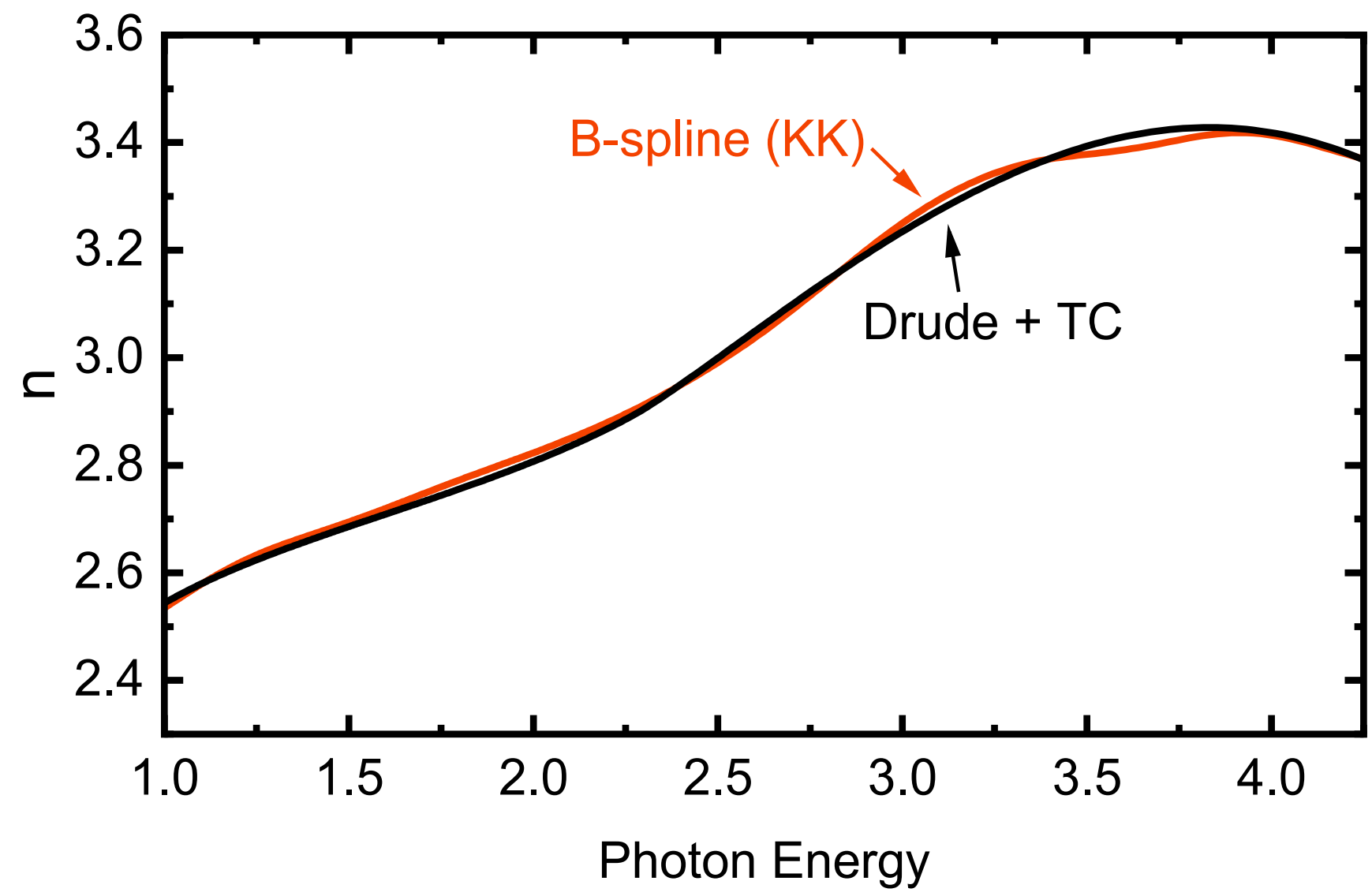

*Figure S38: Comparison of the optical constants of the film grown on $SiO_2$ and annealed at 900 °C for 30min extracted using a Kramers-Kronig compliant B-spline model vs. a one Drude oscillator + one Tauc Lorentz oscillator model.*

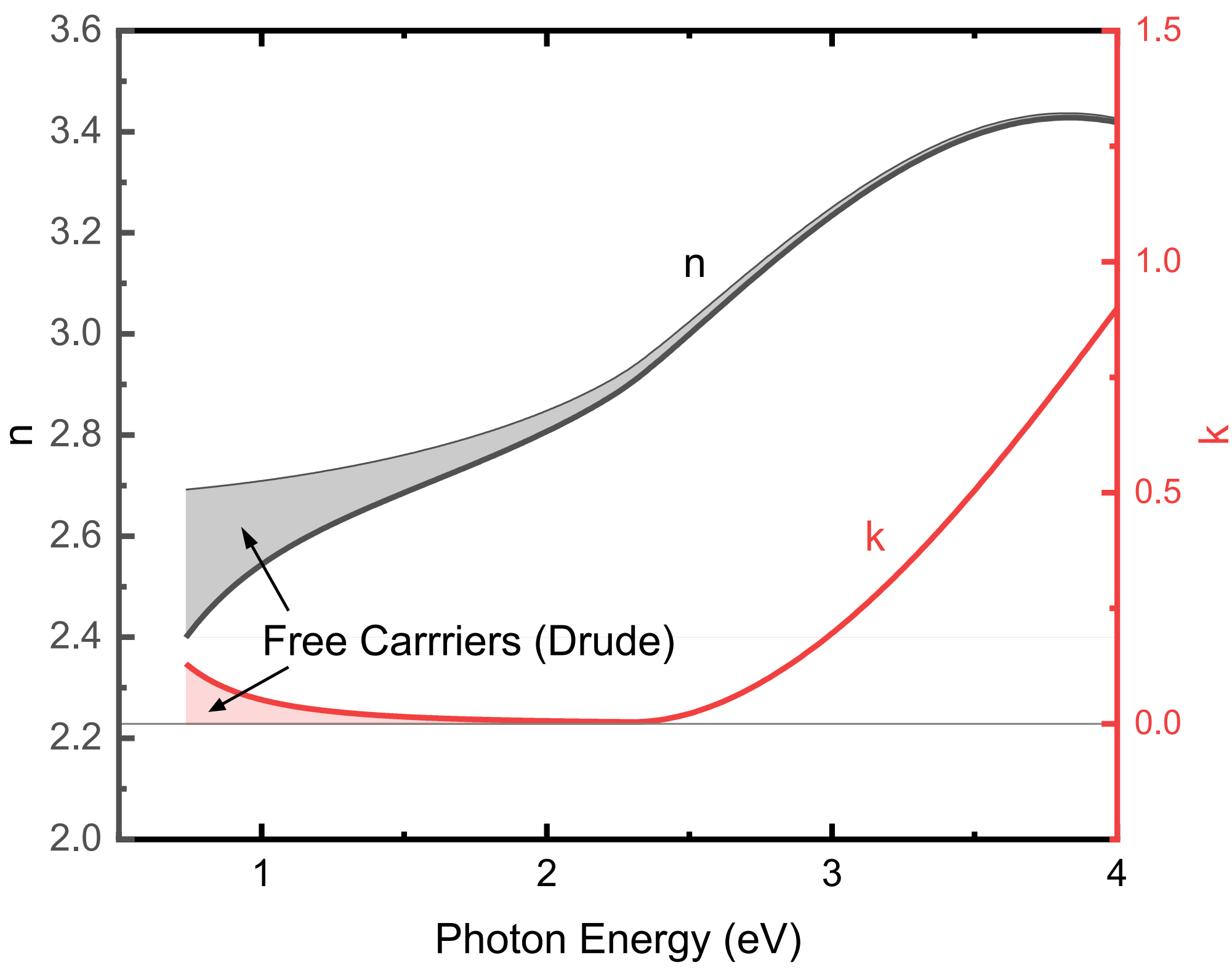

*Figure S39: n and k optical constants of the film grown on $SiO_2$ and annealed at 900 °C for 30min highlighting the specific contribution of free carriers modelled using the Drude model.*